\pdfoutput=1
\documentclass[fleqn, usenatbib]{mnras}

\newif\iflocal
\localfalse

\usepackage[T1]{fontenc}
\usepackage{ae,aecompl}
\usepackage{graphicx}
\usepackage{xspace}
\usepackage{amsmath}
\usepackage{txfonts}
\usepackage{epstopdf}
\usepackage{color}
\usepackage{rotating}
\usepackage{natbib}
\usepackage{ulem}
\usepackage{xspace}
\usepackage{sistyle}
\SIthousandsep{,}

\iflocal
\def\includedir{/Users/benedito/University/docs/latex}
\def\figdir{figs}
\else
\def\includedir{.}
\def\figdir{.}
\fi

\newcommand{\eqmn}[1]{equation~(\ref{#1})\xspace}

\def\gtsima{$\; \buildrel > \over \sim \;$}
\def\ltsima{$\; \buildrel < \over \sim \;$}
\def\prosima{$\; \buildrel \propto \over \sim \;$}
\def\gsim{\lower.7ex\hbox{\gtsima}}
\def\lsim{\lower.7ex\hbox{\ltsima}}
\def\simgt{\lower.7ex\hbox{\gtsima}}
\def\simlt{\lower.7ex\hbox{\ltsima}}
\def\simpr{\lower.7ex\hbox{\prosima}}

\newcommand{\kpch}{{h^{-1}{\rm kpc}}}
\newcommand{\mpch}{h^{-1}{\rm {Mpc}}}

\newcommand{\gyr}{{\rm Gyr}}

\newcommand{\msun}{{M_{\odot}}}
\newcommand{\msunh}{h^{-1} M_\odot}

\newcommand{\mstar}{M_{\ast}}

\newcommand{\tdyn}{t_{\rm dyn}}

\newcommand{\LCDM}{$\Lambda$CDM\xspace}

\def\sparta{\textsc{Sparta}\xspace}
\def\moria{\textsc{Moria}\xspace}

\def\colossus{\textsc{Colossus}\xspace}

\def\rockstar{\textsc{Rockstar}\xspace}
\def\consistenttrees{\textsc{Consistent-Trees}\xspace}
\def\subfind{\textsc{SubFind}\xspace}

\def\wmap{WMAP7\xspace}

\def\erebos{Erebos\xspace}

\def\rmd{{\rm d}}

\def\rmm{{\rm m}}

\def\rmp{{\rm p}}

\def\rmH{{\rm H}}

\def\calA{{\cal A}}

\def\vmax{V_{\rm max}}

\def\mvir{M_{\rm vir}}
\def\rvir{R_{\rm vir}}

\def\mtom{M_{\rm 200m}}
\def\rtom{R_{\rm 200m}}

\def\vtom{V_{\rm 200m}}
\def\ntom{N_{\rm 200m}}

\def\mtoc{M_{\rm 200c}}
\def\rtoc{R_{\rm 200c}}

\def\mfoc{M_{\rm 500c}}

\usepackage{etoolbox}
\makeatletter
\patchcmd{\NAT@citex}
  {\@citea\NAT@hyper@{\NAT@nmfmt{\NAT@nm}\NAT@date}}
  {\@citea\NAT@nmfmt{\NAT@nm}\NAT@hyper@{\NAT@date}}
  {}
  {}
\patchcmd{\NAT@citex}
  {\@citea\NAT@hyper@{%
     \NAT@nmfmt{\NAT@nm}%
     \hyper@natlinkbreak{\NAT@aysep\NAT@spacechar}{\@citeb\@extra@b@citeb}%
     \NAT@date}}
  {\@citea\NAT@nmfmt{\NAT@nm}%
   \NAT@aysep\NAT@spacechar%
   \NAT@hyper@{\NAT@date}}
  {}
  {}
\patchcmd{\NAT@citex}
  {\@citea\NAT@hyper@{%
     \NAT@nmfmt{\NAT@nm}%
     \hyper@natlinkbreak{\NAT@spacechar\NAT@@open\if*#1*\else#1\NAT@spacechar\fi}%
       {\@citeb\@extra@b@citeb}%
     \NAT@date}}
  {\@citea\NAT@nmfmt{\NAT@nm}%
   \NAT@spacechar\NAT@@open\if*#1*\else#1\NAT@spacechar\fi%
   \NAT@hyper@{\NAT@date}}
  {}
  {}
\makeatother

\iflocal
\def\figdir{figs}
\else
\def\figdir{.}
\fi

\defcitealias{diemer_13_scalingrel}{DKM13}
\defcitealias{diemer_14}{DK14}
\defcitealias{diemer_15}{DK15}

\title[Haunted haloes]{Haunted haloes: tracking the ghosts of subhaloes lost by halo finders}

\author[Diemer et al.]{Benedikt Diemer,$^{1}$\thanks{Email: \href{mailto:diemer@umd.edu}{diemer@umd.edu}}
Peter Behroozi,$^{2,3}$
Philip Mansfield$^{4,5,6}$\vspace{1mm}
\\
$^{1}$ Department of Astronomy, University of Maryland, College Park, MD 20742, USA;\\
$^{2}$ Department of Astronomy and Steward Observatory, University of Arizona, Tucson, AZ 85721, USA \\
$^{3}$ Division of Science, National Astronomical Observatory of Japan, 2-21-1 Osawa, Mitaka, Tokyo 181-8588, Japan\\
$^{4}$ Kavli Institute for Particle Astrophysics and Cosmology and Department of Physics, Stanford University, Stanford, CA 94305, USA\\
$^{5}$ SLAC National Accelerator Laboratory, Menlo Park, CA, 94025\\
$^{6}$ Department of Physics, Stanford University, 382 Via Pueblo Mall, Stanford, CA 94305, USA
}

\date{}
\pubyear{2023}

\begin{document}
\label{firstpage}
\pagerange{\pageref{firstpage}--\pageref{lastpage}}
\maketitle


\begin{abstract}
Dark matter subhaloes are key for the predictions of simulations of structure formation, but their existence frequently ends prematurely due to two technical issues, namely numerical disruption in $N$-body simulations and halo finders failing to identify them. Here we focus on the second issue, using the phase-space friends-of-friends halo finder \rockstar as a benchmark (though we expect our results to translate to comparable codes). We confirm that the most prominent cause for losing track of subhaloes is tidal distortion rather than a low number of particles. As a solution, we present a flexible post-processing algorithm that tracks all subhalo particles over time, computes subhalo positions and masses based on those particles, and progressively removes stripped matter. If a subhalo is lost by the halo finder, this algorithm keeps tracking its so-called ghost until it has almost no particles left or has truly merged with its host. We apply this technique to a large suite of $N$-body simulations and restore lost subhaloes to the halo catalogues, which has a dramatic effect on key summary statistics of large-scale structure. Specifically, the subhalo mass function increases by about 50\% and the halo correlation function increases by a factor of two at small scales. While these quantitative results are somewhat specific to our algorithm, they demonstrate that particle tracking is a promising way to reliably follow haloes and reduce the need for orphan models. Our algorithm and augmented halo catalogues are publicly available.
\end{abstract}

\begin{keywords}
methods: numerical -- dark matter -- large-scale structure of Universe
\end{keywords}


\section{Introduction}
\label{sec:intro}

Structure in the Universe forms hierarchically, meaning that small dark matter haloes collapse first and merge to create larger haloes \citep[e.g.,][]{bond_91}. This picture implies that haloes are filled with numerous smaller subhaloes, which has been confirmed in simulations \citep{moore_99_subs, klypin_99_missing, diemand_05, springel_08}. The smallest possible subhalo size is set by the initial power spectrum, which is likely cut off at some scale due to a finite initial temperature (warm dark matter), self-interactions, or quantum mechanics. As long as no such cutoff has been found observationally, however, we must attempt to simulate substructure down to at least the smallest sizes that are expected to form observable galaxies. At the low-mass end, about 30\% of haloes are subhaloes \citep[although this number strongly depends on the definition of the host halo radius,][]{diemer_21_subs}. This large fraction means that there is no hope of accurately predicting statistics such as the galaxy correlation function if subhaloes are not modelled correctly \citep[e.g.,][]{colin_99}.

One important factor determining the abundance of subhaloes is how long they survive in their hosts after infall. The fundamental timescale is the dynamical time, which we define as a crossing time (about $5\ \gyr$ at $z = 0$; see Section~\ref{sec:method:halos}). Unlike the instantaneous merging that is implicit in Press-Schechter type models \citep{press_74, lacey_93}, real subhaloes tend to survive for at least a few orbits. While they lose mass to tidal stripping during each orbit \citep[e.g.,][]{kravtsov_04_satellites, zavala_19}, numerical research has repeatedly confirmed that they can lose most of their mass before entirely disrupting \citep{tormen_98_subs, diemand_06, green_21, errani_21, amorisco_22}.\footnote{Similarly, sharp potential gradients can ``heat'' dynamical systems \citep{ostriker_72_gcs, gnedin_97_gcs, gnedin_99_gcs1, gnedin_99_gcs2} but do not necessarily disrupt them \citep{vandenbosch_18_subs1, green_22}. Given that our dark matter-only simulations do not contain galactic discs, we do not carefully distinguish tidal stripping and tidal heating.} One exception are large subhaloes, roughly speaking those with more than $1/10$ or their host's mass. In this regime, strong dynamical friction can drag subhaloes to the host centre within few orbits \citep[][]{chandrasekhar_43, vandenbosch_99, boylankolchin_08, adhikari_16, naidu_21, vasiliev_22, banik_22}. 

A similar picture applies to the satellite galaxies within subhaloes, whose tightly bound stellar distribution can often survive even stronger tidal forces than their dark matter. Understanding their orbits and evolution is critical because their eventual mergers play a major role in the evolution of the galaxy population \citep[e.g.,][]{toomre_72, barnes_88, hernquist_92_mergers1}. Moreover, the distribution of satellites is thought to be one of the most sensitive tracers of the nature of dark matter \citep[e.g.,][]{cyrracine_16, lovell_20, nadler_21}. The tightest constraints on a potential cutoff in the power spectrum will come from the smallest, ultra-faint satellites that we can observe from our own position near the centre of the Milky Way \citep{drlicawagner_2020}. These objects will inhabit subhaloes with $\approx 10^{-4}$ the mass of the Milky Way \citep{nadler_20}. 

Reliably simulating such tiny, strongly stripped subhaloes near the centres of their hosts is arguably the greatest challenge to $N$-body simulations today. Two separate issues are currently preventing us from achieving this goal: numerical disruption in $N$-body simulations, and halo finders struggling to identify certain substructures. The first issue has long been known as the ``over-merging problem,'' where unphysical effects such as two-body scattering ``heat'' the subhalo and possibly destroy it \citep{white_87, carlberg_94, vankampen_95, moore_96, moore_99_collapse, klypin_99_overmerging}. This problem is far from resolved even in modern simulations, including the ones discussed in this paper \citep{vandenbosch_17, vandenbosch_18_subs1, vandenbosch_18_subs2}. The resulting lack of substructure leads to poor fits to the observed clustering of galaxies \citep{campbell_18}, necessitating ad-hoc solutions such as representing the orbits of disrupted subhaloes by those of their single most bound particle \citep[``orphans,''][]{summers_95, wang_06_orphans, guo_10, moster_10}, multiple particles \citep[``cores,''][]{heitmann_19, heitmann_21, rangel_20, sultan_21, korytov_23}, or analytically integrating orbits forward in time \citep[e.g.,][]{taylor_01_sats, zentner_05, behroozi_19}. We do not tackle the over-merging issue in this paper, but our goal of creating more complete subhalo catalogues is an important step towards a full understanding of how and why numerical disruption occurs.

This second issue is that state-of-the-art halo finders can lose certain subhaloes even though they are composed of significant numbers of particles and are discernible by eye. Most halo finders agree reasonably well on the positions and properties of isolated haloes, but robustly finding subhaloes is much harder because they can blend into the varying background density of their larger host \citep{knebe_11_mad}. This issue can be fixed by using a phase-space friends-of-friends algorithm to group particles not only in position but also in velocity space \citep{davis_85, diemand_06, behroozi_13_rockstar, elahi_19_velociraptor}, but another problem remains: the tidal streams coming off a subhalo can be orders of magnitude more massive than the subhalo itself. The halo finding algorithm must balance a sensitivity to small, bound particle groups against the risk of falsely identifying spurious, unphysical noise. For example, the \rockstar halo finder used in this work tends to fail when a subhalo's tidal tails are gravitationally unbound but dense, a state which naturally occurs during strong tidal stripping \citep{behroozi_13_rockstar}. We emphasise that this issue is not solved by increasing the number of particles: the subhalo mass functions produced by both \rockstar and \subfind \citep{springel_01_subfind} appear to be independent of resolution \citep[e.g.][]{springel_08, nadler_23}, but they are missing heavily stripped haloes at all resolutions.

The root cause of these issues is that haloes are most commonly identified in individual simulation snapshots and then stitched together into merger trees. This means that the same (sub-)haloes need to be identified over and over again, including in challenging situations such as during strong tidal stripping. A better solution would be to identify each halo once when it first emerges and to track its particles forward in time, using the knowledge that the given set of particles was a bound structure in the past. Such an algorithm can be computationally more demanding (e.g., because particle IDs need to be stored), but it automatically creates time-connected merger trees. Some schemes of this nature have been put forward in the literature \citep[e.g.,][]{gill_04_halofinder, poulton_20}, most notably in the HBT+ halo finder \citep{han_12_hbt, han_18_hbt}, which finds subhalo centres from tracked particles and determines membership based on gravitational binding. While the initial results based on these algorithms are encouraging, many questions remain. For example, \citet{springel_21_gadget4} implement the HBT+ algorithm in a combined \textsc{SubFind-HBT} halo finder, but this change seems to have a minimal effect on the abundance of subhaloes (figure 38 in \citealt{springel_21_gadget4}). Moreover, each particle-tracking algorithm has to make a number of choices whose impact is not yet clear. Which particles are initially included in the tracking, and can new particles be added? What are the criteria for removing stripped particles? When is a subhalo deemed to have physically merged?

In this paper, we present a post-processing algorithm to identify and track all particles in all subhaloes in simulations, allowing us to follow the ``ghosts'' of subhaloes that have been lost by the halo finder. Our algorithm is in principle independent of the halo finder used. It features user-adjustable parameters, avoids relying on gravitational boundness criteria, and outputs ghost data into a convenient merger tree format. Besides introducing this technique, the two main purposes of the paper are to understand where and when subhaloes are lost by \rockstar and to investigate to what extent ghosts improve the predictions of simulations. We show that ghosts extend the lives of subhaloes across a vast range of resolutions and that they make significant contributions to basic predictions such as the subhalo mass function and correlation function. Most importantly, however, this work is an exploratory study that opens numerous avenues for further research. For example, we apply our algorithm to dark matter-only simulations although baryons change the abundances of subhaloes significantly \citep[e.g.,][]{garrisonkimmel_17_baryons, richings_20}.

The paper is organised as follows. We describe our simulations and algorithms in Section~\ref{sec:method}. In Section~\ref{sec:results} we investigate how subhaloes lose mass and when they are lost by the halo finder. In Section~\ref{sec:results2} we show the impact of adding ghosts to basic simulation predictions. We further discuss these results in Section~\ref{sec:discussion} and summarise our conclusions in Section~\ref{sec:conclusion}. Our algorithm is implemented in the open-source framework \sparta \citep{diemer_17_sparta}, and the ghost-augmented halo catalogues and merger trees are publicly available \citep{diemer_20_catalogs}.


\section{Methods \& algorithms}
\label{sec:method}
 
We begin by briefly reviewing our simulations (Section~\ref{sec:method:sims}) and halo catalogues (Section~\ref{sec:method:halos}), largely referring the reader to \citet{diemer_20_catalogs} for details. In Section~\ref{sec:method:tracking} we introduce our algorithms to track subhaloes and ghosts via their constituent particles.

\subsection{$N$-body simulations}
\label{sec:method:sims}

Our catalogues are based on the \erebos suite of dissipationless $N$-body simulations \citep{diemer_14, diemer_15}, and we have run our subhalo tracking algorithm on the entire suite \citep{diemer_20_catalogs}. Since variations in cosmology are not important for the purposes of this paper, we focus exclusively on seven simulations of a \wmap \LCDM cosmology, which is the same as that of the {\it Bolshoi} simulation \citep[][$\Omega_{\rm m} = 0.27$, $\Omega_{\rm b} = 0.0469$, $h = 0.7$, $\sigma_8 = 0.82$, and $n_{\rm s} = 0.95$]{komatsu_11, klypin_11}. The initial conditions for the simulations were generated with \textsc{2LPTic} \citep{crocce_06} from power spectra computed by \textsc{Camb} \citep{lewis_00}. The simulations were run with \textsc{Gadget2} \citep{springel_01_gadget, springel_05_gadget2}.

The box sizes of the simulations increase by factors of two from $31.25\ \mpch$ to $2000\ \mpch$ (where the smallest box was run only to $z = 2$). Each box contains $1024^3$ particles, leading to mass resolutions between $2.1 \times 10^6$ and $5.6 \times 10^{11}\ \msunh$. For testing and visualisation, we use a smaller test simulation with $256^3$ particles in a $62.5\ \mpch$ box \citep{diemer_17_sparta}. For the purposes of this work, mass and force resolution are equally important. The force resolutions vary between $0.25$ and $65$ comoving $\kpch$, and they were chosen such that the scale radius of a typical halo with $1000$ particles is resolved with four force resolution lengths at $z = 0$. This equivalence means that the mass and force resolution are well matched in the sense that the lowest resolved halo mass roughly coincides for both criteria (though two-body relaxation complicates this picture as discussed in Appendix A of \citealt{diemer_22_prof1}; see also \citealt{ludlow_19} and \citealt{mansfield_21_resolution}).

\subsection{Halo catalogues and definitions}
\label{sec:method:halos}

We identify haloes and subhaloes using the \rockstar halo finder and \consistenttrees merger tree tool \citep{behroozi_13_rockstar, behroozi_13_trees}. \rockstar uses a phase-space friends-of-friends algorithm that performs well at identifying substructure because it groups particles in velocity as well as position space \citep[e.g.,][]{onions_12}. The resulting catalogues, as well as raw particle data, serve as input to \sparta code, which analyses the orbits of individual particles to compute various halo properties, such as splashback radii \citep{diemer_17_sparta} or the density profiles of only orbiting particles \citep{diemer_22_prof1, diemer_22_prof2}. Eventually, the \moria extension recombines \sparta's output with the \rockstar catalogues to create enhanced catalogues and merger trees both in the original ascii format and in a new hdf5-based format \citep{diemer_20_catalogs}. The ghost subhaloes discussed in the following section were added to the \moria catalogues as if they were normal subhaloes identified by the halo finder. While our algorithm is independent of the halo finder, we emphasise that the results of this paper are specific to \rockstar and \consistenttrees. 

Throughout the paper, we discuss a number of spherical overdensity radius and mass definitions. Specifically, $\rtoc$ is the radius enclosing a mean density of $200$ times the critical density, $\rtom$ encloses $200$ times the mean density, and $\rvir$ indicates a varying overdensity \citep{bryan_98}. The corresponding enclosed masses are denoted $\mtoc$, $\mtom$, and $\mvir$. We further distinguish masses and radii computed including all particles (denoted $M_{\rm 200m,all}$, $R_{\rm 200m,all}$, or simply $\rtom$) and those computed using only gravitationally bound particles according to \rockstar's unbinding algorithm (denoted $M_{\rm 200m,bnd}$ and $R_{\rm 200m,bnd}$). Our catalogues contain all haloes that have reached at least $200$ particles within $R_{\rm 200m,all}$ at some point in their life.

In our \rockstar catalogues, subhaloes are defined to lie within $R_{\rm 200m,bnd}$ of a host halo with larger maximum circular velocity, $\vmax$. We use this assignment throughout unless otherwise mentioned. \moria calculates separate host-subhalo relations for numerous other definitions. We have verified that these relations agree exactly with \consistenttrees if the same radius definition is used. Since we track ghosts regardless of whether they lie inside or outside the host radius, the subhalo definition does not have a large impact on our results. 

We will generally quantify cosmic time in units of the dynamical time, which we define as a halo crossing time. Spherical overdensity radii allow for a convenient definition as the time to cross $\rtom$, 
\begin{equation}
\tdyn \equiv \frac{2 \rtom}{\vtom} = \frac{2 \rtom}{\sqrt{G \mtom / \rtom}} = \frac{t_\rmH(z)\  \Omega_\rmm(z)^{-1/2}}{5} \,,
\end{equation}
about $5\ \gyr$ at $z = 0$ for our cosmology.

\subsection{Tracking subhaloes and ghosts based on their particles}
\label{sec:method:tracking}

\def\panelsize{0.56}
\begin{figure*}
\centering
\vspace{0.4cm}
\includegraphics[trim =  20mm 19.5mm 4mm 4mm, clip, scale=\panelsize]{\figdir/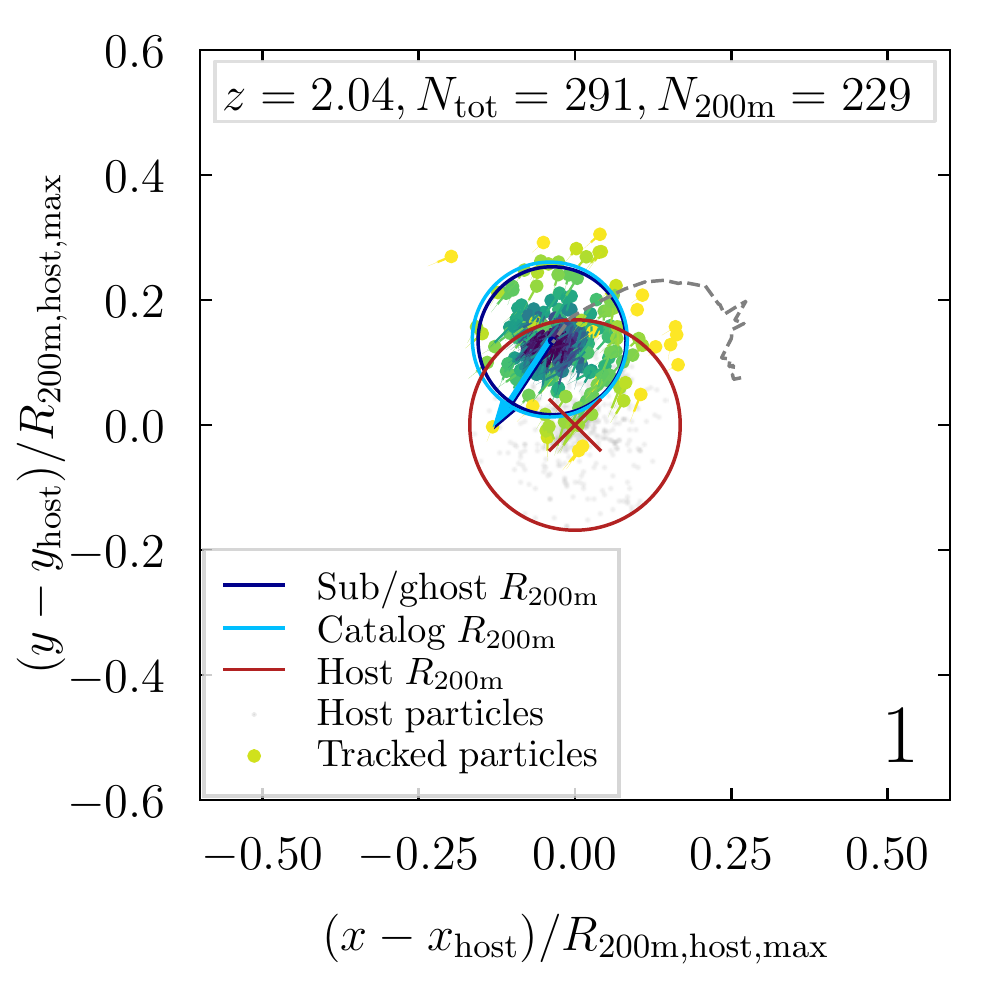}
\includegraphics[trim =  20mm 19.5mm 4mm 4mm, clip, scale=\panelsize]{\figdir/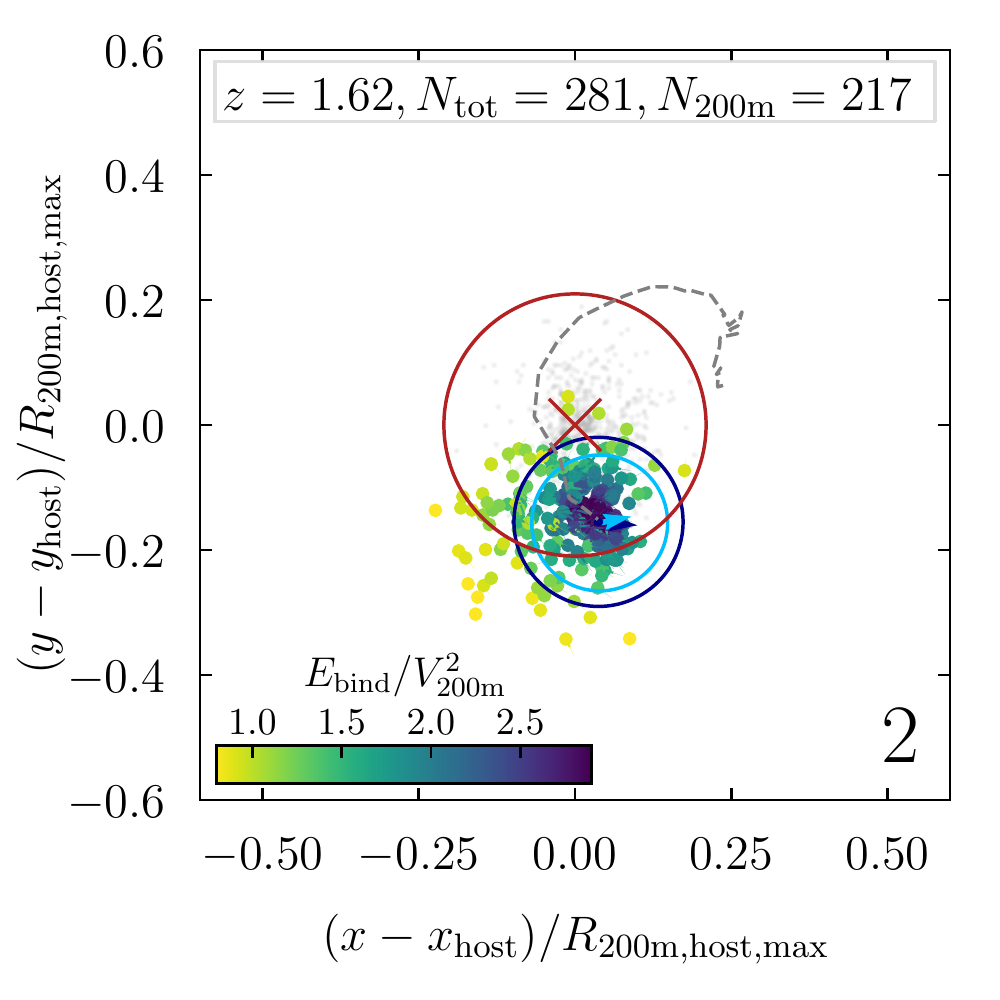}
\includegraphics[trim =  20mm 19.5mm 4mm 4mm, clip, scale=\panelsize]{\figdir/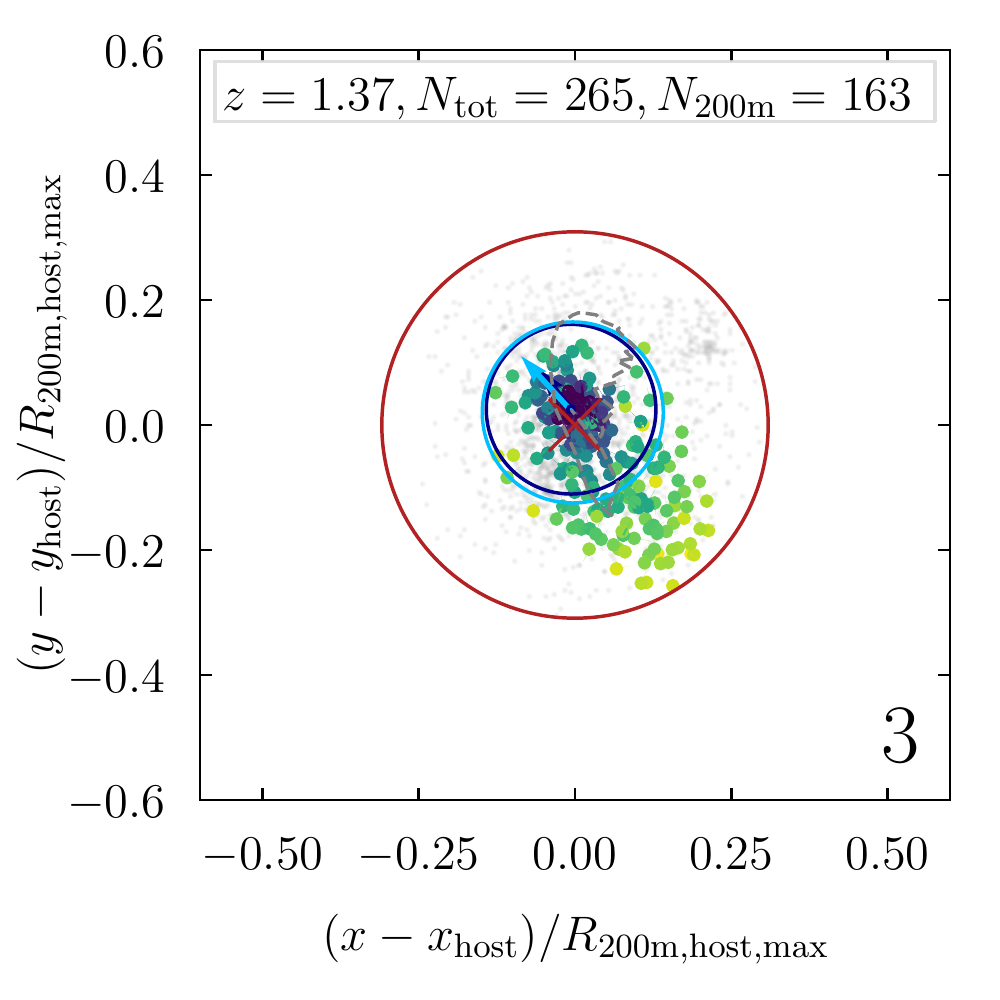}
\includegraphics[trim =  20mm 19.5mm 4mm 4mm, clip, scale=\panelsize]{\figdir/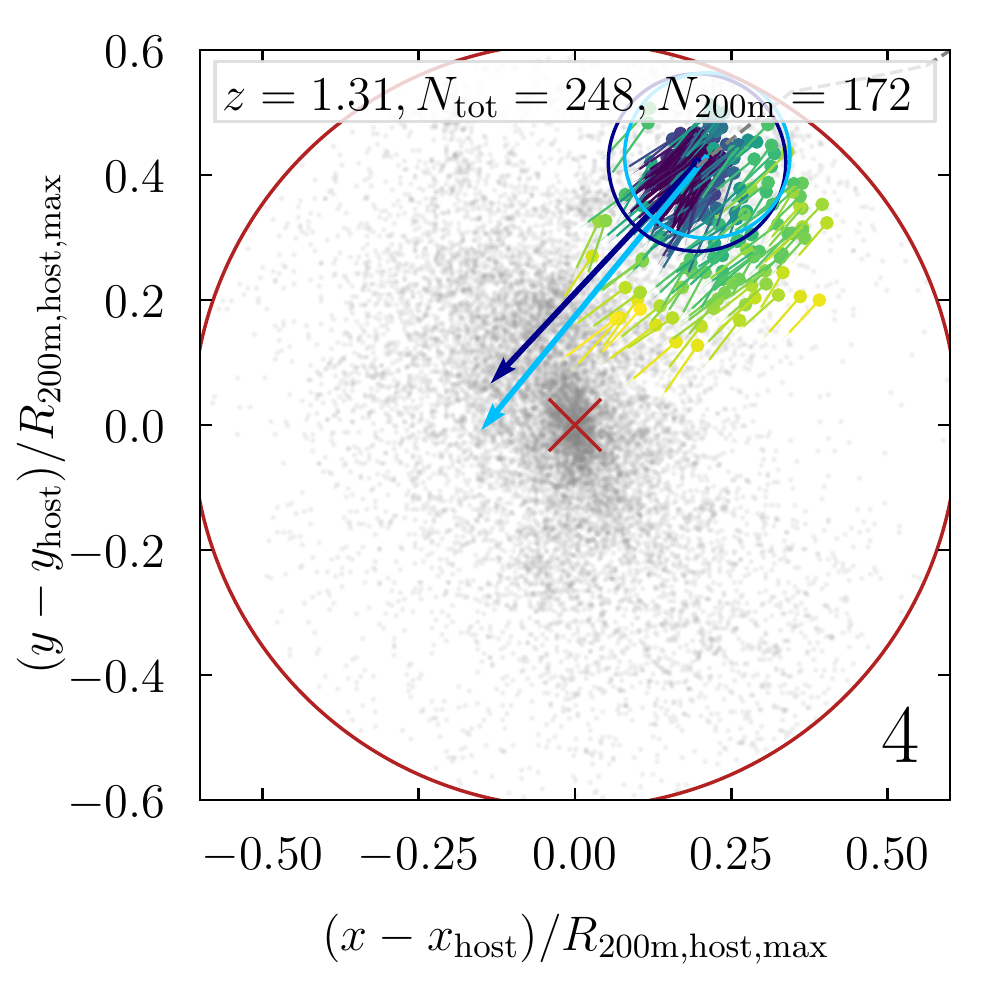}
\includegraphics[trim =  20mm 19.5mm 4mm 4mm, clip, scale=\panelsize]{\figdir/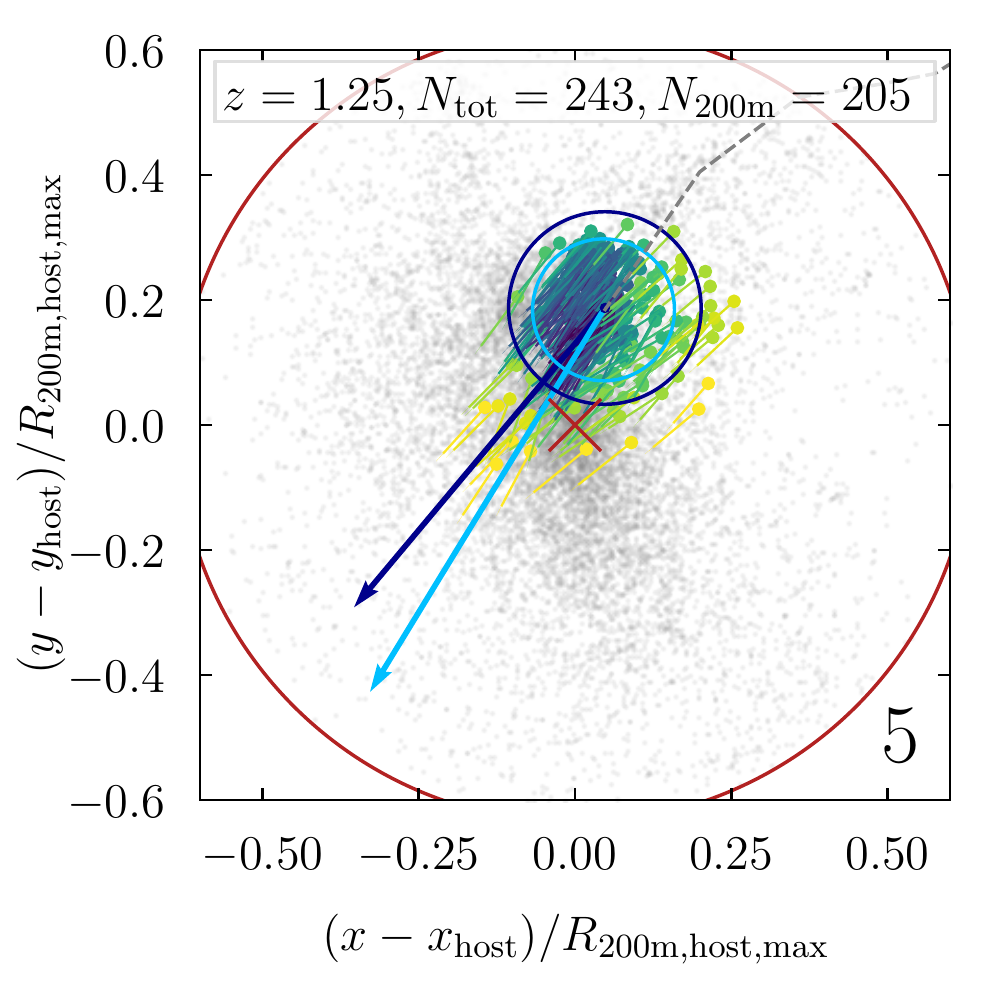}
\includegraphics[trim =  20mm 19.5mm 4mm 4mm, clip, scale=\panelsize]{\figdir/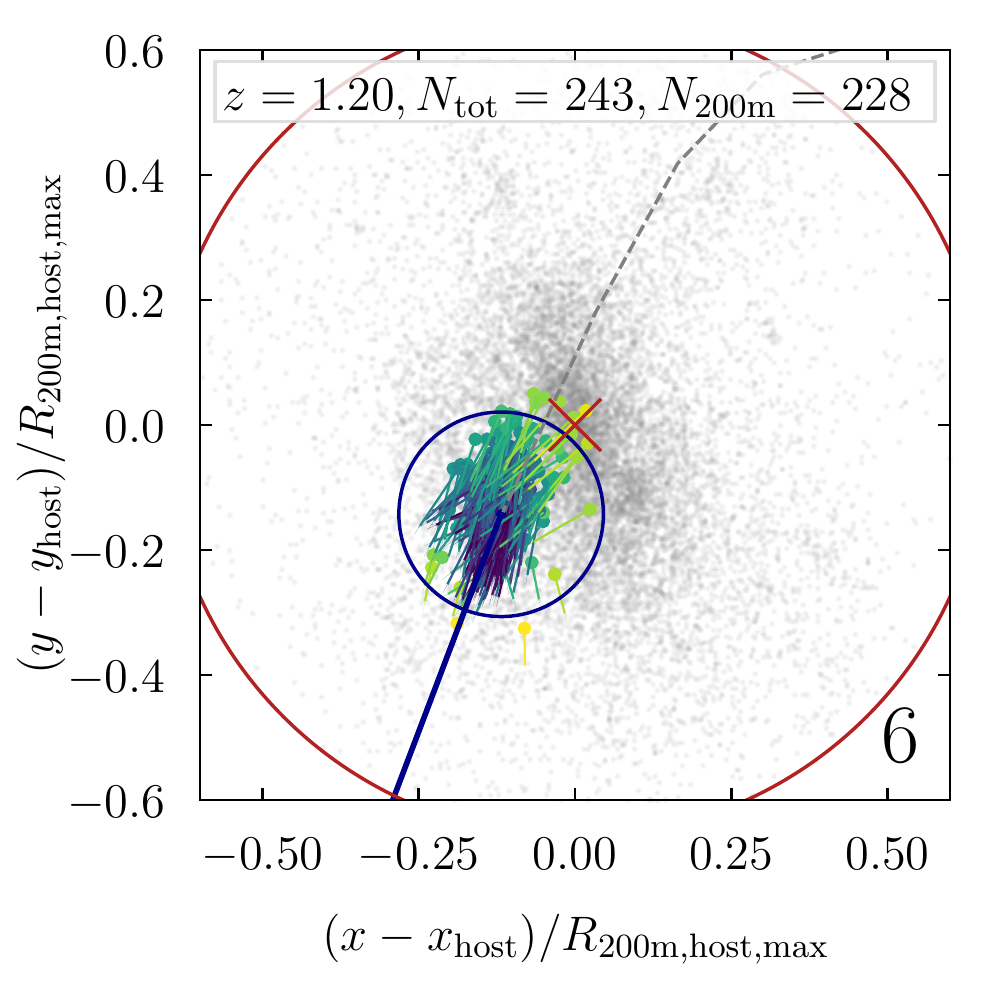}
\includegraphics[trim =  20mm 19.5mm 4mm 4mm, clip, scale=\panelsize]{\figdir/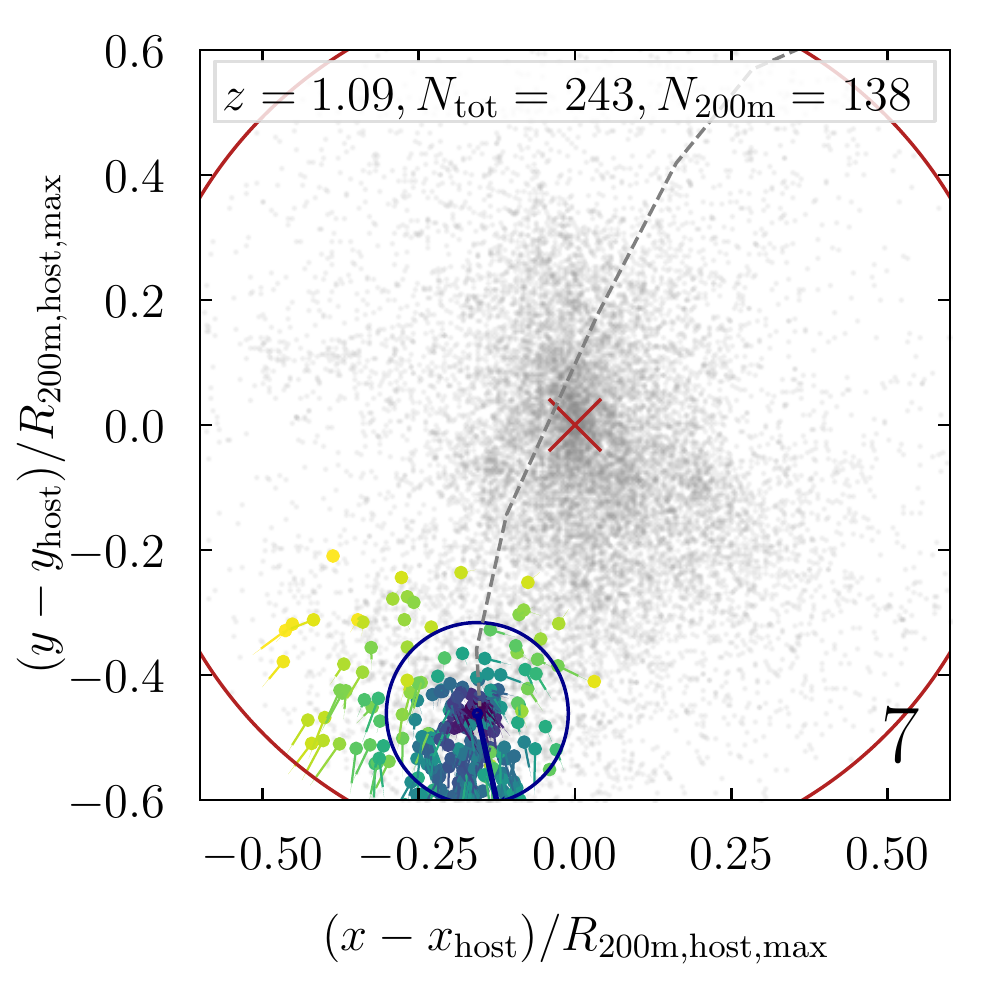}
\includegraphics[trim =  20mm 19.5mm 4mm 4mm, clip, scale=\panelsize]{\figdir/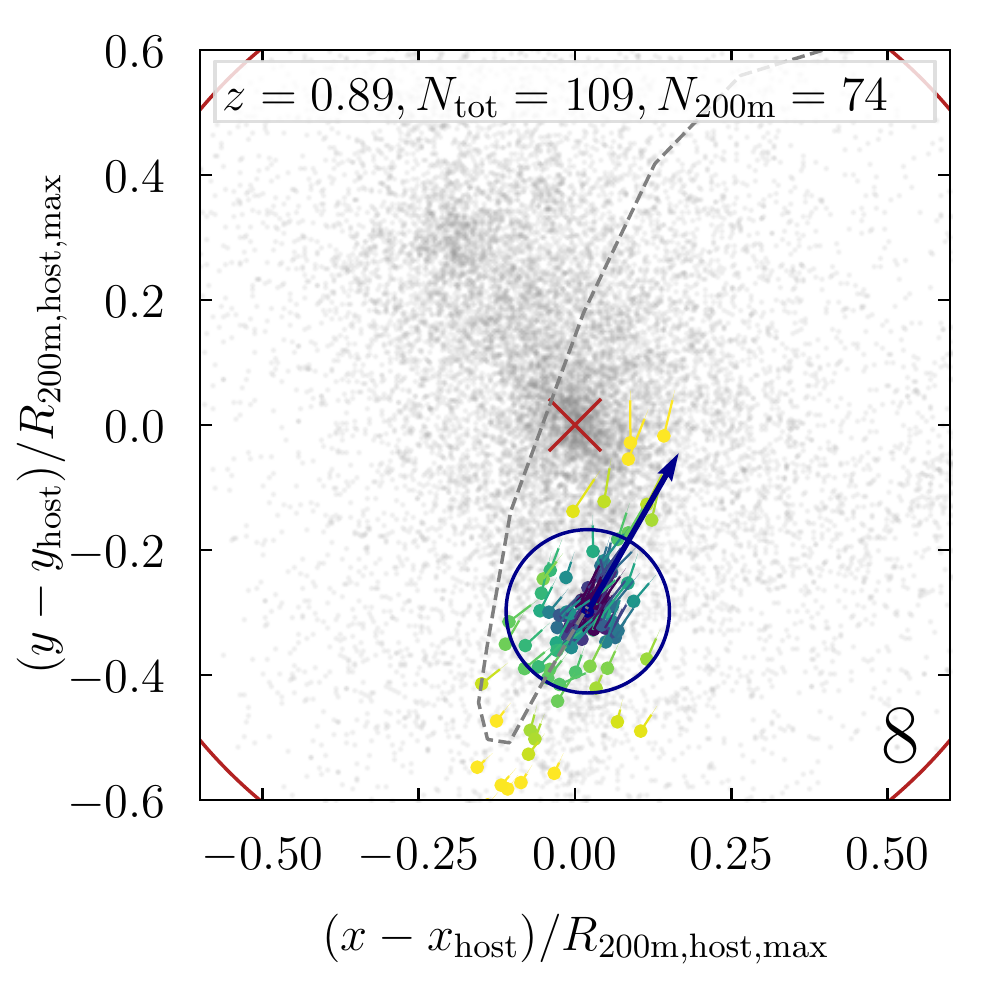}
\includegraphics[trim =  20mm 2mm 4mm 4mm, clip, scale=\panelsize]{\figdir/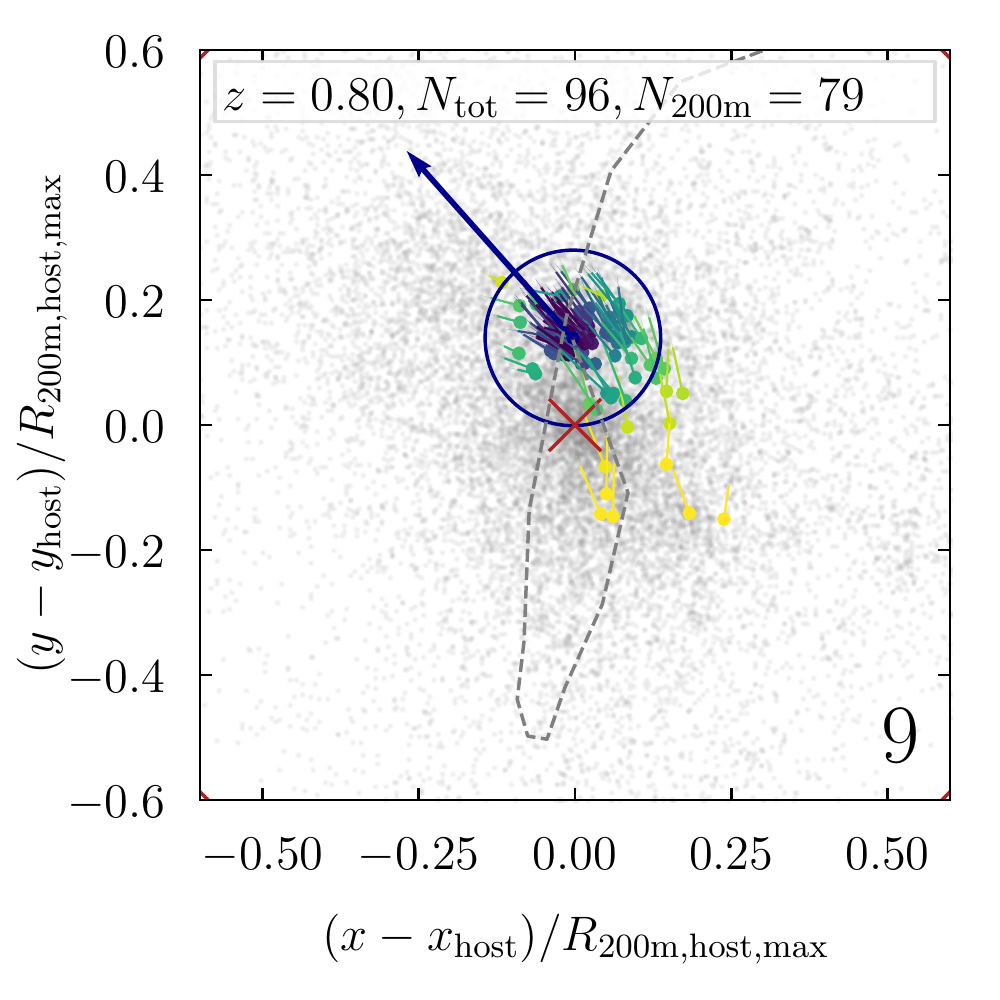}
\includegraphics[trim =  20mm 2mm 4mm 4mm, clip, scale=\panelsize]{\figdir/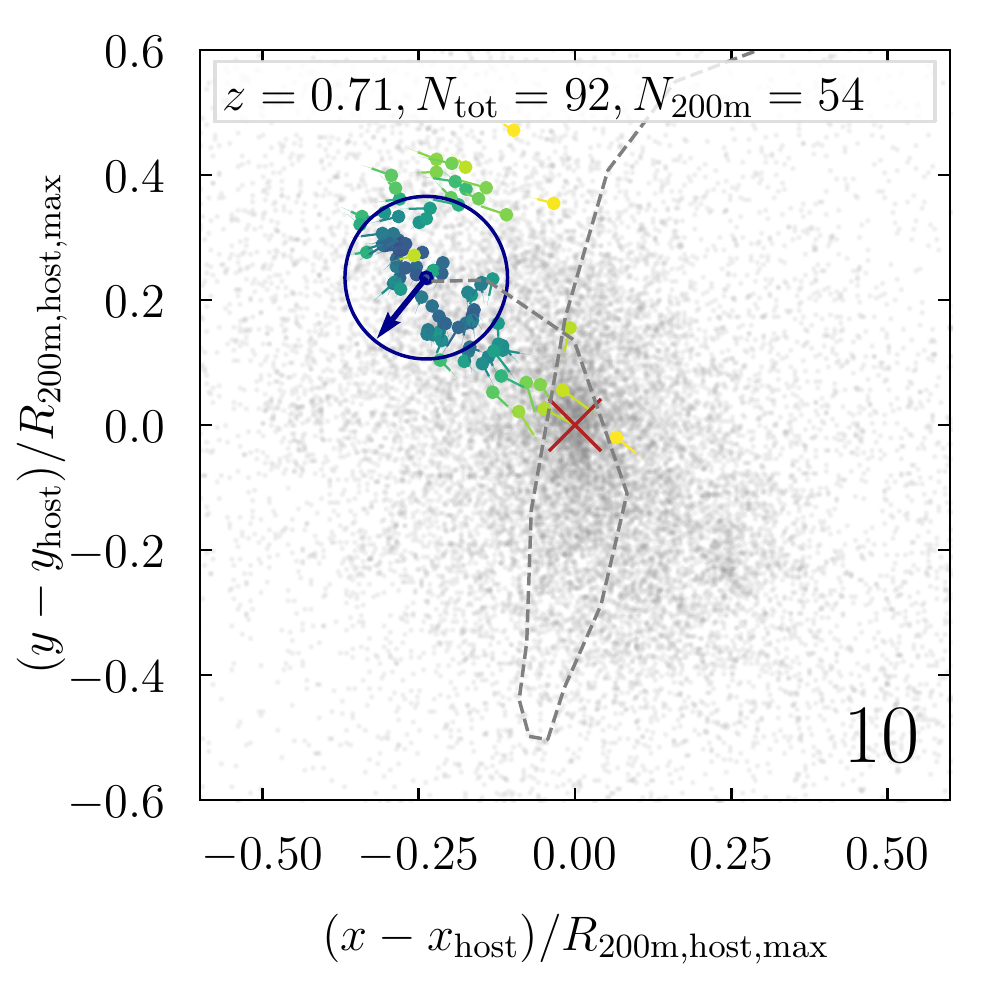}
\includegraphics[trim =  20mm 2mm 4mm 4mm, clip, scale=\panelsize]{\figdir/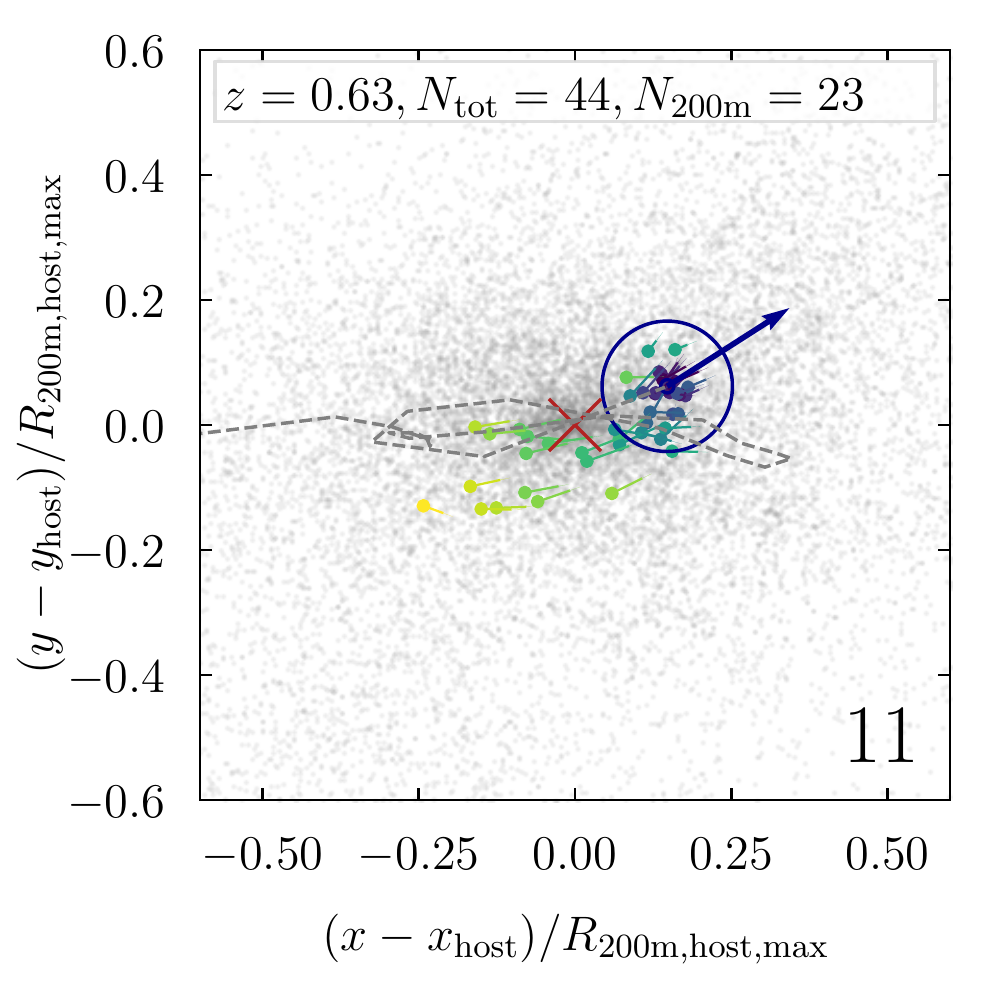}
\includegraphics[trim =  20mm 2mm 4mm 4mm, clip, scale=\panelsize]{\figdir/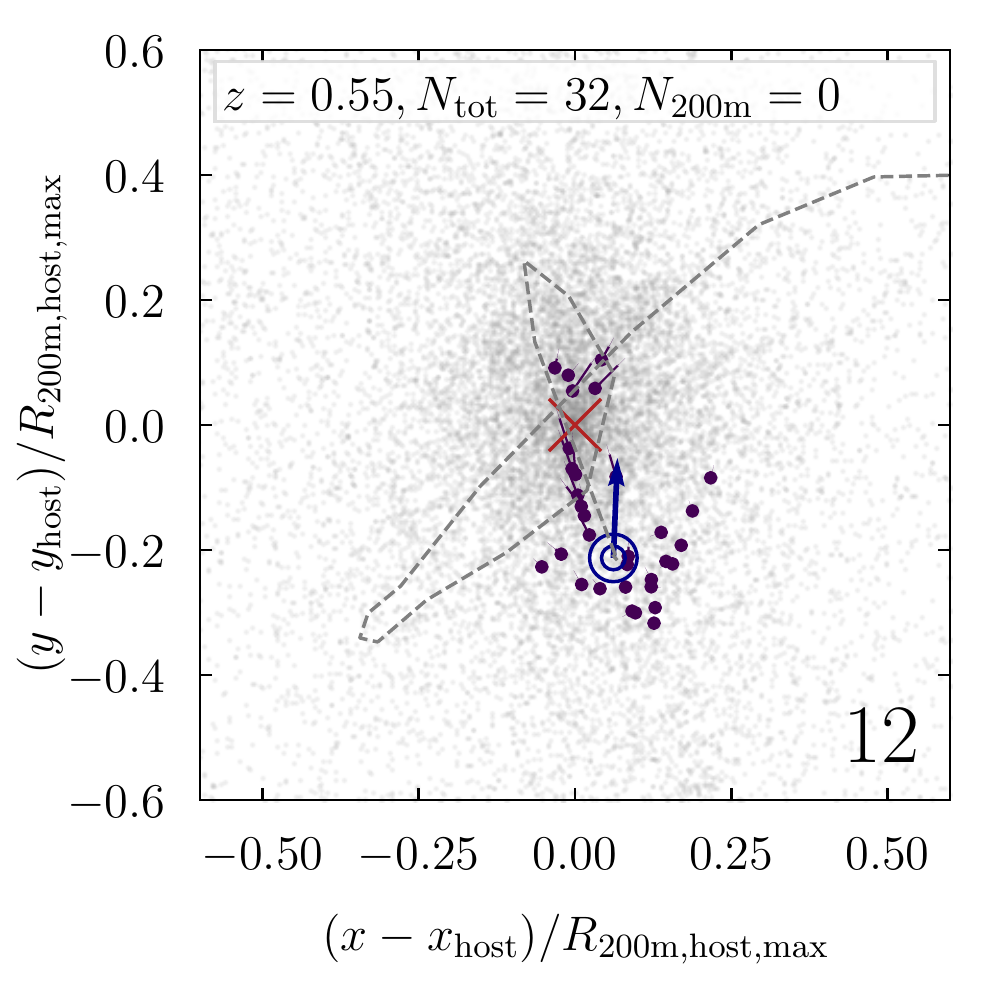}
\caption{Visual illustration of our subhalo tracking algorithm. Each panel shows a snapshot along the history of the same subhalo before and after it is lost by the halo finder. The scale is fixed in physical units. The view is shifted into the host (red) frame and rotated such that the subhalo (dark blue) position and velocity lie in the $x$-$y$ plane. This orientation means that the azimuthal position of the subhalo is somewhat random, but it shows the correct distance and velocity with respect to the host centre. The velocity arrows of subhaloes and particles are scaled such that $1/10$ of the width of the plot would correspond to $200$ km/s and $1000$ km/s, respectively, meaning that the subhalo arrows are five times longer. Gray dots show all particles inside the host radius, whereas coloured points show the tracked subhalo particles (with purple corresponding to strongly bound and yellow to weakly bound members; see colour scale in second panel). The dashed gray lines indicate the subhalo's previous trajectory. The dark blue circles and arrows show $\rtom$ and the relative velocity, calculated only from strongly bound, tracked particles. A smaller dark blue circle shows the uncertainty on the position of the ghost (visible only in the final panel). Before the subhalo disappears from the halo catalogue, its radius is compared to the catalogue position, velocity, and radius (light blue circle and arrow); they agree well. The first panel shows the initial infall of the subhalo into its host, a fairly major merger. After about one orbit (between the 3rd and 4th panels), the host and sub fall into a yet larger host with high velocity. After about half an orbit, the halo finder loses the subhalo (between the 5th and 6th panels). The ghost undergoes more than one full orbit, with strong tidal disruptions at the apocentres (7th to 11th panels). Eventually, there are fewer than $10$ particles within $\rtom$ of the ghost, and we consider it to have been disrupted (12th panel).}
\label{fig:ghost1}
\end{figure*}

\def\panelsize{0.56}
\begin{figure*}
\centering
\vspace{0.4cm}
\includegraphics[trim =  20mm 19.5mm 4mm 4mm, clip, scale=\panelsize]{\figdir/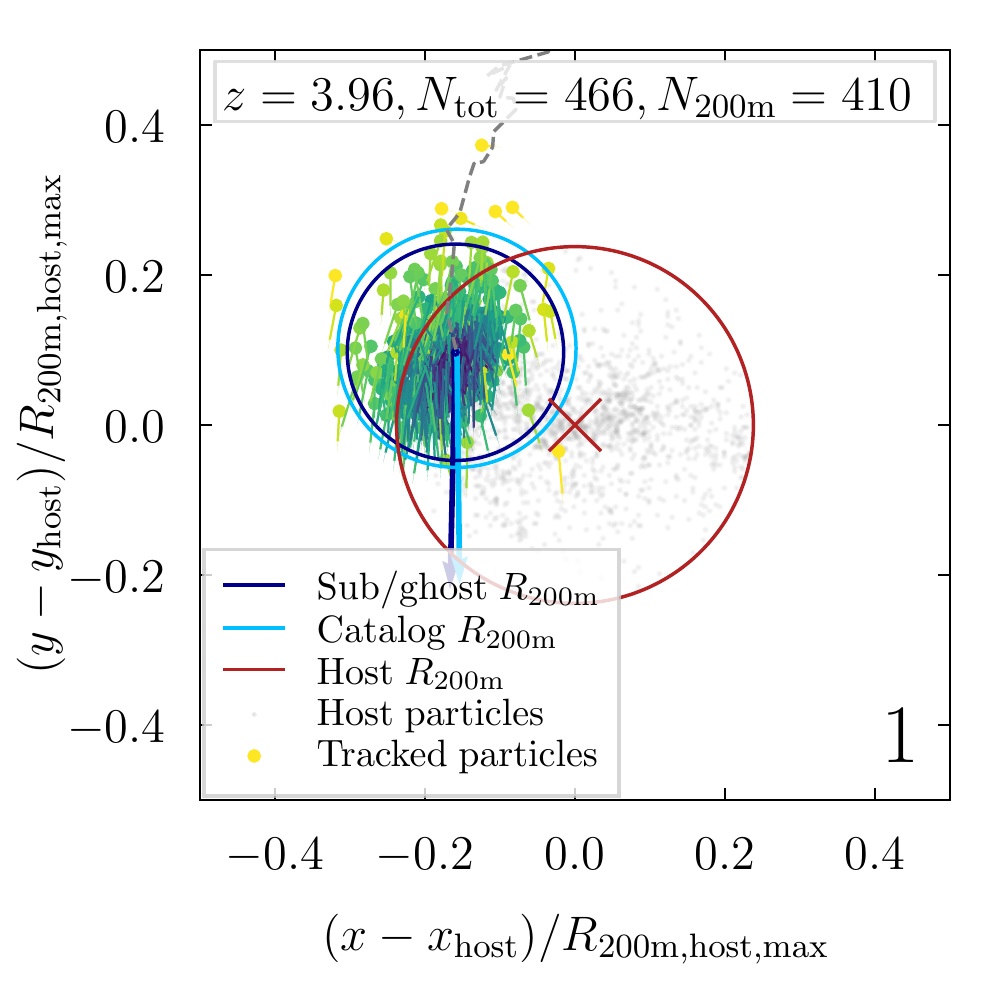}
\includegraphics[trim =  20mm 19.5mm 4mm 4mm, clip, scale=\panelsize]{\figdir/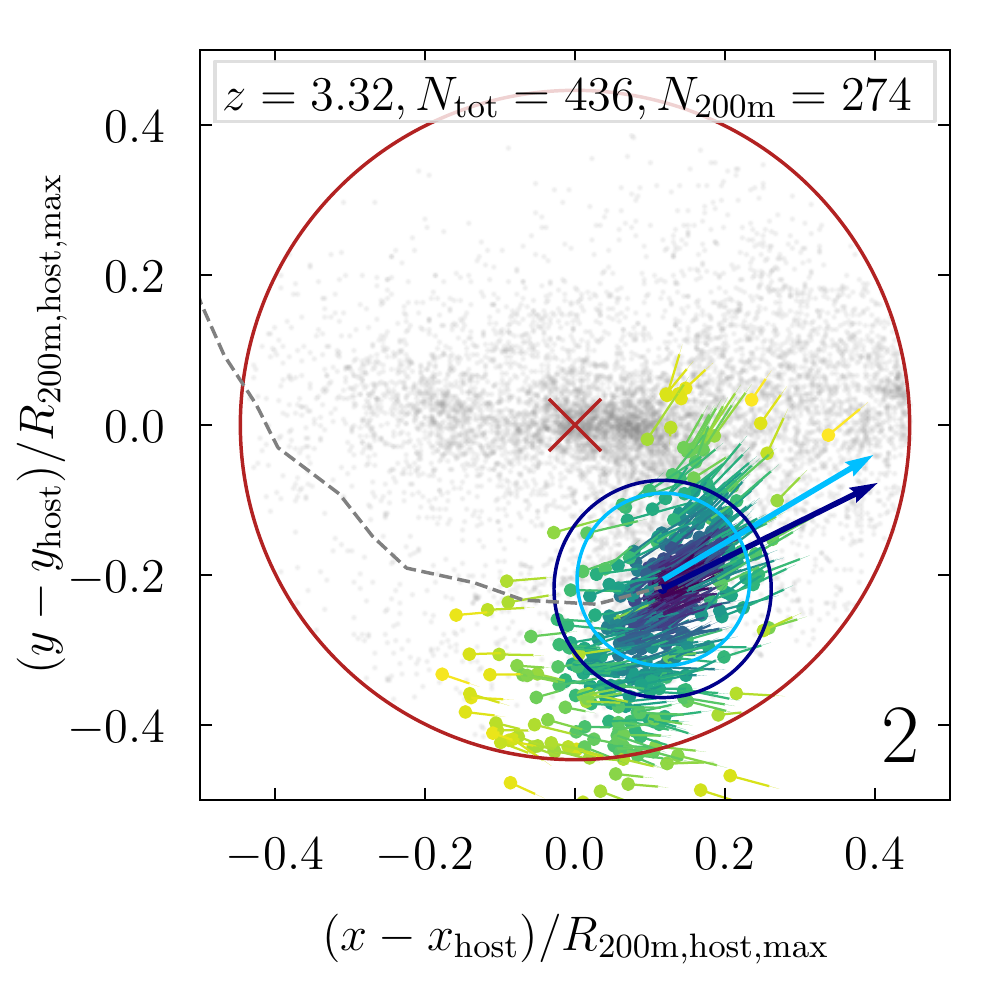}
\includegraphics[trim =  20mm 19.5mm 4mm 4mm, clip, scale=\panelsize]{\figdir/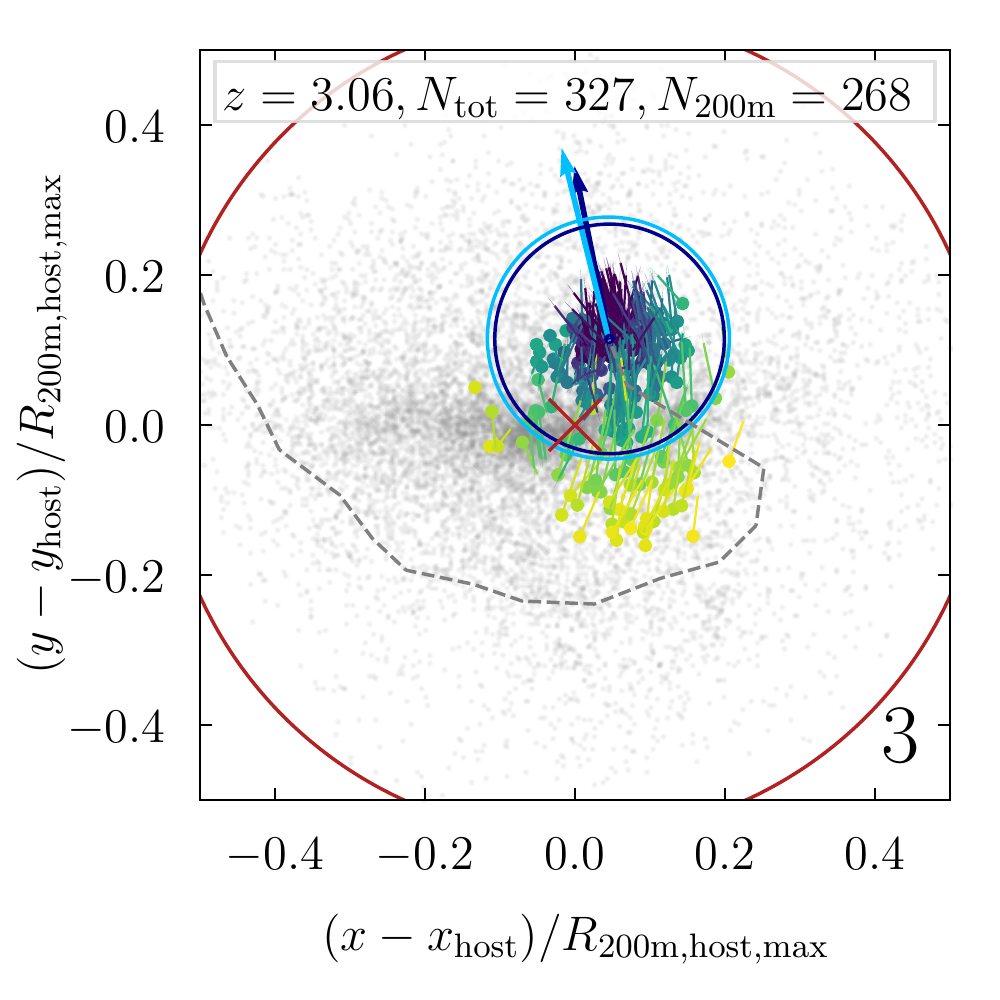}
\includegraphics[trim =  20mm 19.5mm 4mm 4mm, clip, scale=\panelsize]{\figdir/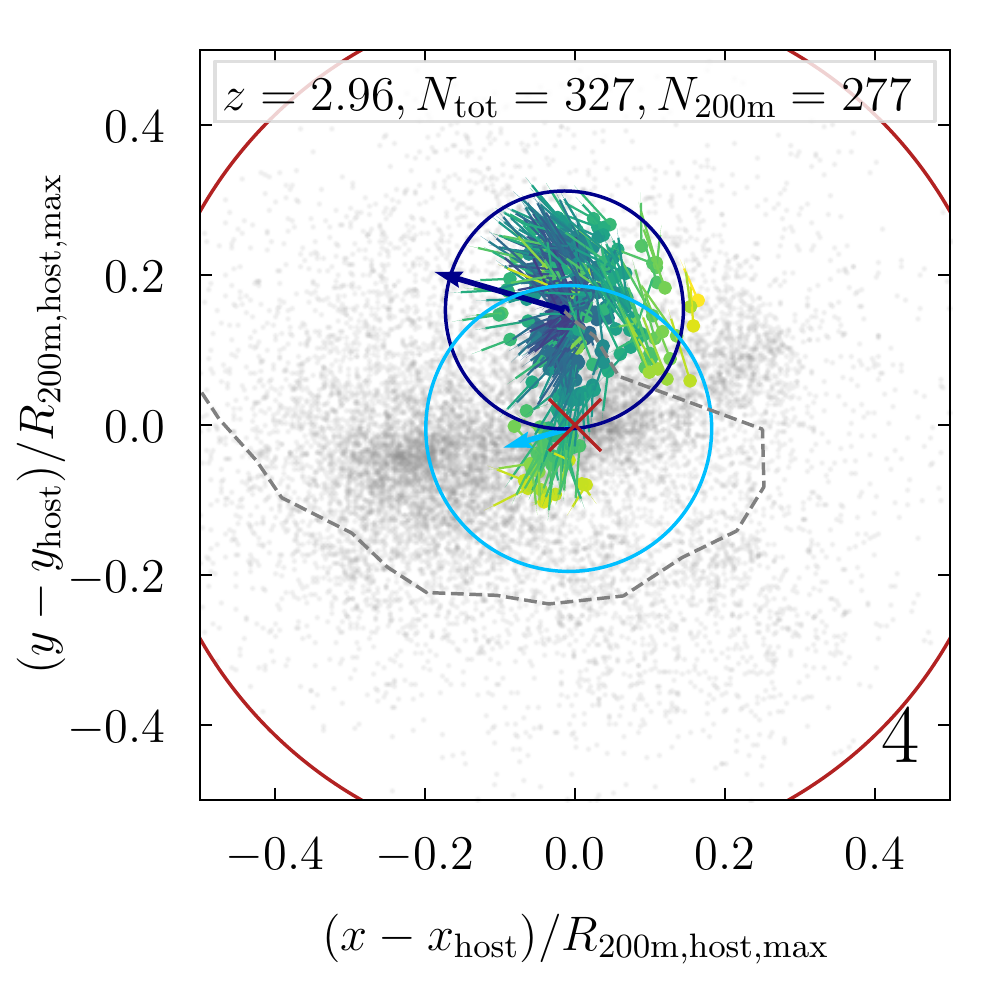}
\includegraphics[trim =  20mm 19.5mm 4mm 4mm, clip, scale=\panelsize]{\figdir/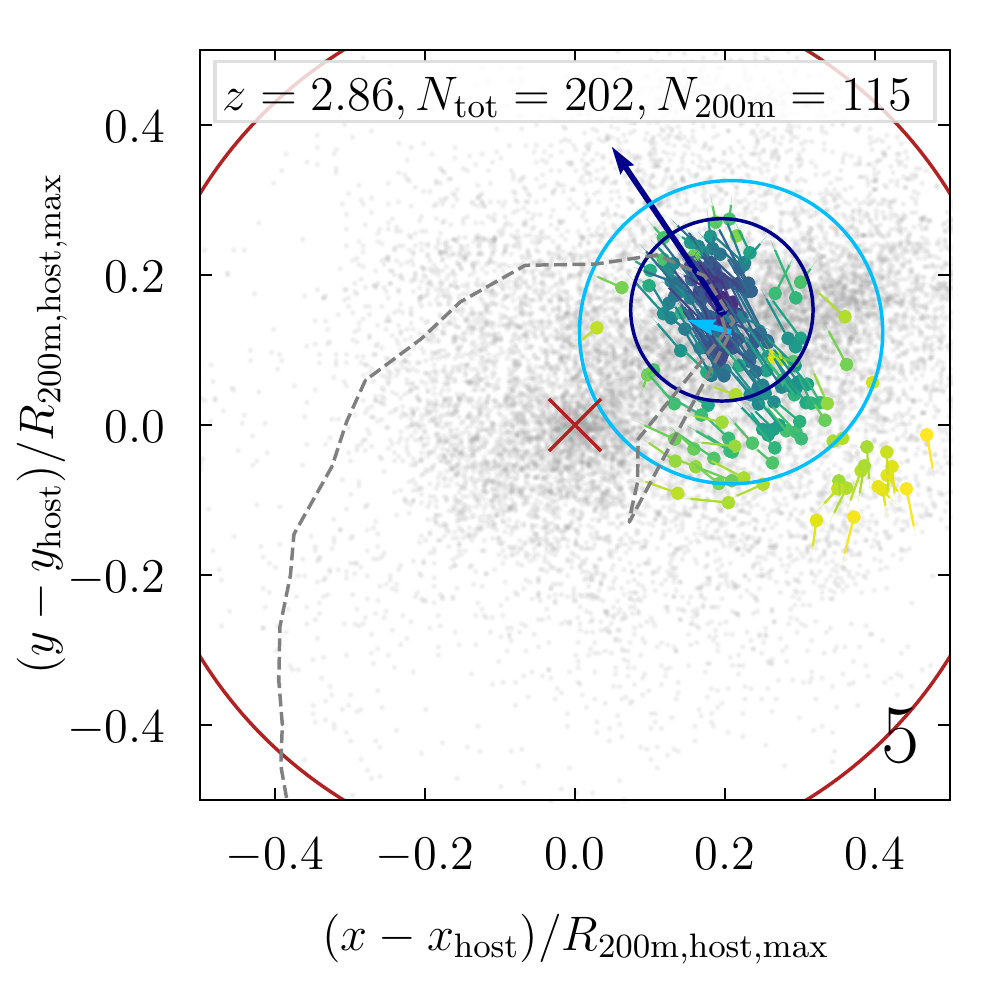}
\includegraphics[trim =  20mm 19.5mm 4mm 4mm, clip, scale=\panelsize]{\figdir/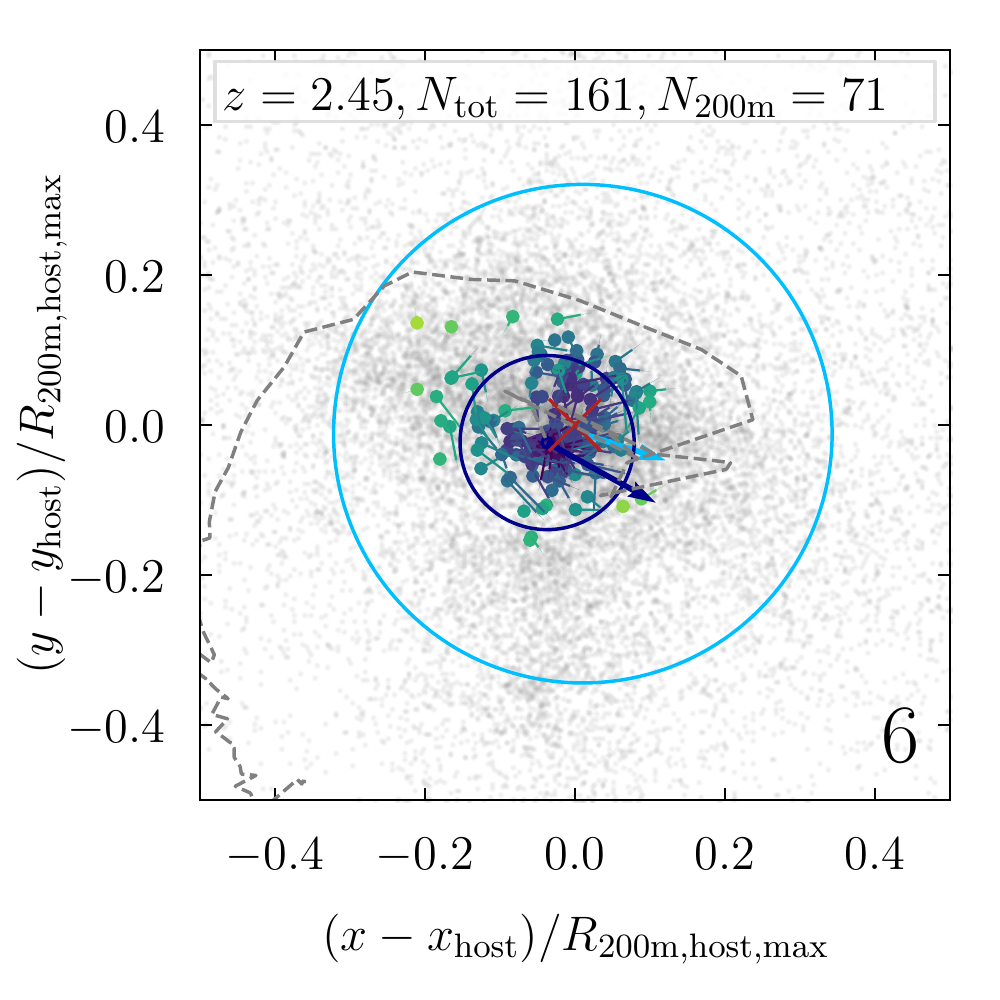}
\includegraphics[trim =  20mm 19.5mm 4mm 4mm, clip, scale=\panelsize]{\figdir/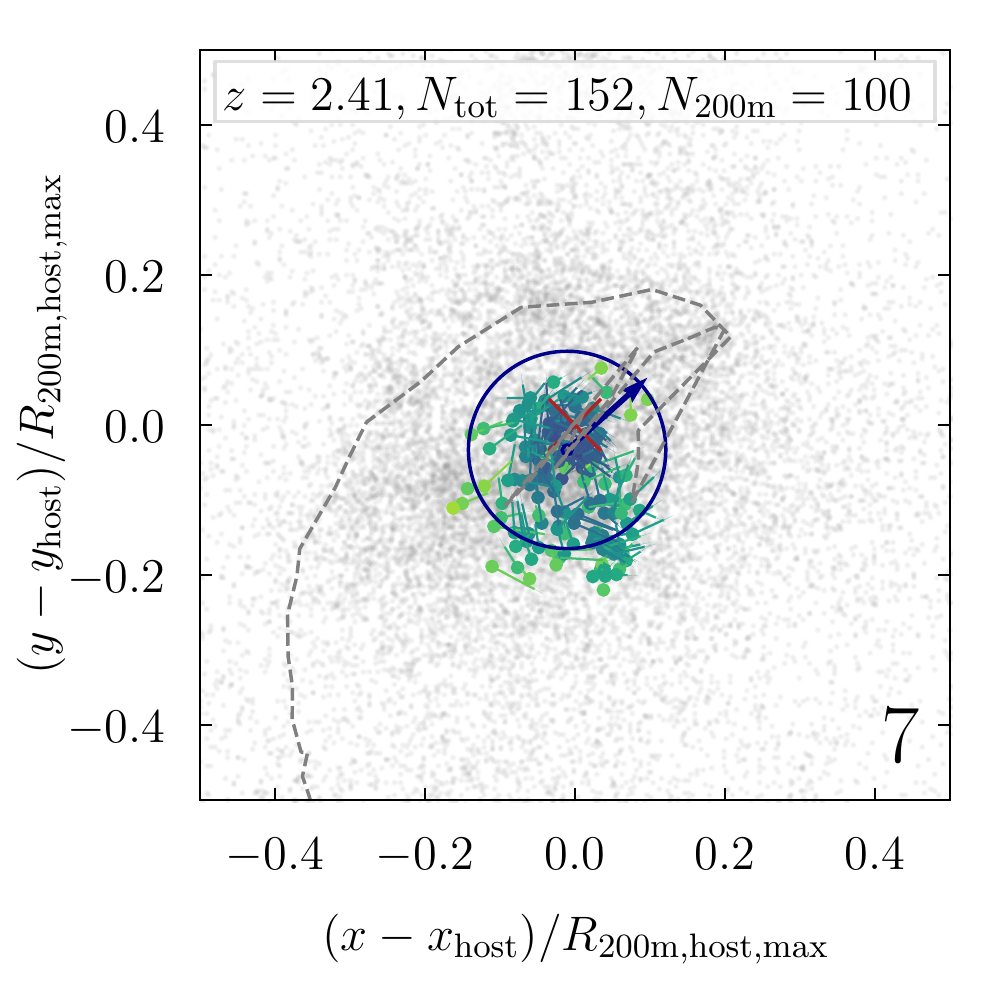}
\includegraphics[trim =  20mm 19.5mm 4mm 4mm, clip, scale=\panelsize]{\figdir/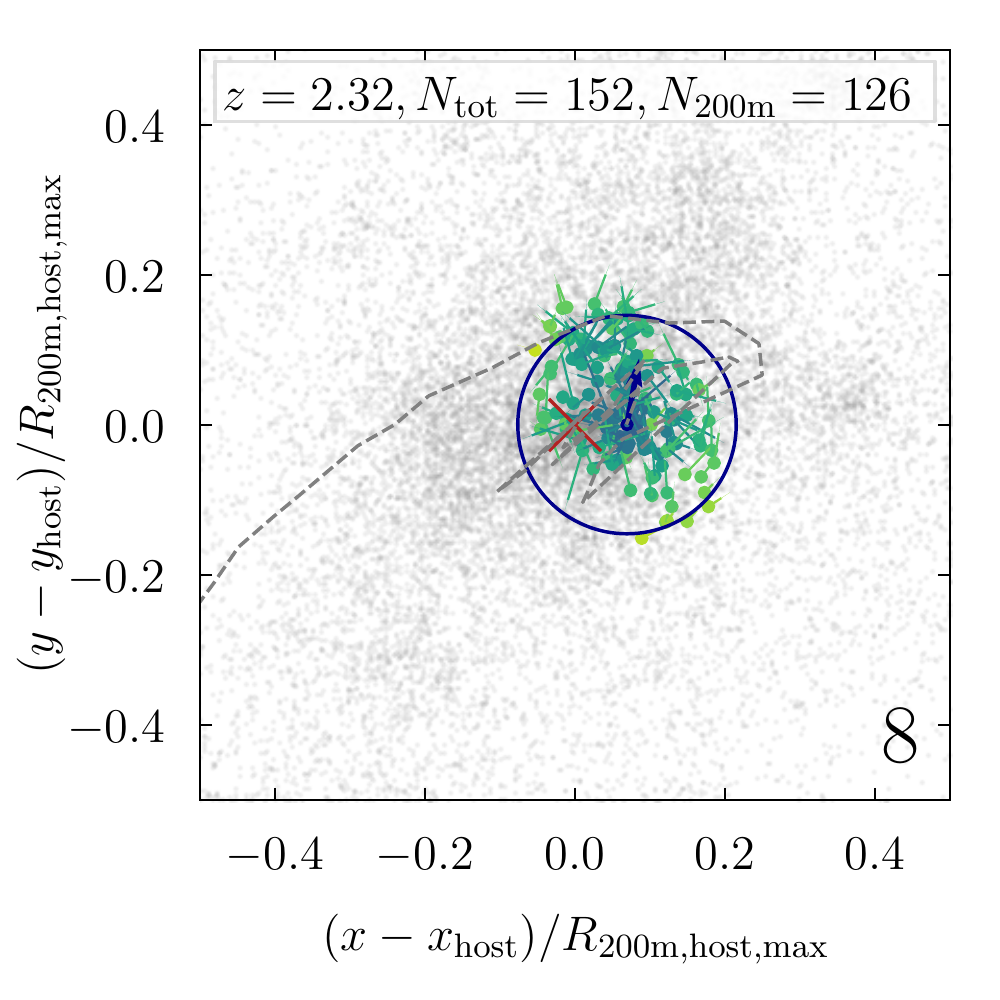}
\includegraphics[trim =  20mm 2mm 4mm 4mm, clip, scale=\panelsize]{\figdir/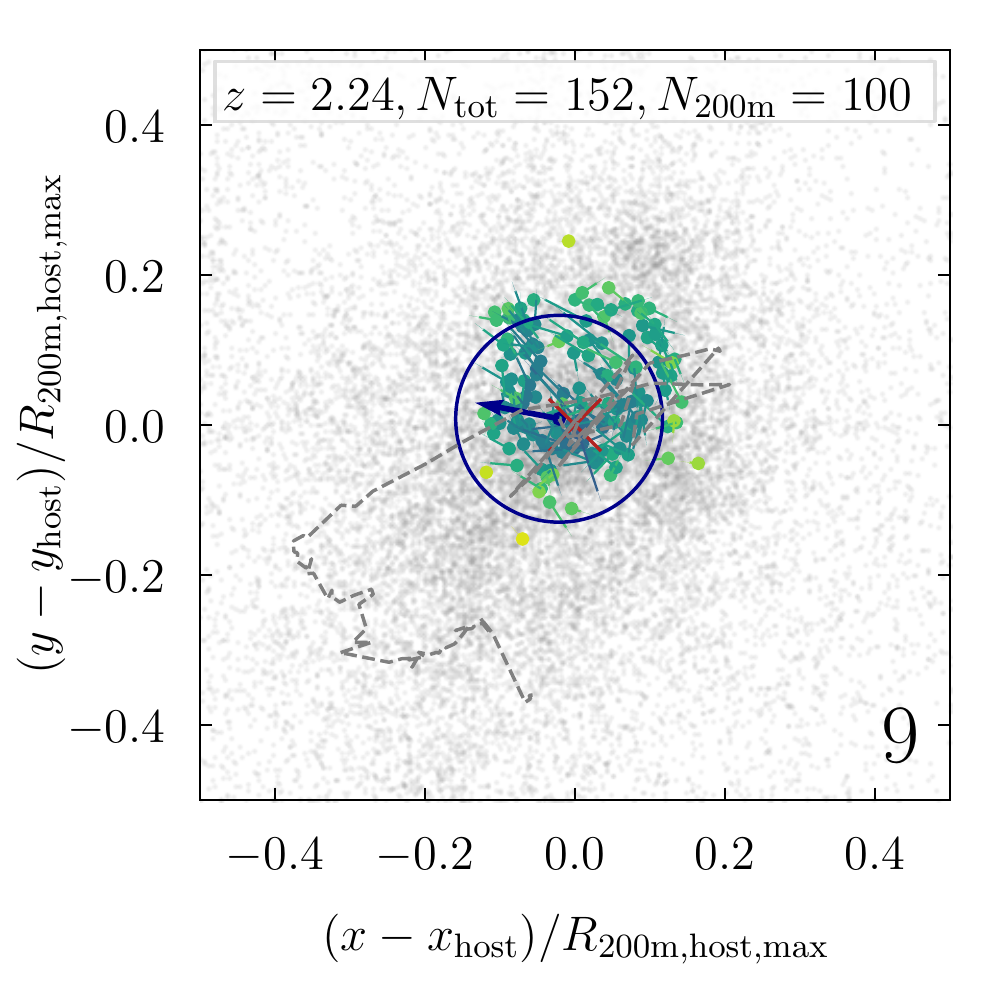}
\includegraphics[trim =  20mm 2mm 4mm 4mm, clip, scale=\panelsize]{\figdir/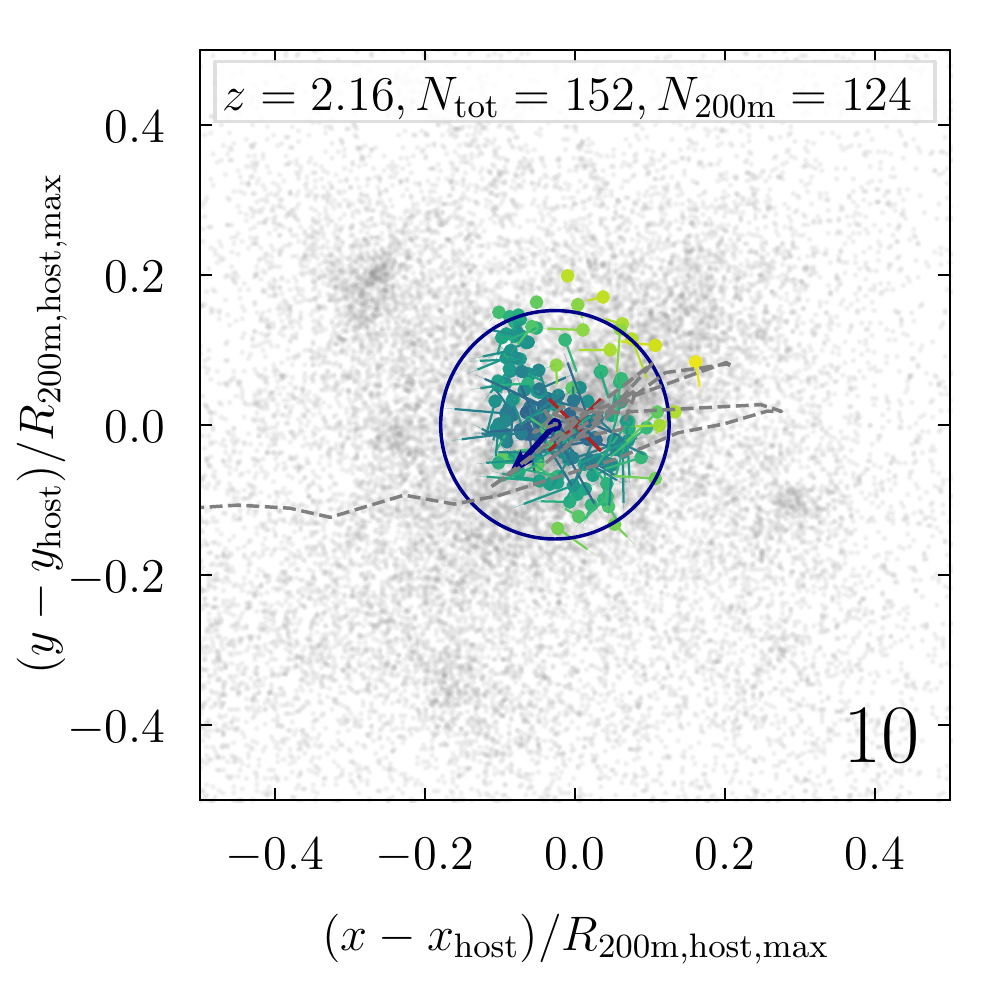}
\includegraphics[trim =  20mm 2mm 4mm 4mm, clip, scale=\panelsize]{\figdir/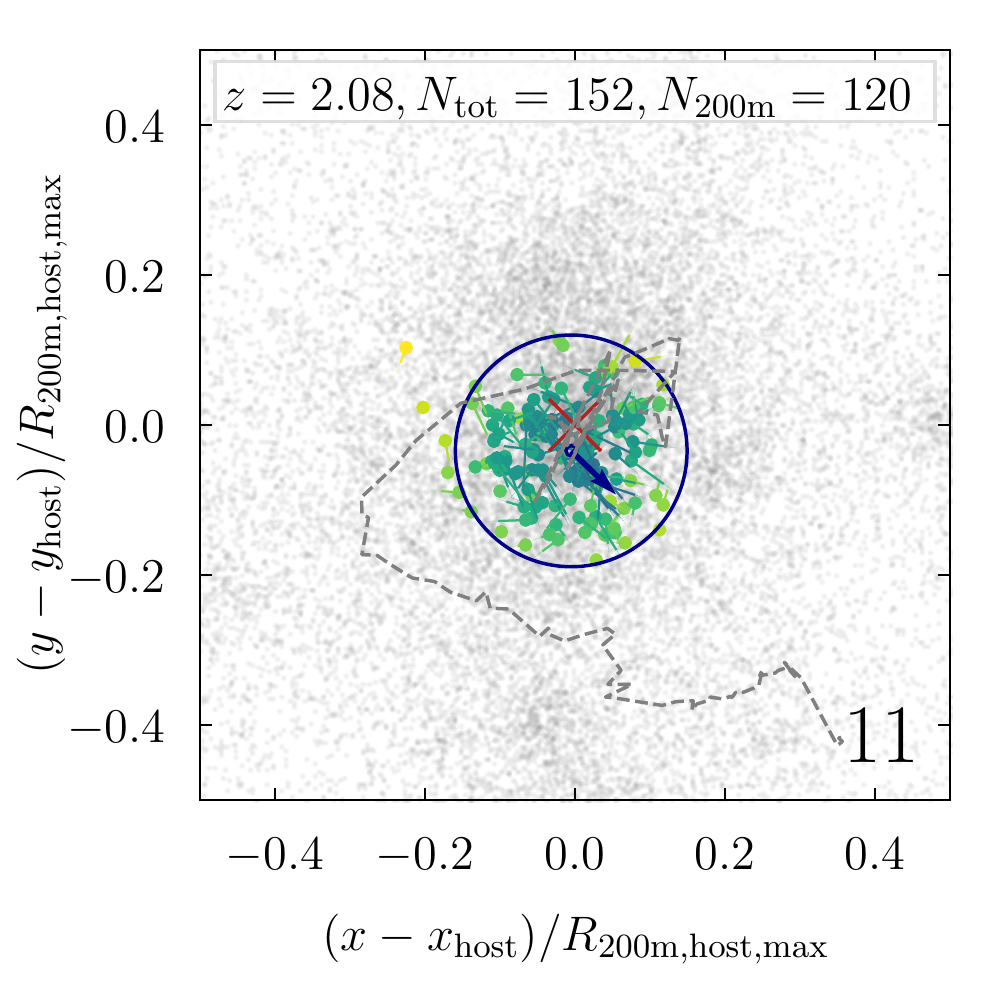}
\includegraphics[trim =  20mm 2mm 4mm 4mm, clip, scale=\panelsize]{\figdir/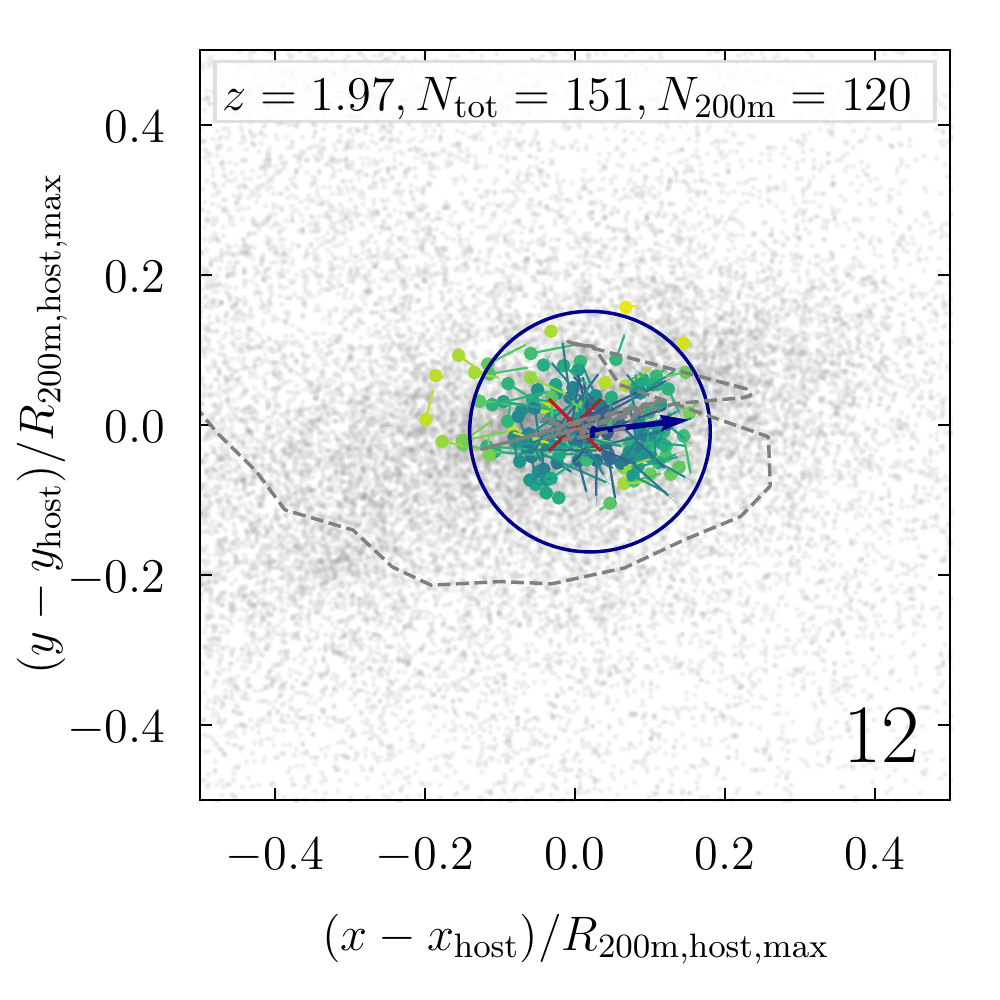}
\caption{Same as Fig.~\ref{fig:ghost1}, but for a ghost that sinks to the host centre after experiencing strong dynamical friction. The subhalo falls in at $z \approx 4$, tidally disrupts to some extent on its first orbit, and sinks close to the host centre (first three panels). For a few snapshots, \sparta's tracked particles and \rockstar's FOF group do not agree in terms of position and velocity (4th and 5th panels). Eventually, they fall into agreement again but \rockstar assigns the halo a larger $\rtom$ (6th panel). Nevertheless, in the next snapshot, the halo is no longer deemed a separate entity by \rockstar and becomes a ghost (7th panel). From that point onward, the particles orbit close to the halo centre and the ghost loses no particles (8th through 11th panels). Finally, \sparta's algorithm ends the ghost because its centre has overlapped the host centre for a significant fraction of a dynamical time (12th panel). In this particular case, it is easy to see why \rockstar merged the two haloes, given that their phase space structure is similar (as indicated by the small relative velocity).}
\label{fig:ghost2}
\end{figure*}

\begin{figure}
\centering
\includegraphics[trim =  15mm 13mm 7mm 7mm, clip, width=0.47\textwidth]{\figdir/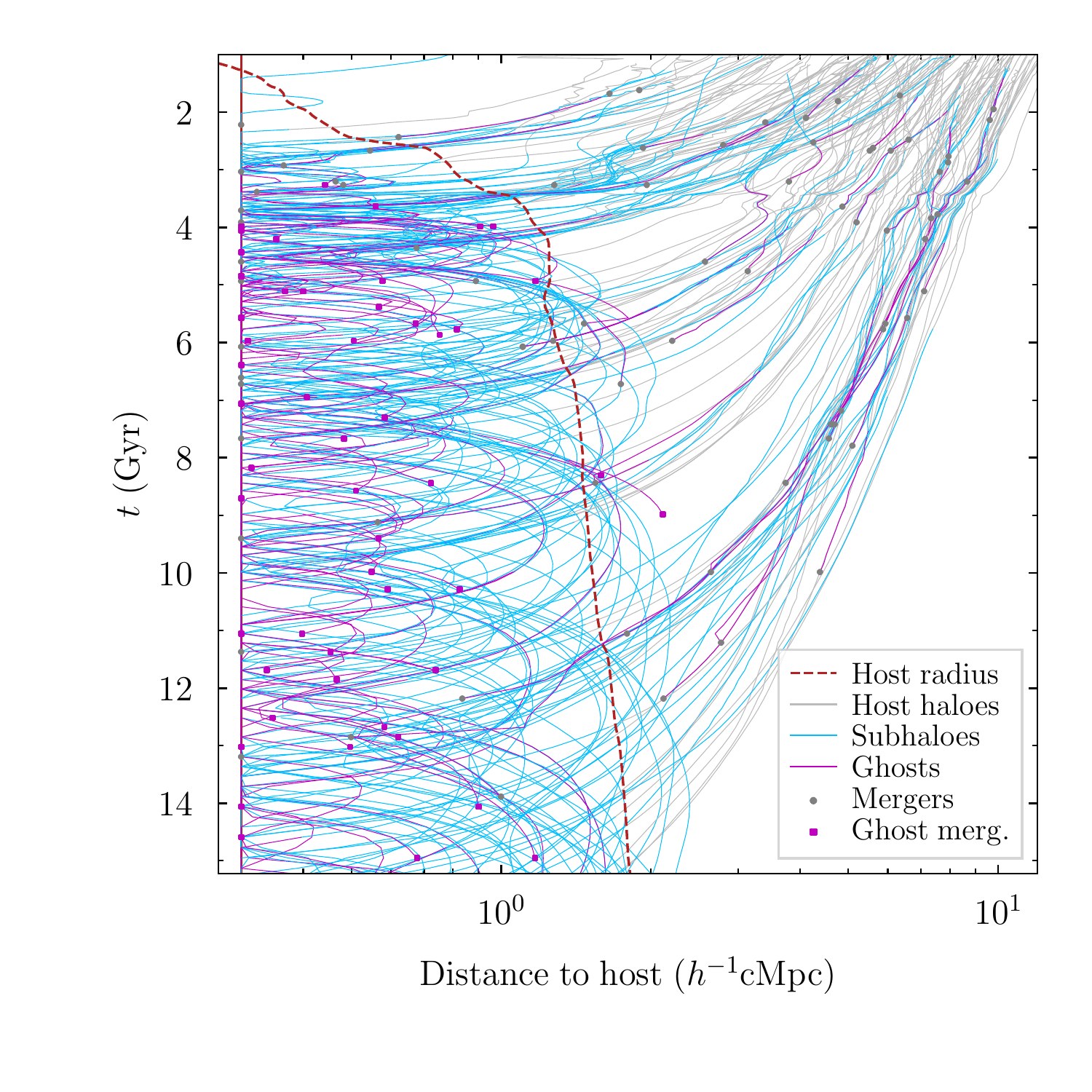}
\caption{Visualization of the merger tree of the largest halo in our test simulation, with ghosts highlighted in magenta. The trajectories reflect the distance from the main halo in comoving units, but they are arbitrarily cut off on the left at a small radius. Each track represents a halo that becomes a subhalo of the main halo, merges into it, or merges into one of its subhaloes. The colour of the lines indicates epochs when those haloes are hosts (gray), subhaloes (light blue), and ghosts (purple). Including ghosts not only adds more subhaloes but also changes the times when mergers occur (gray dots). Ghosts merge into the main across a wide range of radii (highlighted with magenta squares), sometimes even outside of $\rtom$.}
\label{fig:trees}
\end{figure}

Inspecting the \rockstar merger trees for our \LCDM simulations, we find that about half of all halo trajectories reach the end of the simulation and half merge into larger haloes at some previous time (the exact numbers vary depending on box size). Some of the mergers are physical, meaning that the ``real'' subhalo would have been disrupted or mixed with the phase space of its host to the point where it is not a distinct, bound entity any longer; some mergers are due to numerical over-merging, as discussed in Section~\ref{sec:intro}; and some subhalo trajectories end because the halo finder loses track of them. 

To identify and fix instances of the latter, we introduce the concept of a ghost: a subhalo that is no longer detected by the halo finder but that can be identified in the particles that originally belonged to the subhalo at infall. Specifically, our algorithm identifies all particles that make up a subhalo at infall (Section~\ref{sec:method:tracking:identify} and Appendix~\ref{sec:app:members}). It spawns a ghost whenever a subhalo merges according to the merger trees and computes its position and velocity based on the tracked particles (Section~\ref{sec:method:tracking:pos}). It removes (but never adds) particles as the subhalo is being stripped (Section~\ref{sec:method:tracking:evolution}) and it ends the ghost once it has truly merged or contains too few particles to be tracked (Section~\ref{sec:method:tracking:end}). These algorithms are implemented in the \sparta/\moria framework \citep{diemer_17_sparta, diemer_20_catalogs}.

For visual guidance, Figs.~\ref{fig:ghost1} and \ref{fig:ghost2} show examples of the lives of two ghost haloes. Fig.~\ref{fig:trees} shows a visualization of a merger tree in our test simulation with and without ghosts.

\subsubsection{Identifying subhalo member particles}
\label{sec:method:tracking:identify}

The question of subhalo membership is fundamentally ill-posed. We wish to discern particles that physically belong to a subhalo from host particles that happen to be near it, but up to which point before infall can subhaloes accrete new particles? At what point does ``infall'' occur? Can subhaloes add particles once they are inside the host? Our approach is to attempt reasonable answers to these questions and to verify that they do not lead to sudden mass gains or losses when subhaloes cross the (arbitrary) infall boundary. In this section, we briefly summarise our algorithm, referring the reader to Appendix~\ref{sec:app:members} for details. The new algorithm represents a significant improvement over the one of \citet{diemer_17_sparta}, and it was used for all calculations in \citet{diemer_20_catalogs}.

When a halo becomes a subhalo by crossing a larger host halo's $R_{\rm 200m,bnd}$, we consider all particles within $2\ R_{\rm 200m,sub}$ of the subhalo centre as possible members. We require at least one of two criteria to be fulfilled: that the particle entered the subhalo well outside the host's zone of influence, or that the particle is strongly gravitationally bound to the subhalo. Specifically, we consider a particle to belong to a subhalo if it first entered the subhalo's $\rtom$ at a distance at least $2\ R_{\rm 200m,host}$ from the host centre, which excludes host particles that happen to be currently co-located with the subhalo. However, some particles can become physically bound to the subhalo as it travels through the host's outskirts \citep[e.g.,][]{behroozi_14}. Thus, we also consider a particle to belong to a subhalo if its kinetic energy is less than the gravitational binding energy to all particles within $0.5\ R_{\rm 200m,sub}$ of the subhalo centre. We further discuss this algorithm (and a possible third criterion) in Appendix~\ref{sec:app:members}.

Having determined a set of subhalo particles, we impose the condition that the subhalo will not accrete any more particles as long as it is a subhalo according to the halo finder. This condition may not be strictly true, especially in major mergers \citep[e.g.,][]{han_12_hbt}. However, the number of added particles should generally be small compared to the subhalo's initial mass, and deciding subhalo membership at each snapshot would be complicated and computationally expensive. 

\subsubsection{Ghost creation, position, and velocity}
\label{sec:method:tracking:pos}

When a subhalo disappears from the merger tree because \rockstar either deems it to have merged or lost it otherwise, we continue its life a ghost. If the ghost's host merges into another halo itself, the ghost transfers to this new host. We cannot rely on the halo finder for the positions, velocities, and radii of the subhalo any longer, and thus compute them directly from the particle distribution as follows. 

The mean particle position tends to be a poor estimate of the bound core's location since the distribution is often anisotropic and rapidly changing. We thus compute the gravitational self-binding energy of the tracked particles (colour scale in Figures~\ref{fig:ghost1} and \ref{fig:ghost2}) and consider only the most bound quartile of particles. If there are fewer than $10$ particles in that set (or $40$ tracked particles overall), we accept the position of the single most bound particle as the ghost centre. If there are between $10$ and $20$ ($40$ and $80$ particles overall), we take their centre of mass to be the new ghost centre. If there are more than $20$ ($80$ particles total), we still only consider the $20$ most bound particles. Once the centre has been determined, we calculate $\rtom$ from the tracked particles and define the ghost velocity as the mean particle velocity of the tracked particles within $\rtom$ (as opposed to all tracked particles, which extend to $2\ \rtom$). This definition avoids biases due to particles that have strayed far from the ghost centre. 

We have compared our position estimate to \rockstar and find that it generally agrees well. For example, at $z \approx 2$, the majority of subhalo centres are within $0.1\ \rtom$ of each other. The first panels of Figures~\ref{fig:ghost1} and \ref{fig:ghost2} give a fairly typical example of the agreement between the \sparta and \rockstar positions and velocities, although the agreements tends to get worse for haloes with fewer particles. There are also cases of disrupting haloes where \sparta tracks a genuinely different set of particles than those in \rockstar's FOF group, leading to large offsets (e.g., Fig.~\ref{fig:ghost2}).

\subsubsection{Mass and radius evolution}
\label{sec:method:tracking:evolution}

At each snapshot, we locate the tracked particles belonging to each subhalo or ghost and compute its spherical overdensity radii including only tracked particles. We will refer to the corresponding radii and enclosed masses as ``tracer radii'' and ``tracer masses.'' The tracer definition has a number of advantages. First, it is guaranteed not to include any host material. Second, it avoids spikes and other rapid changes as the particle distribution is compressed or stretched by tidal forces (e.g., first six panels of Fig.~\ref{fig:ghost2}). Third, not all tracked particles that count into the tracer mass need to be gravitationally bound, which avoids ill-defined gravitational boundness criteria that can lead to a noisy subhalo mass evolution \citep[Appendix~\ref{sec:app:members},][]{vandenbosch_17}. Tracer radii are allowed to be larger than the distance to the farthest tracked particle, a case that arises for compact particle distributions (typically near pericentre). The calculation fails only if the subhalo density is too low to reach the threshold at its centre. 

After each snapshot, we permanently remove particles that have drifted beyond a maximum radius of $2 \rtom$ (computed from bound particles in normal subhaloes and from tracer particles in ghosts). Tracer masses typically decrease with time as particles are stripped, but they can increase if particles that were outside a given overdensity radius return to smaller distances. Tracer radii also grow through pseudo-evolution as the overdensity threshold decreases with cosmic time \citep{diemer_13_pe}. For example, the radius in the first panel of Fig.~\ref{fig:ghost1} appears to be about the same as in the 10th panel, even though the particle number has decreased from $229$ to $54$. 

We find that the initial subhalo particle tagging does matter for the ghost masses. Erroneously tagged particles should quickly drift away from the subhalo or ghost, but nonetheless the initial assignment can have an effect on tracer masses. For example, simply assigning all particles within $R_{\rm 200m,sub}$ at infall significantly inflates the average tracer masses until the ghosts merge into the host's centre. We conclude that the absolute tracer masses are, to some extent, dependent on our definition of subhalo membership. We compare tracer and bound masses in detail in Section~\ref{sec:results:tracer}.

\subsubsection{Ending a ghost}
\label{sec:method:tracking:end}

The final step in our algorithm is to check whether a ghost is still a meaningful unit, whether it has been disrupted (physically or numerically), or whether it has merged with its host. The case of disruption is easy to detect: we abandon a ghost if $\ntom < 10$, a limit the user can adjust (Fig.~\ref{fig:ghost1}). However, a significant fraction of ghosts sink to the host centre and remain there indefinitely \citep[see also][]{han_18_hbt}. As all of their particles are on low-radius orbits, they can maintain $\ntom > 10$ even though their particles are not meaningfully distinct from their host's any more. We thus end a ghost when it has been within $0.05\ R_{\rm 200m,host}$ of the host centre for $0.5$ current dynamical times\footnote{The noise in the measurement of the ghost centre can exceed $0.05\ \rtom$. We avoid artificially keeping the ghost alive by subtracting the uncertainty in the ghost position from the distance. This uncertainty is estimated as $\sigma_{\rm x} = \sigma_{\rm r} / \sqrt{N}$, where $\sigma_{\rm r}$ is the standard deviation in the radius of the $N$ particles used to determine the centre. This estimate is approximate, particularly if we take the position of the most bound particle as the ghost centre. Further simplifying the expression, we find that $\sigma_{\rm r} \approx 0.5\ \rtom$ for the vast majority of haloes, leading to the simple estimate of $\sigma_{\rm x} \approx 0.5\ \rtom / \sqrt{N}$. } (Fig.~\ref{fig:ghost2}; the user can change these parameters). When a ghost ends, we assign its host (as previously determined by \consistenttrees) as the halo into which the ghost has merged. If the host itself merges into another halo, the ghost can live on in that new host and eventually merge with it (Fig.~\ref{fig:ghost1}).


\def\figsize{0.65}
\begin{figure*}
\centering
\includegraphics[trim =  2mm 0mm 0mm 0mm, clip, scale=\figsize]{\figdir/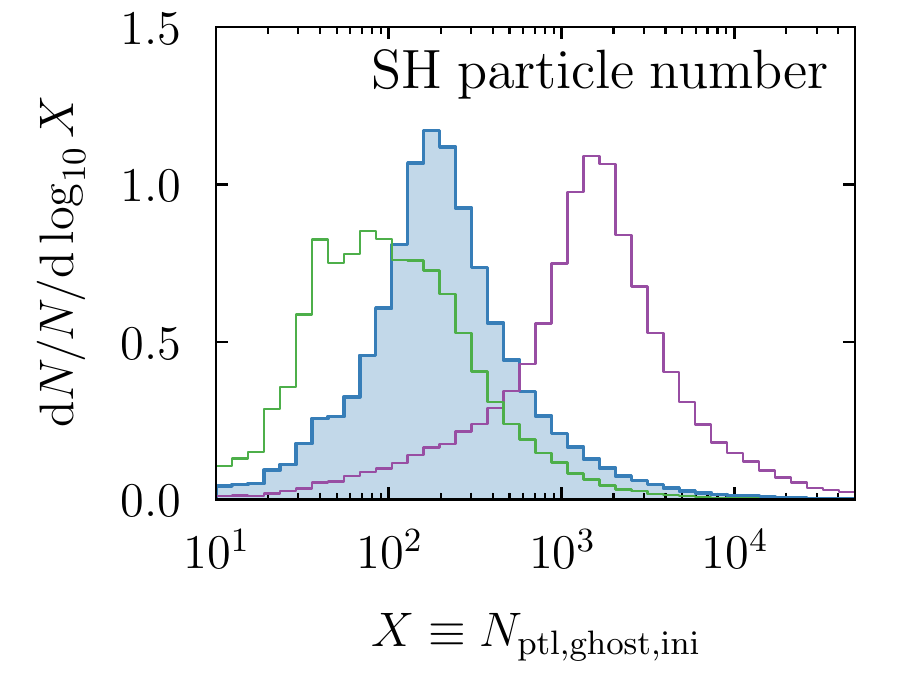}
\includegraphics[trim =  2mm 0mm 0mm 0mm, clip, scale=\figsize]{\figdir/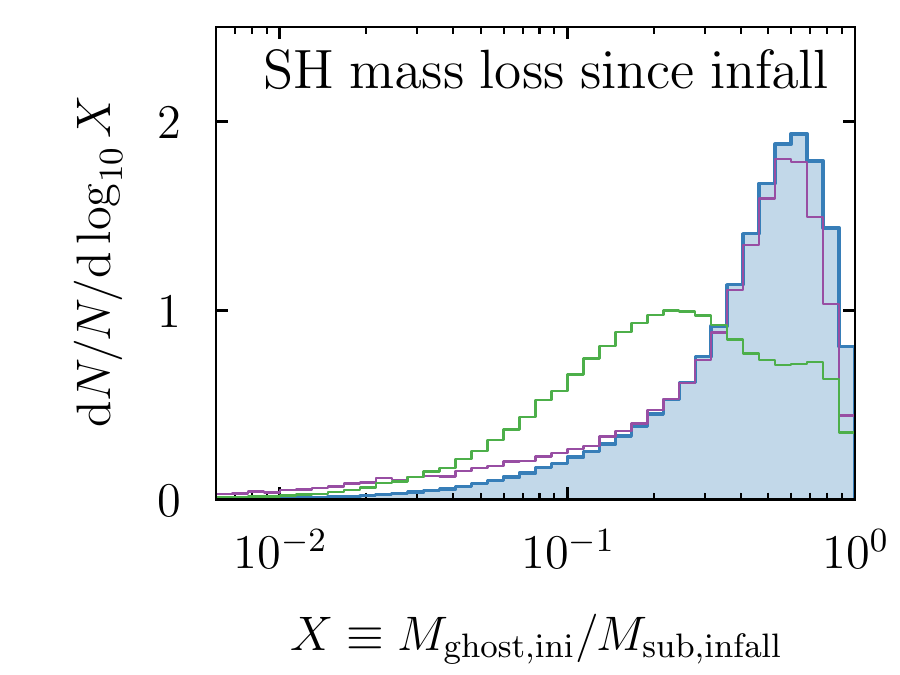}
\includegraphics[trim =  2mm 0mm 0mm 0mm, clip, scale=\figsize]{\figdir/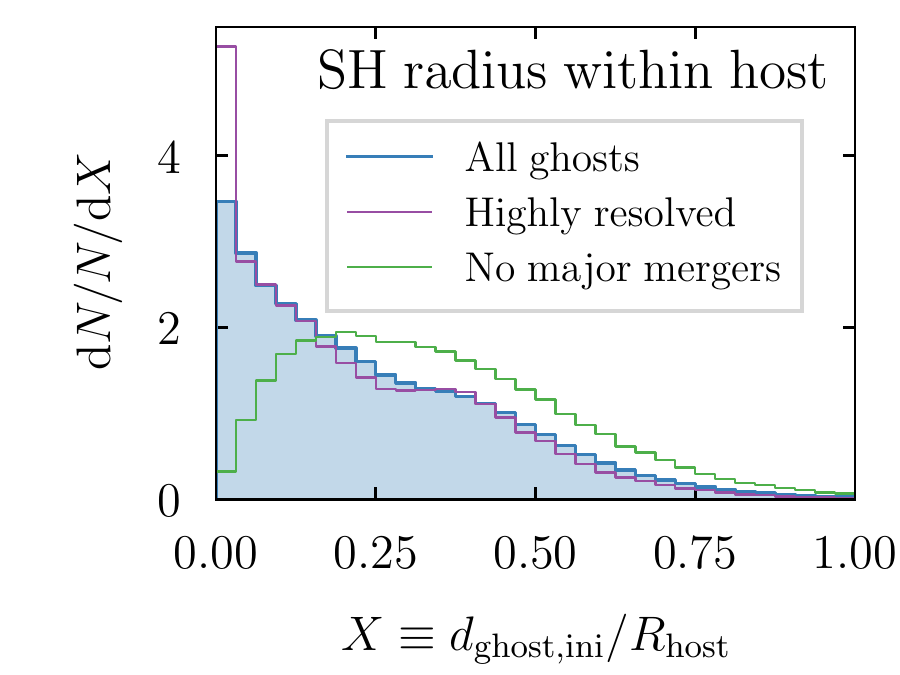}
\includegraphics[trim =  2mm 0mm 0mm 0mm, clip, scale=\figsize]{\figdir/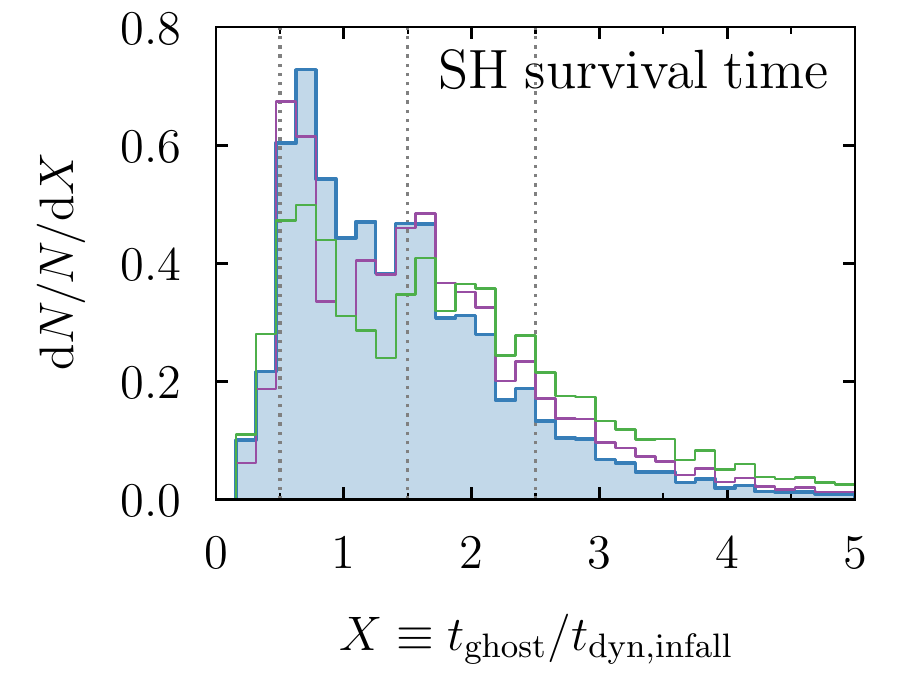}
\includegraphics[trim =  2mm 0mm 0mm 0mm, clip, scale=\figsize]{\figdir/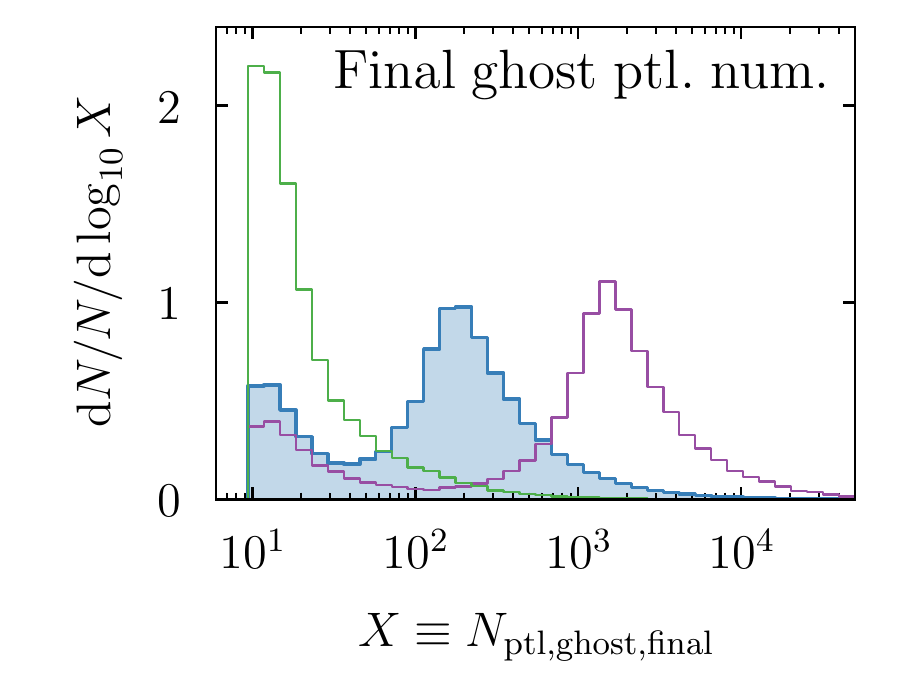}
\includegraphics[trim =  2mm 0mm 0mm 0mm, clip, scale=\figsize]{\figdir/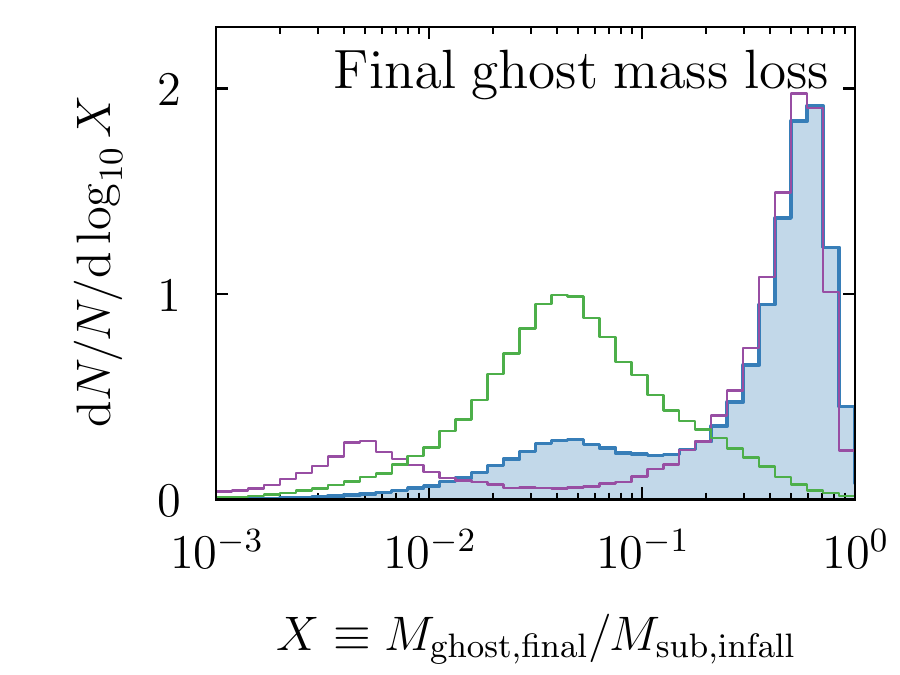}
\caption{Properties of subhaloes when they merge or are lost by \rockstar. The first four panels show the final properties of subhaloes (and thus the initial properties of ghosts), whereas the last two panels show the properties of ghosts when they are abandoned. We compare all ghosts in the \wmap sample (blue), only those where the subhalo contained at least $\ntom \geq 2000$ particles at infall (purple), and only those with $M_{\rm sub} / M_{\rm host} < 0.1$ at infall (green). We combine all simulations and redshifts because the differences between them do not change the overall picture. {\it Top left:} Subhaloes are lost at a wide range of particle numbers. The distribution strongly depends on the initial resolution, demonstrating that a low particle number is not the main reason for subhalo loss. {\it Top centre:} The typical mass loss experienced prior to becoming a ghost does not depend on resolution and peaks at about $40\%$, which reaffirms that sheer loss of particles cannot be the dominant factor. {\it Top right:} Subhaloes are much more likely to be lost near the host centre, but many of those cases are major mergers that have sunk quickly due to dynamical friction. Minor-merger subhaloes tend to get lost at intermediate radii. {\it Bottom left:} Peaks in the survival time at $1/2$, $3/2$, and $5/2$ crossing times indicate losses near pericentre. However, the peaks significantly extend to later times (especially for minor mergers), indicating that losses can occur near apocentre following a pericentric passage. {\it Bottom centre:} The double-peaked distribution of final ghost particle counts corresponds to the two criteria for ending a ghost, namely that it has lost all but $10$ particles (the majority of minor mergers) or that it has physically merged into the host centre (the typical outcome of major mergers). {\it Bottom right:} The double-peaked structure is also visible in the distribution of final ghost masses compared to subhalo mass at infall. The mass loss can reach arbitrarily small fractions, limited only by the resolution of the simulation.}
\label{fig:stats}
\end{figure*}

\section{Results I: Subhalo evolution and loss}
\label{sec:results}

In this section we investigate the evolution and loss of subhaloes. We begin by quantifying when and where subhaloes become ghosts in Section~\ref{sec:results:stats}. We compare their tracer masses to conventional bound masses in Section~\ref{sec:results:tracer} and consider the evolution of these masses in Section~\ref{sec:results:subevo}. We investigate the impact of ghosts on various summary statistics for Section~\ref{sec:results2}.

The following figures are based on combined subhalo trajectories from all redshifts and from seven simulations with different box sizes (Section~\ref{sec:method:sims}). We have confirmed that results from the individual simulations broadly agree, but there are trends with halo mass, redshift, and resolution that we average over. We include all haloes that reach a mass $\mtom$ of at least $200$ particle masses at some point in their history. The resulting sample contains about $1.5$ million subhalo trajectories, about \num{700000} of which result in a ghost (while the rest of the subhaloes reaches the end of the respective simulation). A single halo can contribute multiple subhalo trajectories if it intermittently becomes a host again. Unless otherwise mentioned, we use $R_{\rm 200m,bnd}$ to separate subhaloes from hosts (Section~\ref{sec:method:halos}).

\subsection{When, where, and why do subhaloes get lost?}
\label{sec:results:stats}

Fig.~\ref{fig:stats} shows histograms of the properties of ghosts at their creation (at the epoch when subhaloes are lost, first four panels) and when they end (last two panels). To isolate resolution-dependent trends, we compare to highly resolved haloes ($\ntom > 2000$ at infall, purple) and minor mergers ($\mu \equiv M_{\rm 200m,bnd,sub} / M_{\rm 200m,bnd,host} \leq 0.1$). The overall sample (blue) is dominated by major mergers with $\mu > 0.1$. The main conclusion is that most subhalo losses occur not due to low particle resolution but due to tidal deformations around (but not necessarily close to) the host centre. \rockstar sometimes drops subhaloes in such situations because more than half of their particles are gravitationally unbound. Simply not imposing such a limit is not a solution because it would lead to numerous tidal tails being spuriously be classified as haloes \citep{behroozi_13_rockstar}.

The top left panel of Fig.~\ref{fig:stats} shows the number of tracked particles in subhaloes when they disappear from the halo catalogues. The vast majority of ghosts are created with more than $25$ particles, which is roughly the limit to which \rockstar can identify haloes \citep{behroozi_13_rockstar}. Moreover, the distribution shifts to higher numbers when we require the subhalo to have had a higher initial particle count, demonstrating that there is no ``magic'' number of particles at which subhaloes are lost \citep[see also][]{behroozi_19}. Strikingly, there also does not seem to be a clear upper cutoff, with some subhaloes being lost when they still contain \num{500000} particles. Some such events are major mergers that may have physically joined the host, but some are still a distinct, moving substructure when they are lost. We highlight one such example in Section~\ref{sec:discussion:major}. The distribution of mass loss rates (top centre panel of Fig.~\ref{fig:stats}) leads to similar conclusions as the particle number, namely that there is no preferred fractional mass loss that leads to missed subhaloes. The distribution peaks around a remaining mass of 60\% compared to infall ($20\%$ for minor mergers) but it is broad and spans from 1\% to unity. The tail towards strongly stripped subhaloes is more prominent in the high-resolution sample because mass losses of 99\% cannot be resolved in subhaloes that have fewer than $2500$ particles (assuming a lower limit of $25$ particles for haloes to be detected).

We can gain further insight into the causes of subhalo losses from the locations and times where and when they occur. The top right panel of Fig.~\ref{fig:stats} shows that losses peak near the host centre, but the distribution is broad and reaches all the way to the host's $\rtom$. We expect that physical mergers, where the subhalo has truly joined the host's phase space, should be strongly peaked in the innermost bins. Notably, the distribution of small subhalo losses (green) is peaked around a quarter of the host radius. For all samples, we find that about half of subhalo losses occur at or near pericentre (within about $30\%$ of the smallest distance we record). This finding is mirrored in the time distribution (bottom left panel of Fig.~\ref{fig:stats}), which shows broad peaks at $1/2$ and $3/2$ crossing times (near pericentre) that extend towards $1$ and $2$ crossing times (near apocentre). 

In summary, a picture arises where some large subhaloes physically merge at small radii and some are lost prematurely, while small subhaloes are lost across a wide range of radii that includes their apocentres. We speculate that losses tend occur due to tidal forces that are strongest at pericentre. However, much of the mass loss and deformation in physical space can occur near apocentre (e.g., Fig.~\ref{fig:ghost1} and Section~\ref{sec:results:subevo}). Such situations may lead to a large fraction of formally unbound particles and the subhalo being dropped, spreading out the distribution of loss radii and times.

The bottom centre and bottom right panels of Fig.~\ref{fig:stats} show the final particle number and mass loss fractions of ghosts. These distributions reflect our criteria for when a ghost cannot be tracked any longer or has merged (Section~\ref{sec:method:tracking:end}). Most ghosts end at substantial particle numbers because they have merged, with a distribution that depends on the number of particles at infall (purple). The second group, namely those ended because their particle number decreases to $10$ or below, contains the majority of minor mergers and exists independently of how many particles the initial subhalo contained. The bottom right panel demonstrates that our algorithm can track ghosts to arbitrarily small fractions of the initial subhalo mass.

\subsection{Tracer vs. bound masses}
\label{sec:results:tracer}

\def\figsize{0.56}
\begin{figure}
\centering
\includegraphics[trim =  4mm 6mm 3mm 1mm, clip, scale=\figsize]{\figdir/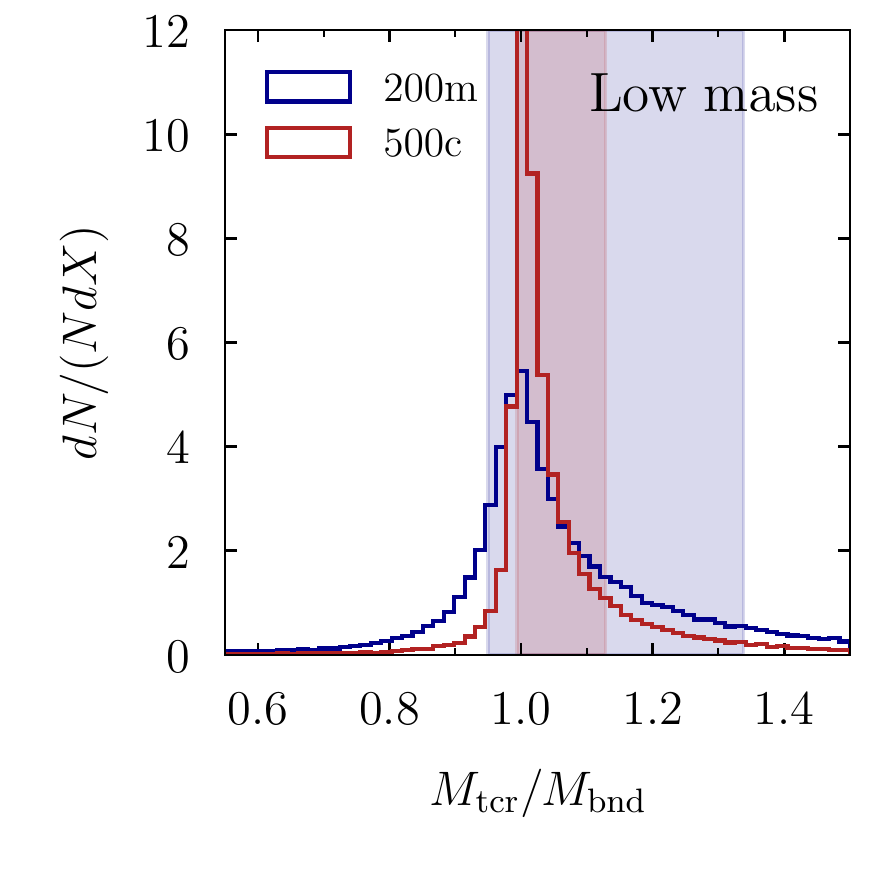}
\includegraphics[trim =  21mm 6mm 3mm 1mm, clip, scale=\figsize]{\figdir/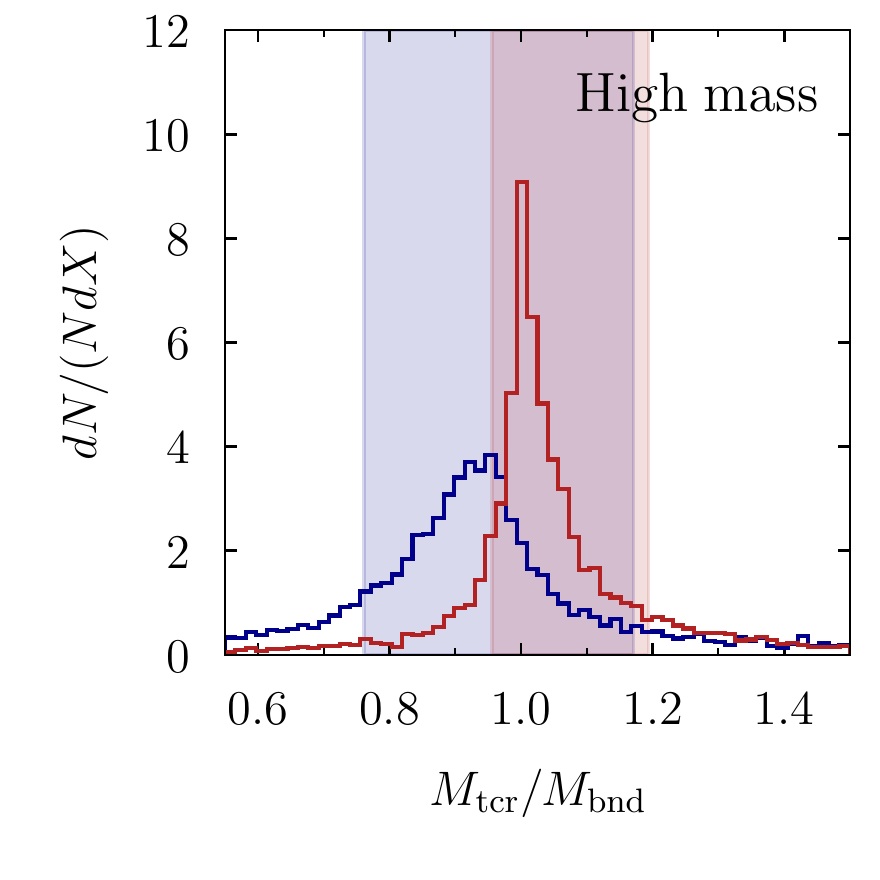}
\caption{Ratio of tracer to bound masses for low-mass (left) and high-mass (right) subhaloes in the \wmap sample at $z = 0$. While the distribution has large tails, the 68\% and 95\% intervals  (shaded areas) are relatively tight and overlap with unity. High-threshold definitions such as $\mfoc$ (red) lead to better agreement than low thresholds such as $\mtom$ (blue) because they include mostly particles that are unambiguously bound. The peak of the $\mfoc$ histogram in the left panel is cut off, highlighting the striking agreement (1-$\sigma$ range of just over 10\%).}
\label{fig:ratio_tracer_lin}
\end{figure}

As described in Section~\ref{sec:method:tracking:pos}, we measure tracer radii and masses that include only tracked particles, side-stepping the ill-defined question of gravitational boundness. Before we can meaningfully consider the evolution of these masses for ghosts, we need to check how similar they are to bound-only definitions from \rockstar (Section~\ref{sec:method:halos}). Fig.~\ref{fig:ratio_tracer_lin} shows histograms of the mass ratio for subhaloes in the \wmap sample at $z = 0$, including only subhaloes that currently have at least $200$ particles within $R_{\rm 200m,bnd}$. We focus on the overall sample, neglecting differences between subhaloes that have experienced strong stripping, populations at different radii, and so on. We split the sample into two coarse mass bins with $1.4 \times 10^{10} < M < 3.2 \times 10^{12}\ \msunh$ (peak heights between $0.5$ and $1$, left) and $3.4 \times 10^{13} < M < 1.4 \times 10^{14}\ \msunh$ (peak heights between $1.5$ and $2$, right). We further restrict the plot to only the low and high density thresholds of $\mtom$ and $\mfoc$ because the intermediate $\mvir$ and $\mtoc$ definitions compare similarly. The corresponding ratios of spherical overdensity radii are smaller by a factor of $R \propto M^{1/3}$.

The 68\% intervals (shaded areas in Fig.~\ref{fig:ratio_tracer_lin}) demonstrate the close agreement between tracer and bound masses, especially for high-density definitions such as $\mfoc$. Here, the interval varies between 14\% at low masses and $z = 0$ to about 40\% at high mass and $z \approx 2$. For $\mtom$ the interval remains between 40\% and 55\%, with the same trend that the agreement gets slightly worse with redshift. Overall, 95\% of subhaloes have mass ratios within a factor of two. The tails are somewhat asymmetric, with fewer ratios below unity than above. 

As evidenced by the lack of strong tails to low ratios, it is rare for the subhalo tagging to miss a significant number of particles that are gravitationally bound. Differences could arise due to particles that are re-bound after previously having left the subhalo \citep[e.g.,][]{han_12_hbt}. The halo finder would include such particles, but also host particles that become temporarily ``bound'' because they happen to roughly match the subhalo velocity. The more substantial tails towards higher ratios, especially in $\mtom$ at low masses, reflect cases where particles are not technically bound but deemed to belong to the subhalo according to our tagging scheme. Overall, the good agreement gives us confidence that the tracer mass determination is sensible and roughly corresponds to bound mass.

\subsection{Mass loss of subhaloes and ghosts}
\label{sec:results:subevo}

Having convinced ourselves that tracer masses approximate bound mass, we can now compare the mass evolution of subhaloes and ghosts over time. Fig.~\ref{fig:evo} visualises the average evolution using two different metrics, namely the fraction of mass remaining and the mass loss rate. We consider epochs both before and after infall, which we define as the final snapshot before the halo becomes a subhalo (vertical gray line in Fig.~\ref{fig:evo}). We rescale time by the dynamical time at infall, which allows us to combine trajectories from all redshifts. We consider all subhaloes in the \wmap simulations that live for at least $5$ snapshots in total and whose trajectory includes at least half a dynamical time before and one dynamical time after infall. We emphasise that the results in this figure are significantly biased by subhaloes reaching the resolution limit of $10$ particles and dropping out of the average trajectories. We have quantified the effects of this bias by restricting the figure to subhaloes with at least $2000$ particles at infall. The general shape and relative amplitude of the curves remains roughly the same, but after a few dynamical times they asymptote to a higher remaining mass fraction (by about a factor of two at the final time plotted). In the following, we focus on qualitative interpretations that do not rely on the exact amplitude of the asymptotic mass loss.

\def\figsize{0.61}
\begin{figure}
\centering
\includegraphics[trim =  3mm 22mm 0mm 1mm, clip, scale=\figsize]{\figdir/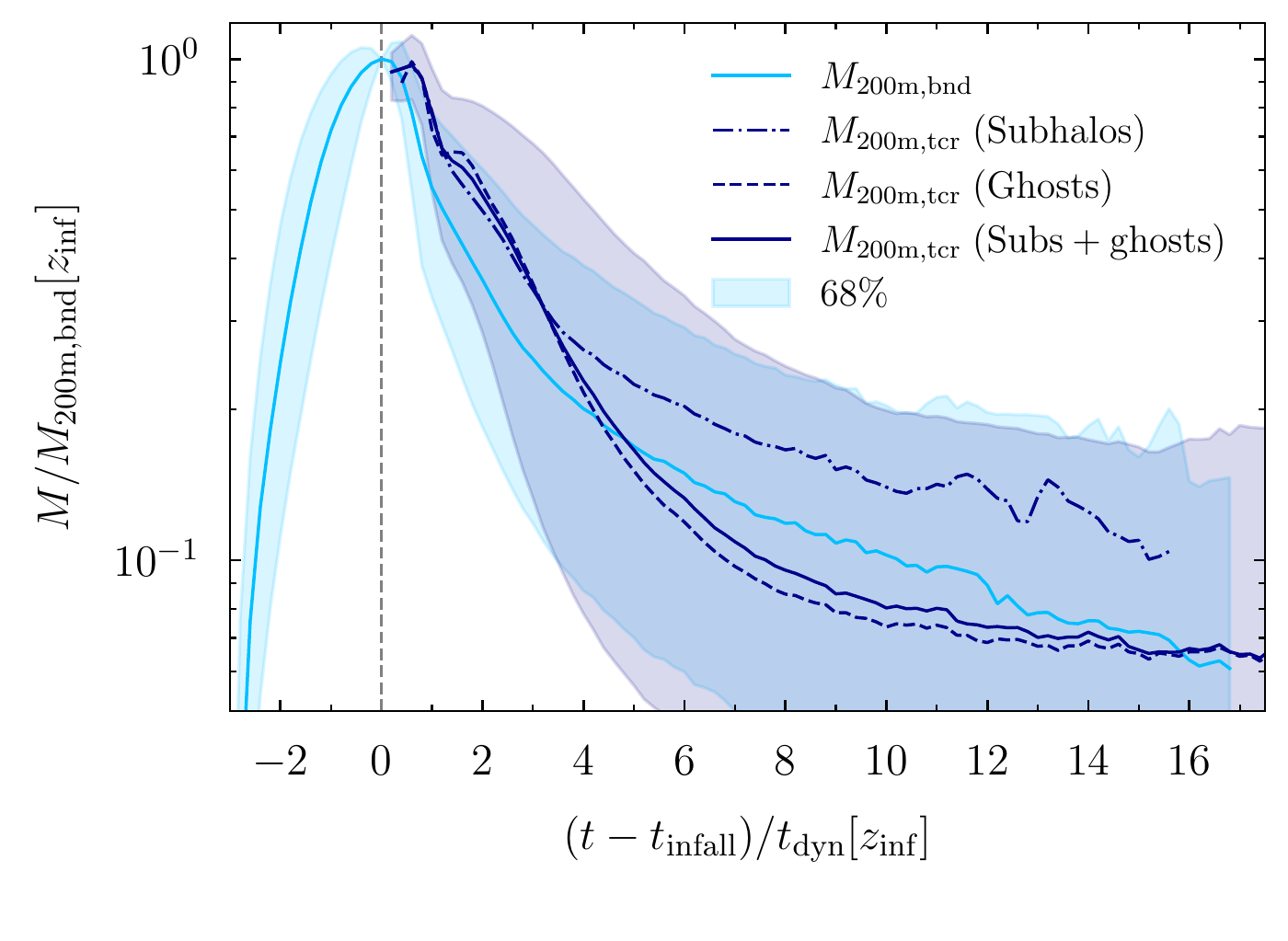}
\includegraphics[trim =  3mm 7mm 0mm 1mm, clip, scale=\figsize]{\figdir/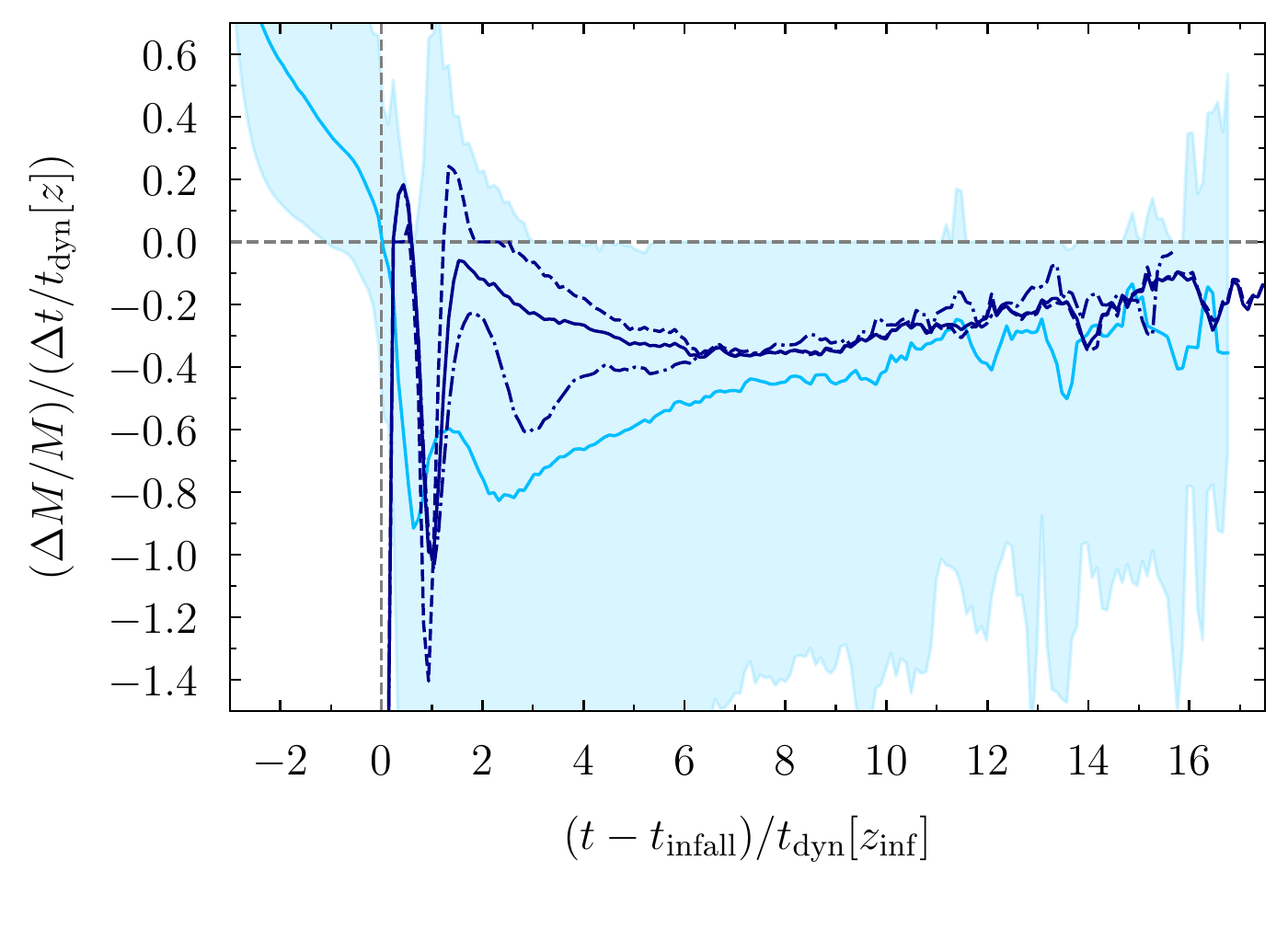}
\caption{Mass evolution before and after infall (top) and logarithmic mass loss rate per dynamical time (bottom) of subhaloes and ghosts according to the bound-only and tracer definitions (using $\mtom$, but other definitions behave similarly). The light blue curves show the bound mass as computed by \rockstar, and the dark blue lines show the tracer mass for all haloes (solid), non-ghosts (dot-dashed), and ghosts (dashed). The shaded areas show the 68\% contours around the bound mass and tracer mass for all subhaloes (the latter only in the top panel to avoid crowding the figure). The trajectories become more and more biased towards higher-mass haloes because haloes near the resolution limit successively fall out of the sample. After a few dynamical times, the sample becomes dominated by ghosts.  {\it Top panel:} Tracer and bound masses evolve similarly, with the tracer mass capturing slightly more particles on average. {\it Bottom panel:} The mass loss rate exhibits extremely large scatter, highlighting that stripping sensitively depends on the individual orbits of subhaloes. The median mass loss rate varies strongly and is highest on average after one crossing time (at first apocentre).}
\label{fig:evo}
\end{figure}

The top panel of Fig.~\ref{fig:evo} shows that tracer and bound masses experience overall similar evolutions, confirming the conclusions of Section~\ref{sec:results:tracer}. The bound mass decreases slightly more rapidly directly after infall. The difference in the average mass persists until about four dynamical times after infall, indicating that there is a population of particles that become formally unbound but do move with the subhalo. The tracer masses exhibit a slight spike directly after infall, which we suspect to correspond to a compression of subhaloes. Given that $M \propto R^3$, the corresponding changes in radius are negligibly small. Ghosts follow roughly the same trajectory as normal subhaloes until about three dynamical times after infall, at which point they lose mass more rapidly and come to dominate the overall sample. After about $10$ dynamical times, the median ghost mass decreases only very little. Here, the sample is dominated by the long-lived remnants of major mergers that have sunk to the centre of their hosts, with their lifetime determined by the criteria for ending ghosts (Section~\ref{sec:method:tracking:end}).

The bottom panel of Fig.~\ref{fig:evo} shows the logarithmic mass loss rate per dynamical time. The scatter in tracer mass losses is comparable to that in the bound mass but is omitted to avoid crowding. A key first impression is that the scatter is enormous, highlighting that mass loss rates depend sensitively on the individual orbits and density structure of subhaloes. On average, the strongest mass loss occurs after one crossing time, which may seem surprising since the strongest tidal forces should occur at pericentre rather than at apocentre. However, the actual mass loss does not necessarily occur where the tides are strongest: subhaloes can pass the host centre without losing much mass before spreading out at apocentre (Fig.~\ref{fig:ghost1}). The differences in the mass loss rates of ghosts and subhaloes are partially a selection effect, given that the orbits of subhaloes influence whether and when they become a ghost. After about six crossing times (three full orbits), the differences disappear. The bound mass behaves somewhat differently directly after infall, as discussed above. 

The mass loss rate in the units of Fig.~\ref{fig:evo} has been described by a number of theoretical models. One popular assumption has been that the orbit-averaged mass loss rate is independent of the time since infall. For example, \citet{jiang_16} describe it as 
\begin{equation}
\label{eq:mass_loss}
\frac{\rmd M_{\rm sub}}{\rmd t} \frac{t_{\rm dyn}(z)}{M_{\rm sub}} = - \calA \frac{4}{\pi} \left( \frac{M_{\rm sub}}{M_{\rm host}} \right)^\zeta
\end{equation}
with $\calA = 0.86$ and $\zeta = 0.07$, meaning that it is only a weak function of the sub-to-host mass ratio $\mu \equiv M_{\rm sub} / M_{\rm host}$ \citep[see also][]{taylor_01_sats, zentner_03, vandenbosch_05, han_16_subs}. The factor of $4 / \pi$ arises due to a slightly different definition of the dynamical time. The model predicts mass loss rates between $-0.7$ and $-1.1$ for mass ratios between $10^{-3}$ and unity, which roughly matches the strongest loss rates we find during the first few orbits. However, it is clear from Fig.~\ref{fig:evo} that the average loss rate varies strongly with time, a conclusion that does not change if we bin the subhaloes by the mass ratio $\mu$. While the mass loss rate after many dynamical times is affected by the aforementioned bias due to small subhaloes being removed, the strongest differences in loss rate occur soon after infall.

There are a number of possible reasons for the disagreement between the model of \citet{jiang_16} and our data. Most notably, \eqmn{eq:mass_loss} was calibrated on the evolution of idealised NFW subhaloes on their first orbit. The model parameters were tuned to match simulated subhalo mass functions and thus implicitly account for the effects of artificial disruption (which removed subhaloes rather than reducing their mass, and is thus not included in Fig.~\ref{fig:evo}). Moreover, the model only accounts for haloes inside of $\rvir$, whereas numerous ``backsplash'' haloes orbit outside of that radius \citep{diemer_21_subs}. The large scatter in the mass loss rate demonstrates that more accurate models need to take the orbital evolution into account. Such models have been proposed, for example by \citet[][see also \citealt{jiang_21}]{green_21}, whose predictions were trained on high-resolution simulations of individual subhaloes \citep{ogiya_19}. We cannot easily evaluate their model, however, because it relies on the density profiles of both hosts and subhaloes at all times.

In summary, we find that subhaloes and ghosts typically lose the majority of their mass within the first two orbits (four crossing times). Thereafter, the mass loss rate slowly decreases. We have also tried to quantify the mass loss rate as a function of radius rather than time, but this relation demands a careful reconstruction of the poorly time-resolved subhalo trajectories near pericentre.


\begin{figure*}
\centering
\includegraphics[trim =  0mm 1mm 1mm 1mm, clip, scale=0.75]{\figdir/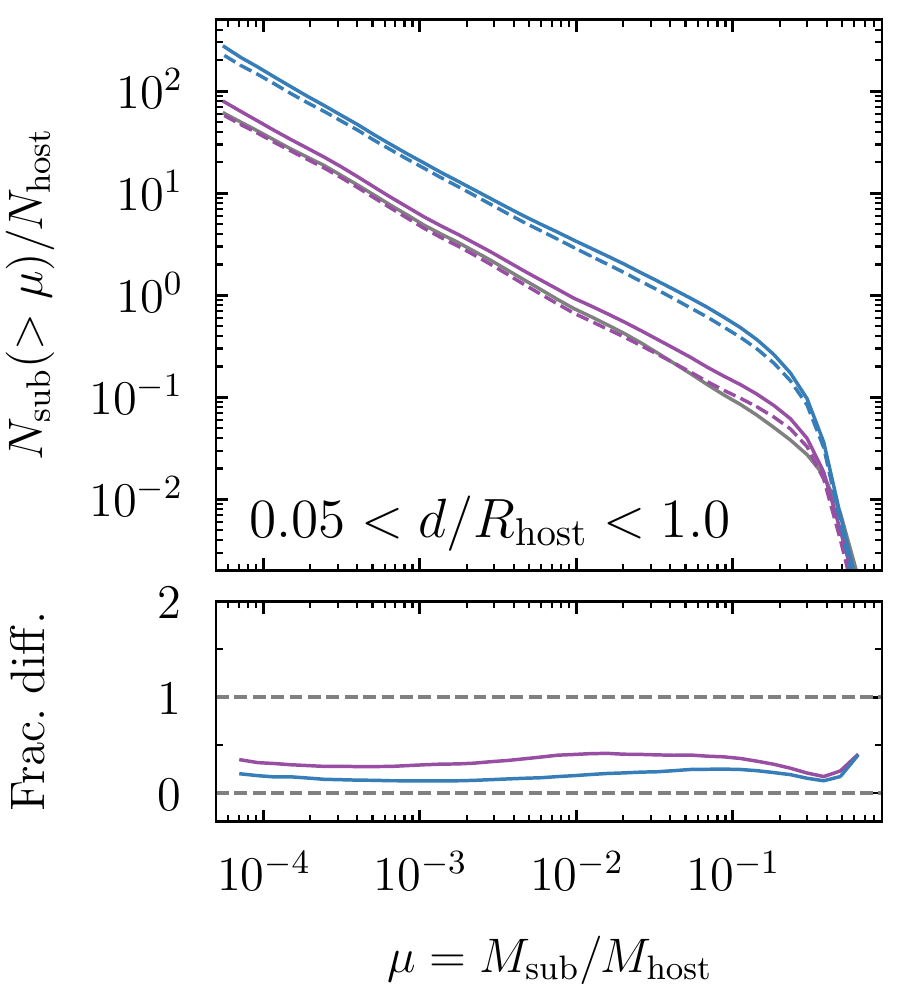}
\includegraphics[trim =  19mm 1mm 1mm 1mm, clip, scale=0.75]{\figdir/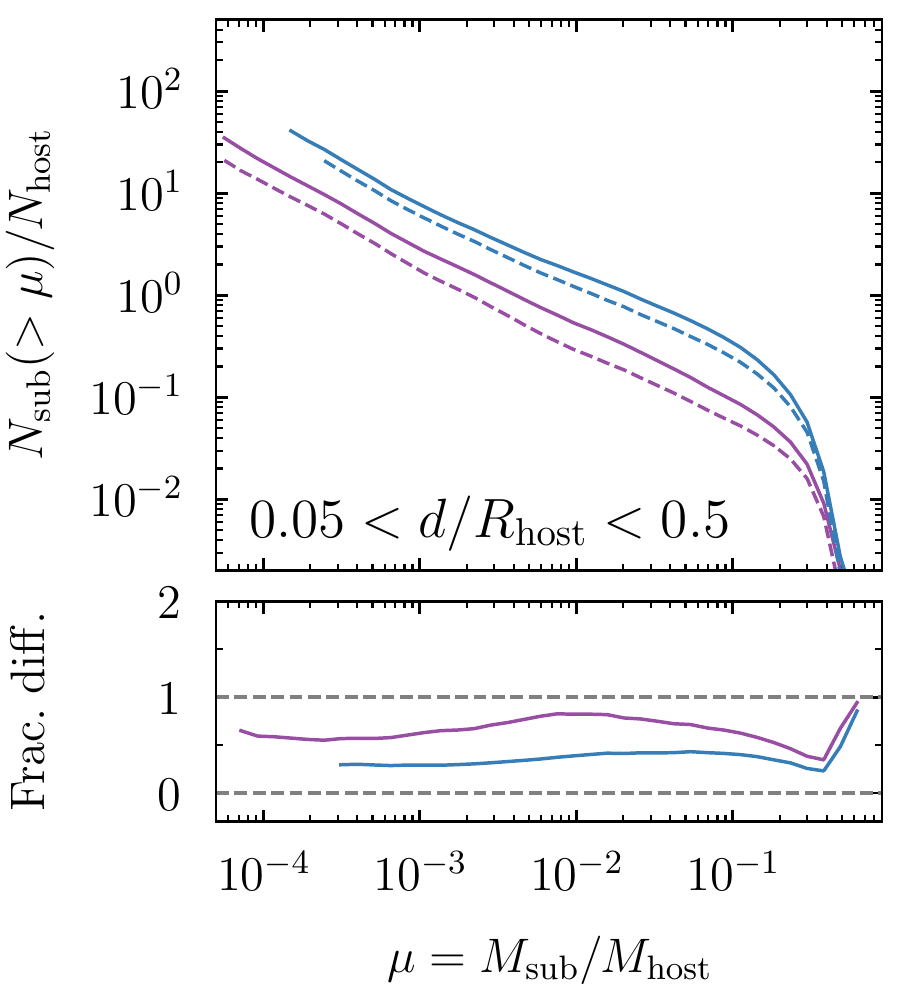}
\includegraphics[trim =  20mm 1mm 1mm 1mm, clip, scale=0.75]{\figdir/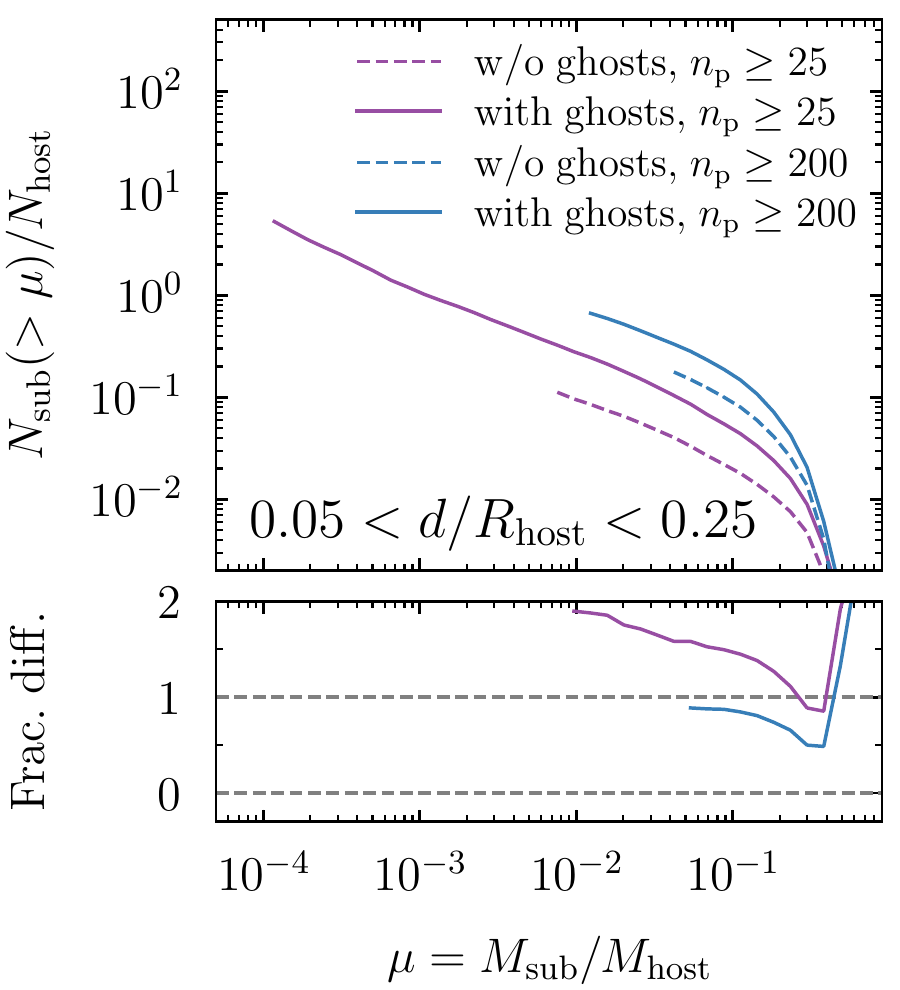}
\caption{Cumulative subhalo mass functions per host for the co-added simulations in the \wmap sample at $z = 0$. We select host haloes with $3.4 \times 10^{13} < \mvir < 1.4\times 10^{14} \msunh$ and sufficient numbers of particles to resolve subhaloes of a given mass ratio $\mu$ with at least $25$ (purple) or $200$ particles (blue). The latter requirement is sufficient for the mass functions to converge between simulations. Adding ghosts (solid lines) significantly increases the abundance of subhaloes. We exclude backsplash haloes by limiting the radius to $\rvir$ or fractions thereof, and we excise ghosts in the innermost 5\% of $\rvir$ to avoid a dependence of the SHMF on the algorithm for ghost termination. The solid gray line in the left panel corresponds to the purple dashed line but using bound masses according to \rockstar instead of tracer masses; the differences are negligible.}
\label{fig:shmf}
\end{figure*}

\section{Results II: Impact on simulation predictions}
\label{sec:results2}

The loss of subhaloes affects key predictions of $N$-body simulations. In this section, we investigate to what extent the addition of ghosts can mitigate these errors, specifically subhalo mass functions (Section~\ref{sec:results:shmf}) and correlation functions (Section~\ref{sec:results:corr}). For the sake of simplicity, we bin only by mass or particle number, glossing over the complex dependencies of subhalo populations on host properties such as concentration and formation time \citep[e.g.,][]{gao_04_subs}.

\subsection{Subhalo mass functions}
\label{sec:results:shmf}

Perhaps the most straightforward metric to quantify the abundance of subhaloes is their mass function (SHMF). In this section, we quantify how much the SHMF increases due to the addition of ghosts. We emphasise that the results do depend on our chosen halo finder, our criteria for which particles are subhalo members, how they are removed, and when ghosts end. Fig.~\ref{fig:shmf} shows the cumulative SHMF as a function of the sub-to-host mass ratio $\mu$ at $z = 0$. For hosts, we use the bound-only $\mvir$ and $\rvir$ from \rockstar. For subhaloes, we can measure mass functions either from bound or tracer masses, but the results are virtually identical (left panel of Fig.~\ref{fig:shmf}).

To obtain converged results in different simulations, we impose a minimum number of particles per subhalo (either $25$ or $200$ in Fig.~\ref{fig:shmf}). For a given bin in $\mu$ and the particle mass $m_\rmp$ of a given simulation, we compute a minimum host halo mass. We then construct the differential mass function, $d N_{\rm sub} / d \log_{10}(\mu) / N_{\rm host}$, by counting the number of subhaloes in sufficiently resolved hosts. We obtain the cumulative mass function by adding all differential bins above the minimum resolved $\mu$. The cumulative and differential mass functions end up looking similar because the lowest $\mu$ bins dominate. We add the halo counts from the different simulations in each bin, which means that the contributions are automatically number-weighted. By overplotting mass functions from individual simulation boxes with different mass resolutions, we conclude that a limit of $N \geq 200$ particles per subhalo is sufficient for congruent results. The $N \geq 25$ SHMFs in Fig.~\ref{fig:shmf} are decidedly not converged. They appear lower because, in each simulation, subhaloes near the resolution limit drop out, which lowers the SHMF at a given mass ratio. We show the unconverged SHMFs to highlight the effect of ghosts near the resolution limit, which is slightly stronger than in the converged sample. 

We focus on a particular host mass bin with $3.4 \times 10^{13} < \mvir < 1.4\times 10^{14} \msunh$ (corresponding to peak heights between $1.5$ and $2$). This bin allows for a large range of $\mu$ to be explored with our simulations: too many subhaloes fall below the resolution cutoff for lower host masses, and the number of possible hosts sharply decreases at higher host masses. We additionally filter by the subhaloes' distances from the host centre in units of $d_{\rm sub} / R_{\rm vir,host}$. We consider three radial ranges in Fig.~\ref{fig:shmf}, which highlight that the addition of ghosts leads to the strongest increase in the SHMF near the host centre. We do not count ghosts within the innermost 5\% of $\rvir$ because their continued existence may depend on the criteria for ghost disruption (Section~\ref{sec:method:tracking:end}).

Overall, Fig.~\ref{fig:shmf} demonstrates that ghosts increase the SHMF by significant factors. If we include subhaloes anywhere in the host (left panel), the converged and unconverged SHMFs increase by about 20--30\% and about 40\%, respectively. Within half the host radius, this increase goes up to 25--50\% and a factor of almost two for the unconverged SHMF. Within $0.25$ host radii, ghosts can dominate the sample depending on $\mu$. We find similar trends at higher redshift and for different host masses, although more massive hosts have more subhaloes at fixed $\mu$ (due to the shallower power spectrum at larger scales in \LCDM). If we included ghosts near the very halo centre, the mass functions increase even more. Interestingly, the increase in the number of subhaloes is comparable to the roughly 40\% of satellites that are artificially added as orphans in the \textsc{UniverseMachine} model \citep{behroozi_19}. This agreement is a hint that using ghost-augmented catalogues might drastically reduce the need for orphans.

\subsection{Correlation functions}
\label{sec:results:corr}

\begin{figure}
\centering
\includegraphics[trim =  2mm 15mm 0mm 1mm, clip, scale=0.8]{\figdir/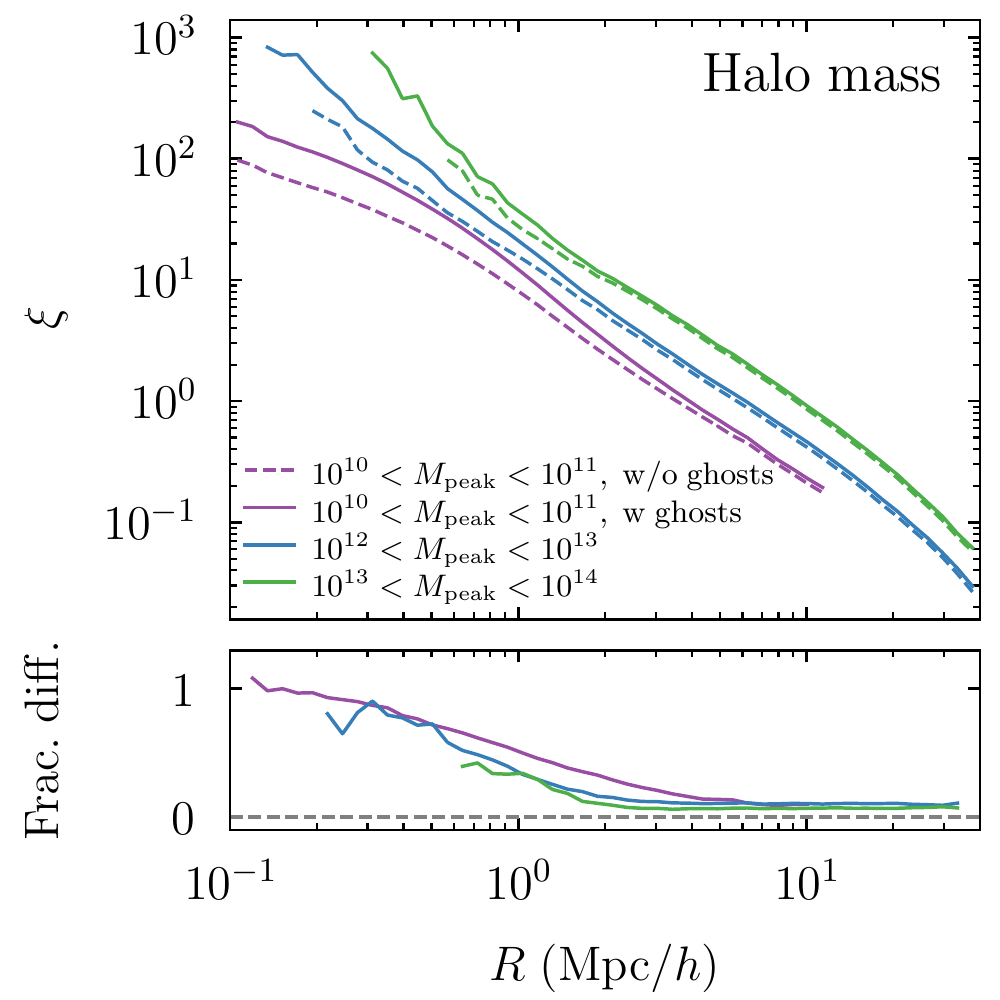}
\includegraphics[trim =  2mm 0mm 0mm 1mm, clip, scale=0.8]{\figdir/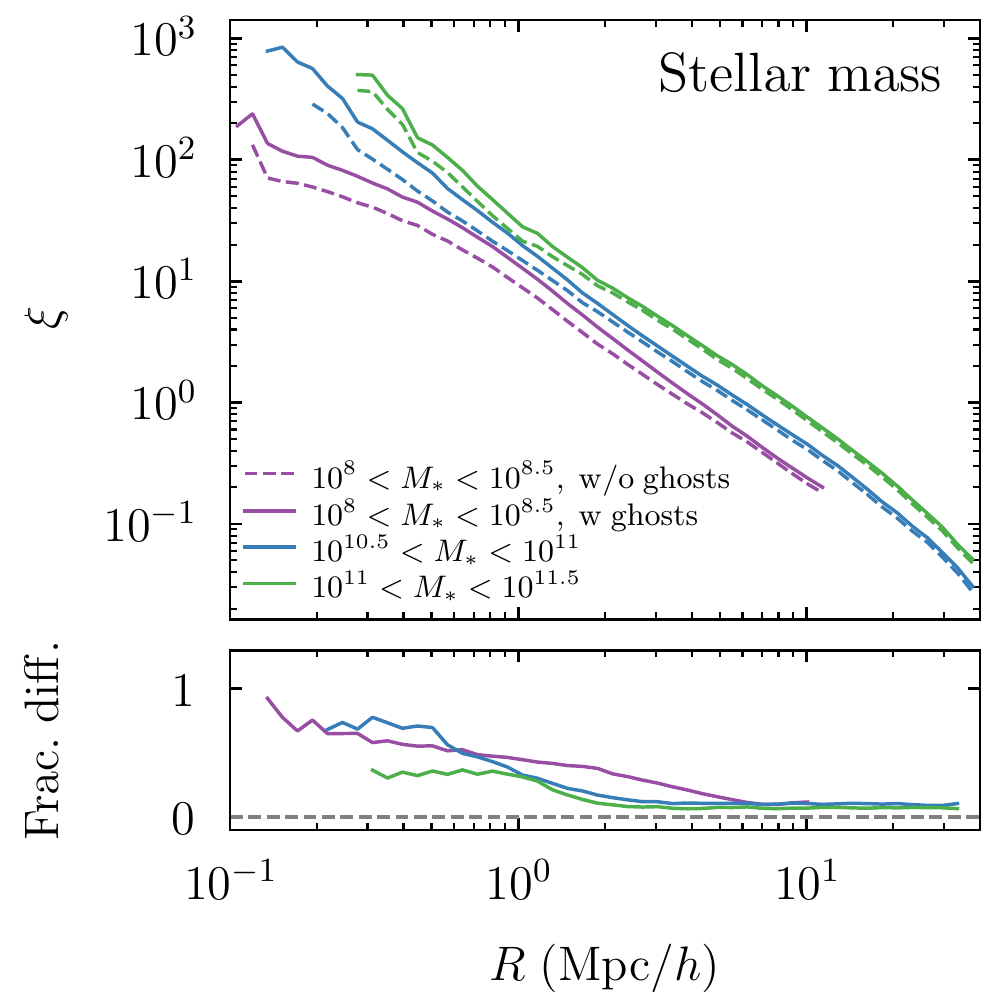}
\caption{Correlation function for halo samples binned by peak mass (top panel) and approximate stellar mass (third panel) at $z = 0$, computing without (dashed) and with ghosts (solid). The smaller panels show the fractional increase in $\xi$ due to ghosts. The lowest mass bin (purple) is resolved only in the smaller simulation boxes. While the absolute level of correlation depends on halo mass, the increase due to ghosts is roughly mass-independent and reaches a factor of two at small radii. In the bottom panels, halo masses were approximately converted to stellar masses using the \textsc{UniverseMachine} model (note that $\mstar$ is given in units of $\msun$ and halo masses in $\msunh$). The trends with stellar mass are similar to those with halo mass.}
\label{fig:corr}
\end{figure}

\def\panelsize{0.76}
\begin{figure*}
\centering
\vspace{0.4cm}
\includegraphics[trim =  20mm 19.5mm 4mm 4mm, clip, scale=\panelsize]{\figdir/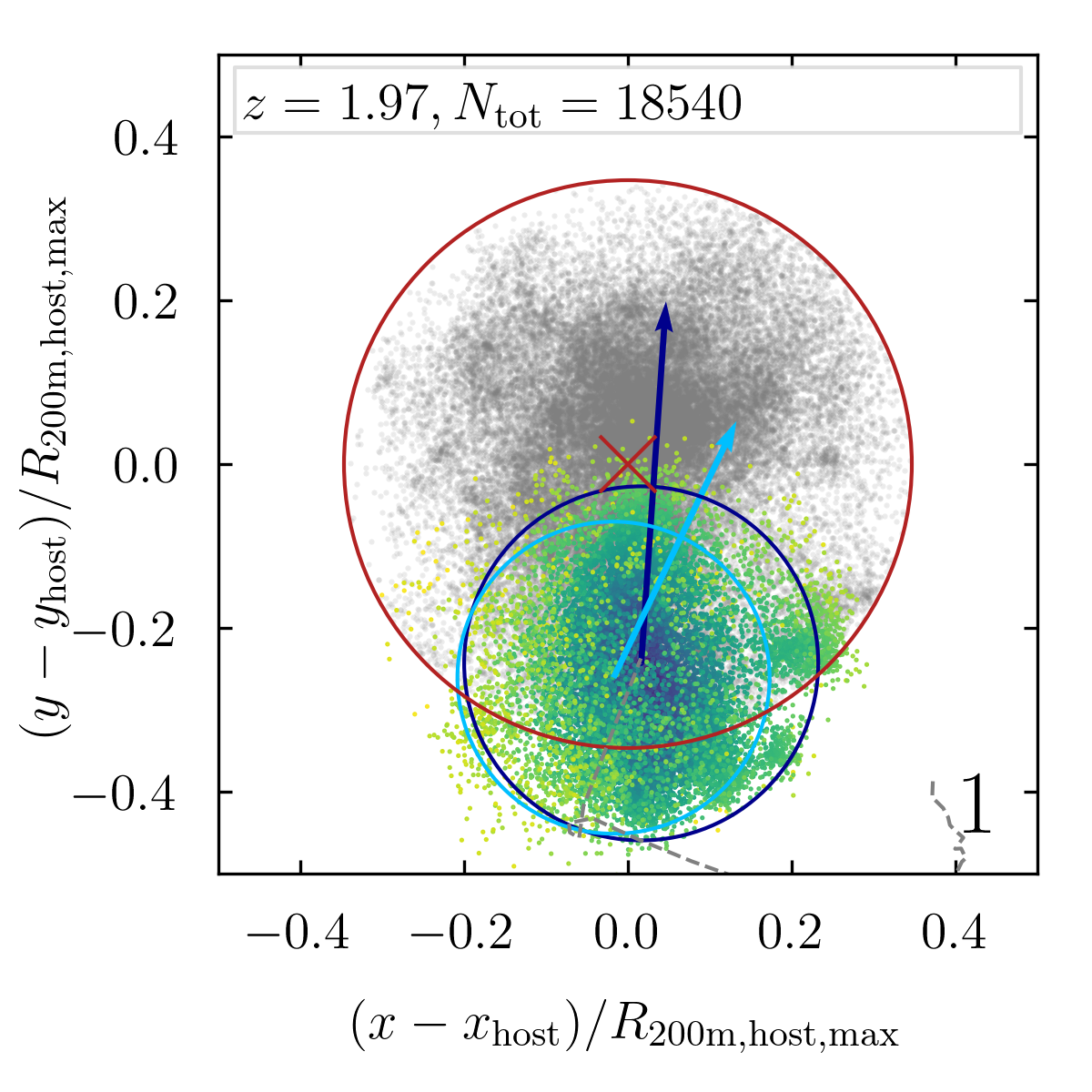}
\includegraphics[trim =  20mm 19.5mm 4mm 4mm, clip, scale=\panelsize]{\figdir/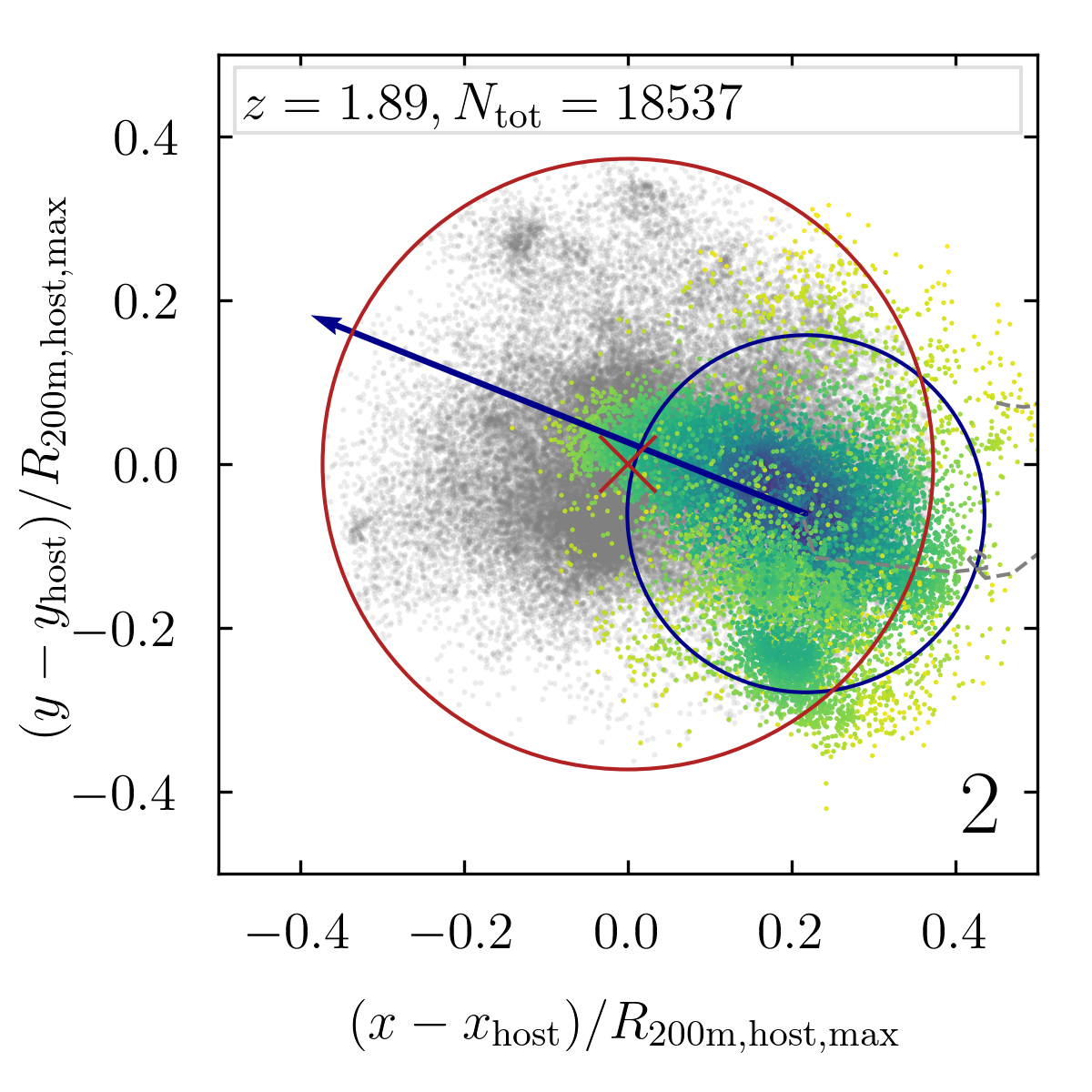}
\includegraphics[trim =  20mm 19.5mm 4mm 4mm, clip, scale=\panelsize]{\figdir/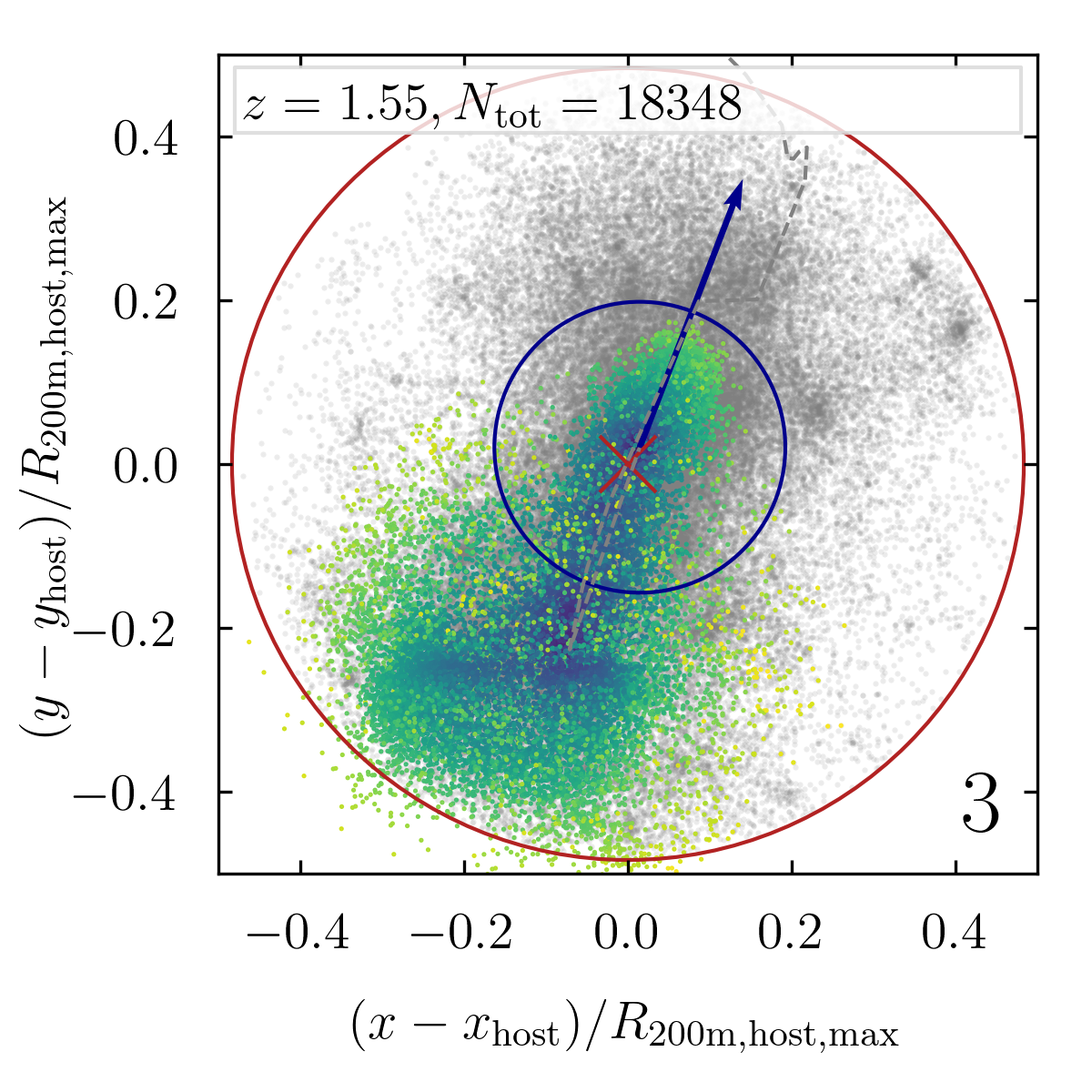}
\includegraphics[trim =  20mm 3mm 4mm 4mm, clip, scale=\panelsize]{\figdir/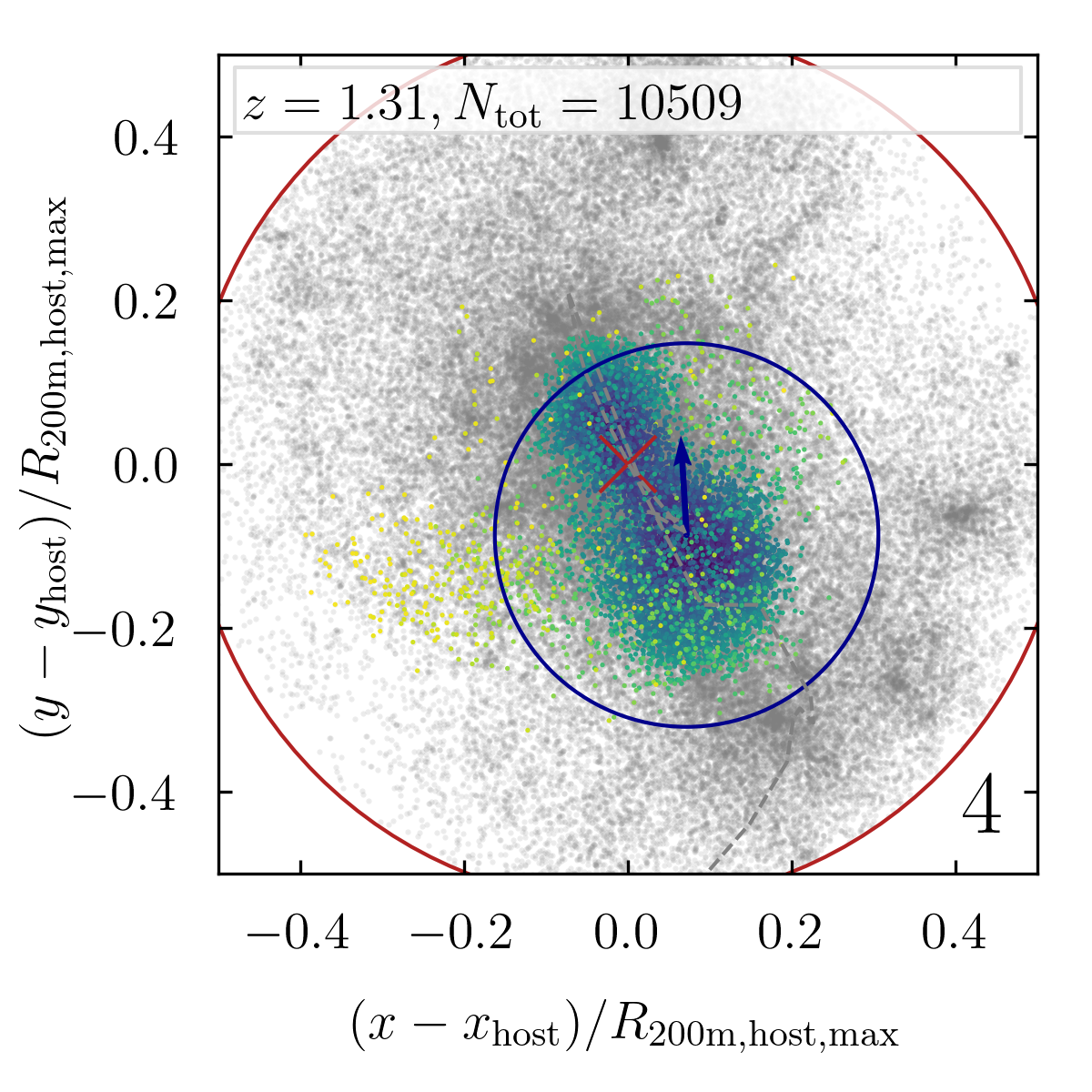}
\includegraphics[trim =  20mm 3mm 4mm 4mm, clip, scale=\panelsize]{\figdir/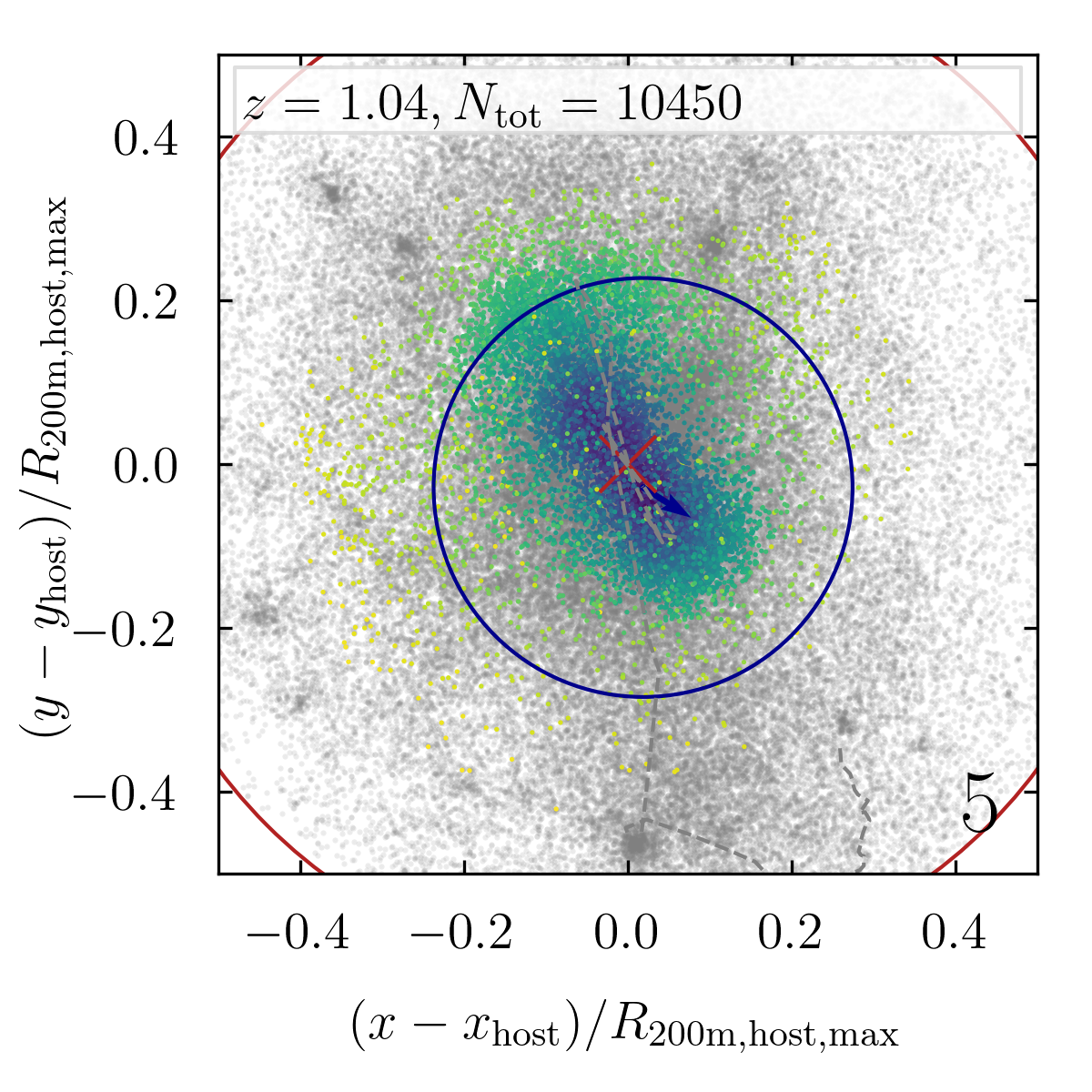}
\includegraphics[trim =  20mm 3mm 4mm 4mm, clip, scale=\panelsize]{\figdir/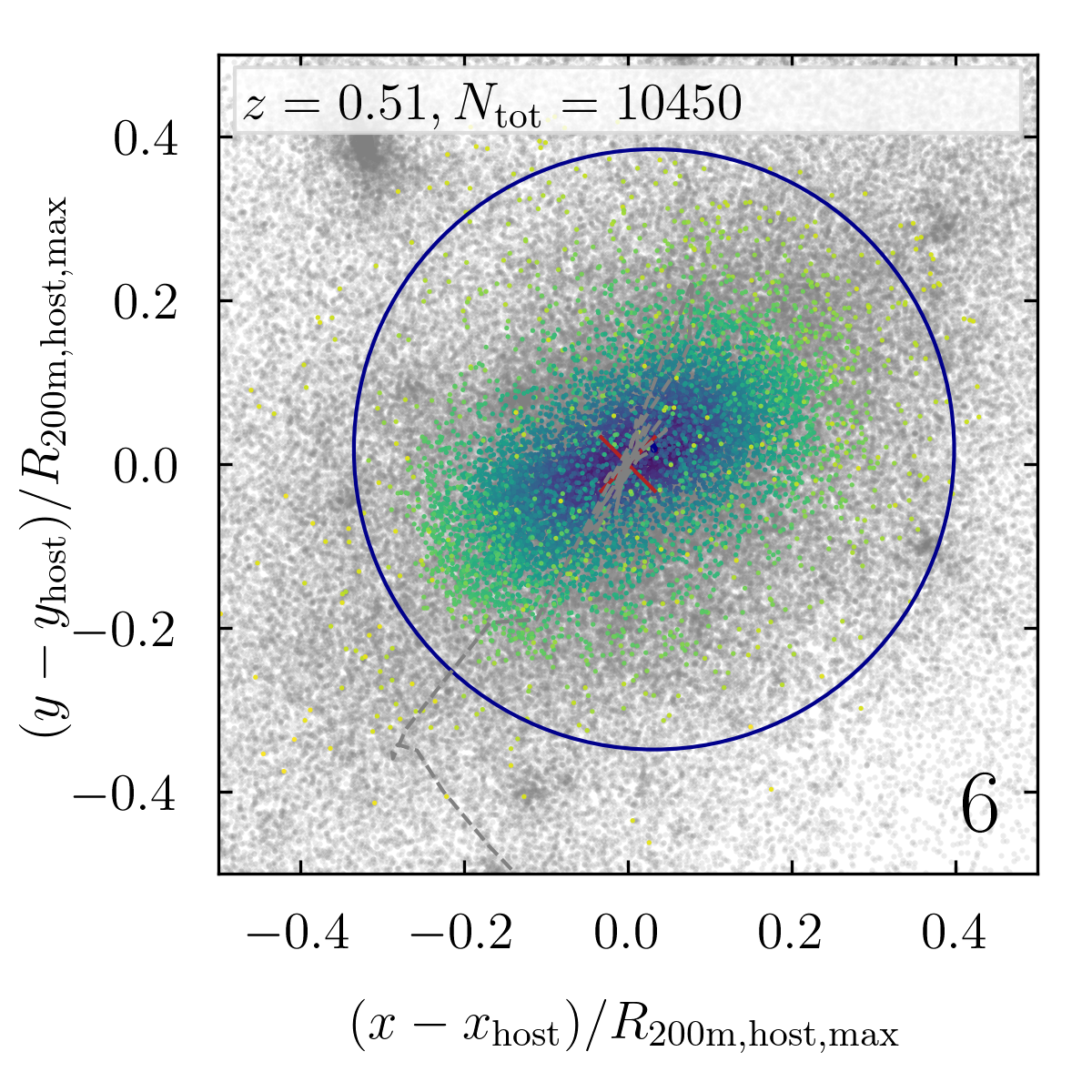}
\caption{Same as Fig.~\ref{fig:ghost1}, but for the largest lost subhalo lost in our test simulation. For clarity, we omit the ghost particles' velocity vectors and reduce the point size. The binding energy (point colour) refers to an arbitrary scale between its minimum and maximum in each panel. The halo finder and \sparta roughly agree on the subhalo properties at infall ($z = 2$, panel 1), but one snapshot later, \rockstar considers the halo to have merged. The reason is the strong tidal disruption experienced by the subhalo, which manifests in an irregular particle distribution (panel 2). Thereafter, the ghost is orbiting with a fairly short period and loses significant material (panel 3). Even after a number of orbits, the ghost still contains about $10^4$ particles and has appreciable velocity with respect to the host centre (panels 4 and 5). Only at $z = 0.5$ does \sparta decide that the ghost has truly merged because its relative velocity has become negligible.}
\label{fig:ghost3}
\end{figure*}

The correlation function of haloes (and thus of galaxies) is one of the key summary statistics of large-scale structure \citep[e.g.,][]{zehavi_02}. Subhalo losses have the pernicious effect of reducing the predicted clustering signal in a scale-dependent manner, with the strongest effect at small scales. We quantify the impact of adding ghosts in Fig.~\ref{fig:corr}, which shows the auto-correlation function of haloes within certain mass ranges. We have selected haloes by the indicated ranges in their peak $\mvir$ ever attained (top panel) and computed the correlation function using the \textsc{CorrFunc} code \citep{sinha_20}. Once again, we have checked that the correlation functions from the individual \wmap simulations agree before combining them in each bin (weighted by the number of halo pairs).

Fig.~\ref{fig:corr} demonstrates that ghosts increase the correlation functions more or less independently of halo mass, while the absolute level of clustering increases with mass. The increase asymptotes to a fixed, relatively small level at large scales but ramps up to a factor of two at megaparsec scales. In practice, correlation functions are generally measured for galaxies selected by stellar mass. We convert peak mass to stellar mass using an approximate stellar mass-halo mass relation based on the \textsc{UniverseMachine} framework \citep{behroozi_19}. We add a log-normal scatter of $0.2$ dex to the stellar mass, although this uncertainty has little effect on the results. When selecting by stellar mass, the difference due to ghosts largely persists. The effect tops out at an increase of about $75\%$ at small scales.

These results highlight that, given the accuracy of modern galaxy surveys, subhalo loss is a roadblock on the way towards accurate predictions from simulations. We leave it to future work to investigate whether ghosts alone can reconcile simulations with observations \citep[see, e.g.,][]{campbell_18}.


\section{Discussion}
\label{sec:discussion}

We have analysed under what conditions phase-space halo finders lose subhaloes, presented an algorithm to keep following such subhaloes, and shown that adding the resulting ghosts has significant effects on basic predictions of $N$-body simulations. In Section~\ref{sec:discussion:major} we further discuss the surprising finding that subhaloes can be lost even though they contain a large number of particles. In Section~\ref{sec:discussion:future} we ponder the numerous limitations of this work and lay out paths towards possible solutions.

\subsection{Loss of subhaloes with many particles}
\label{sec:discussion:major}

When we investigated the particle numbers at which subhaloes are lost (and ghosts created), we found that the tails of the distribution reach seemingly arbitrary numbers (Fig.~\ref{fig:stats}). While we argued that losses are related to the phase-space structure of subhaloes rather than to their particle number, it remains counter-intuitive that such large objects would not show up as an obviously bound entity. 

To understand this question better, Fig.~\ref{fig:ghost3} visualises the evolution of the lost subhalo with the greatest particle number (\num{18000} at infall) in our test simulation. The format follows Figs.~\ref{fig:ghost1} and \ref{fig:ghost2}, but we omit the particles' velocity arrows and decrease the point size to avoid crowding. The particular merger shown in Fig.~\ref{fig:ghost3} occurs at $z \approx 2$ with host and subhalo tracer masses of $\mtom = 8.3 \times 10^{13}$ and $1.7 \times 10^{13}\ \msunh$, respectively. The resulting mass ratio of $0.2$ makes this a ``major'' merger (given our definition of $\mu > 0.1$). Accordingly, strong dynamical friction completes the merger process within a few orbits. If we picked a fractionally smaller subhalo, it would typically spend more time orbiting before its disruption.

Major mergers are known to present challenges to halo finders \citep[e.g.,][]{behroozi_13_unbound, behroozi_15_mergers}. Already at infall, \rockstar and \sparta disagree slightly regarding the direction of the subhalo's velocity vector, but they agree that it has significant, mostly radial velocity. After only one snapshot, \rockstar abandons the subhalo, which is noticeably fractured (panel 2), but \sparta identifies a bound, coherently moving unit (dark blue colours in particle distribution). The ghost remains alive for about $12$ dynamical times and thus orbits the host a number of times. During the first few orbits, it loses a significant amount of mass to tidal disruption (namely, between panels 3 and 4). The merger becomes particularly difficult to follow at this time because the particle distribution appears to split into multiple, somewhat bound centres. Thereafter, the mass remains more or less constant at $\approx \num{10000}$ particles, and it slowly settles into the centre of the host via dynamical friction. However, even at $z = 1$ (panel 5), our algorithm still detects significant, coherent motion with respect to the host centre. At $z = 0.5$, \sparta stops tracking the ghost because its motions have become negligible.

The example of Fig.~\ref{fig:ghost3} demonstrates that the presumed main mechanism for subhalo loss (tidal distortion) can operate regardless of how many particles make up a subhalo. While distorted subhaloes pose a genuinely difficult, and possibly ambiguous, challenge to halo finders, particle tracking is a viable solution to follow such events.

\subsection{Limitations and future directions}
\label{sec:discussion:future}

Given the magnitude of the issues in simulating and detecting subhaloes, we see this work as an exploratory study. We have presented a particular algorithm, evaluated it based on particular parameter choices, and shown that it significantly improves simulation predictions. However, it is beyond the scope of this paper to fully understand the (numerical or physical) disruption of subhaloes, to systematically test the full array of possible algorithmic choices, or to realistically model the resulting galaxy observables.

Throughout the paper, we have developed an understanding of why, when, and where subhaloes are typically lost by the halo finder. While tidal distortion seems to be the likely main culprit, we cannot exclude the existence of other pathological configurations that may pose particular challenges to particle-tracking algorithms. One way to make progress could be to connect the likelihood of loss to the details of a subhalo's orbit, such as its radial and tangential velocities and the host's density profile \citep[e.g.,][]{wetzel_11}.

Perhaps the most important limitation of this work is that we had to, for practical purposes, focus on one particular algorithm to track subhalo particles. While we hope that our choices seem sensible, they are by no means unique. We have made the strong assertion that subhaloes can only lose particles, which may be a poor assumption during major mergers or subhalo-subhalo encounters \citep{behroozi_15_mergers, vandenbosch_17}. \citet{han_18_hbt} did allow the reintegration of subhalo particles that were initially tracked and found this effect to be somewhat significant. We leave a detailed comparison to their HBT+ code to future work. Similarly, it would be valuable to compare our ghosts to orphan models (e.g., those where a single particle is tracked) or the few-particle cores of \citet{heitmann_19}. 

Even within our algorithm, it would be valuable to further study the impact of the free parameters on statistics such as the correlation function. While we have performed numerous test runs and visual inspection to optimise the parameter values, creating full merger trees for a large set of parameters would be a substantial computational effort. One major algorithmic uncertainty is when to end ghosts. We defined reasonable criteria in Section~\ref{sec:method:tracking:end}, but different parameters would lead to shorter or longer ghost lifetimes (although restricted to ghosts near the centres of hosts). In the context of predicting galaxy statistics, the model would have to take into account the baryonic physics of how long satellite galaxies survive relative to their haloes. One path forward could be to train machine learning algorithms on merger outcomes in baryonic simulations \citep[e.g.,][]{petulante_21}.

Our algorithm has the advantage that it generalises to any halo finder, but subhalo identification should not have to rely on post-processing tools such as \sparta. In the long term, integrating particle tracking into a halo finder would allow for easier use and for more self-consistency. For example, one of the main benefits of tracking all particles in subhaloes (as opposed to orphans or cores) is that we can measure their tracer mass. However, this mass definition is currently applied only to subhaloes, which leads to an inconsistency across the infall boundary in time and space (Fig.~\ref{fig:evo}). A comprehensive halo finder based on particle tracking would maintain a set of particles deemed to belong to a certain structure at all times, regardless of host-subhalo relationships. First steps in this direction have already been taken in a number of codes \citep{han_18_hbt, springel_21_gadget4}.


\section{Conclusions}
\label{sec:conclusion}

We have studied the conditions under which friends-of-friends halo finders can lose subhaloes. We have proposed an algorithm to track all of a subhalo's particles and continue it as a ``ghost'' halo if it is dropped from the halo finder's output. We have shown that adding such ghosts to the halo catalogues has order-unity effects on key summary statistics of large-scale structure. While our study focuses on \rockstar, we expect it to generally extend to other halo finders. Our main conclusions are as follows.
\begin{enumerate}

\item Subhaloes can end because they merge into the phase space of their host, because they disrupt numerically or physically, or because the halo finder loses them. Large subhaloes (more than 10\% of the host mass) experience a mixture of physical mergers and loss, whereas the majority of smaller subhaloes disrupt or are eventually lost across a wide range of radii. Such losses are not a numerical error in the underlying simulation and persist to arbitrary numbers of particles.

\item While tidal distortion near the host centre is likely a key driver, the loss of subhaloes does not necessarily occur at their orbital pericentre.

\item We have presented a post-processing algorithm to track all particles in subhaloes, which become so-called ghosts if the subhalo is lost.

\item We introduce tracer masses computed only from tracked particles as a robust mass definition and show that they take on values similar to gravitationally bound masses, although they can contain formally unbound particles.

\item Restoring lost subhaloes to halo catalogues significantly changes basic predictions of $N$-body simulations, including the subhalo mass function (by about 40\%) and the halo correlation function (by up to a factor of two at small scales). The added ghosts should at least partially alleviate the need for artificially adding orphan satellites.

\end{enumerate}
All code and data used in this project have been made publicly available in the hope that they will stimulate further research into the effects of subhalo loss and improved algorithms for halo finding.


\section*{Acknowledgements}

We are grateful to Andrew Hearin, Katrin Heitmann, Fangzhou Jiang, Alexie Leauthaud, Frank van den Bosch, and Risa Wechsler for productive discussions. This research was supported in part by the National Science Foundation under Grant numbers AAG 2206688 and PHY-1748958. Our computations were run on the \textsc{Midway} computing cluster provided by the University of Chicago Research Computing Center as well as the \textsc{DeepThought2} cluster at the University of Maryland. This research made extensive use of the python packages \textsc{NumPy} \citep{code_numpy2}, \textsc{SciPy} \citep{code_scipy}, \textsc{Matplotlib} \citep{code_matplotlib}, and \colossus \citep{diemer_18_colossus}.


\section*{Data Availability}

The \sparta code is publicly available in a BitBucket repository, \href{https://bitbucket.org/bdiemer/sparta}{bitbucket.org/bdiemer/sparta}. An extensive online documentation can be found at \href{https://bdiemer.bitbucket.io/sparta/}{bdiemer.bitbucket.io/sparta}. The \moria catalogue files (which contain ghosts) are available in an hdf5 format at \href{http://erebos.astro.umd.edu/erebos/sparta}{erebos.astro.umd.edu/erebos/moria}. A Python module to read these files is included in the \sparta code. The full particle data for the \erebos $N$-body simulations are too large to be permanently hosted online, but they are available upon request. 


\bibliographystyle{../../../latex/citestyle_mnras.bst}
\bibliography{\includedir/bib_mine.bib,\includedir/bib_general.bib,\includedir/bib_structure.bib,\includedir/bib_galaxies.bib,\includedir/bib_clusters.bib}


\appendix

\section{Identifying subhalo member particles}
\label{sec:app:members}

\def\panelsize{0.69}
\begin{figure*}
\centering
\vspace{0.4cm}
\includegraphics[trim =  0mm 15mm 2mm 2mm, clip, scale=\panelsize]{\figdir/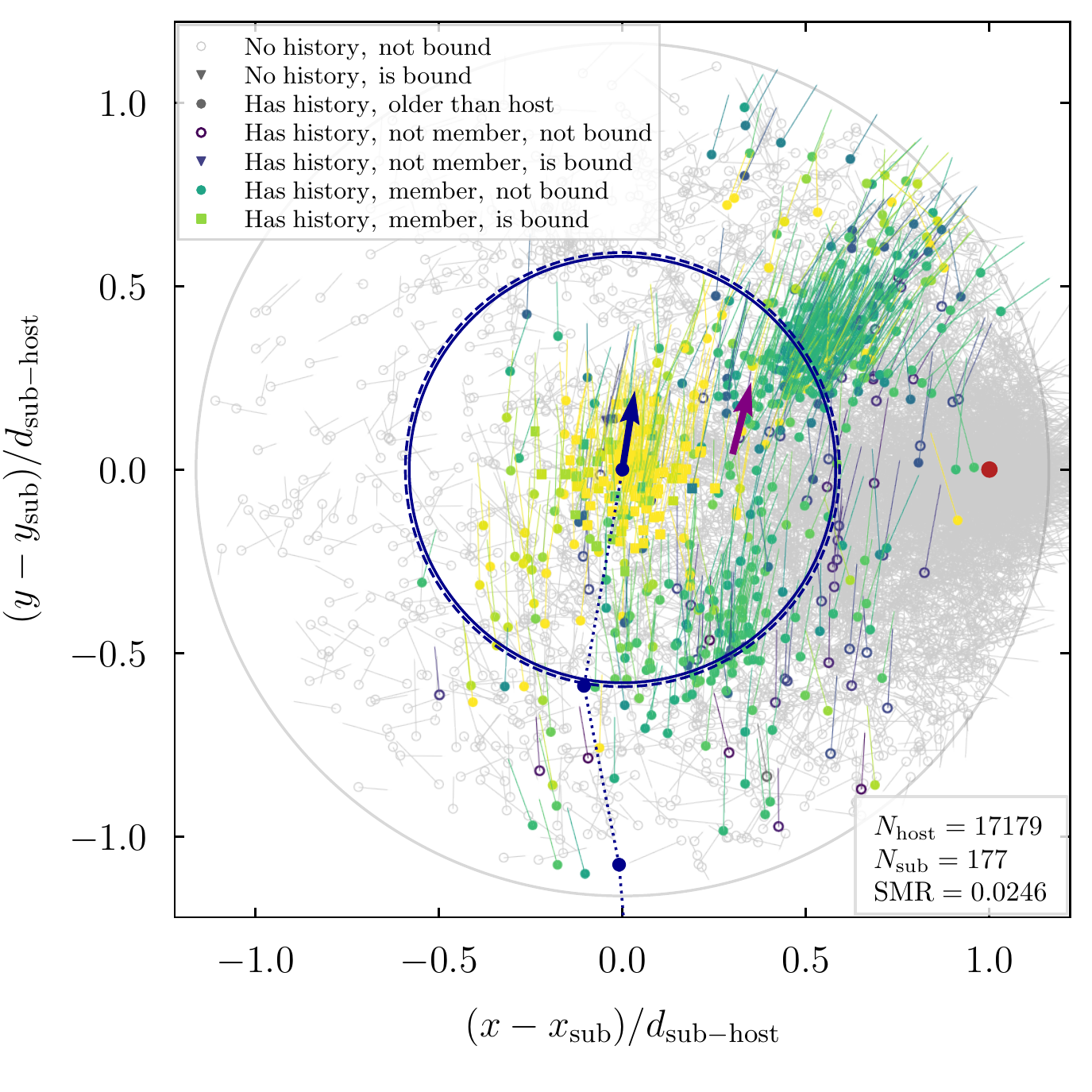}
\includegraphics[trim =  21mm 15mm 2mm 2mm, clip, scale=\panelsize]{\figdir/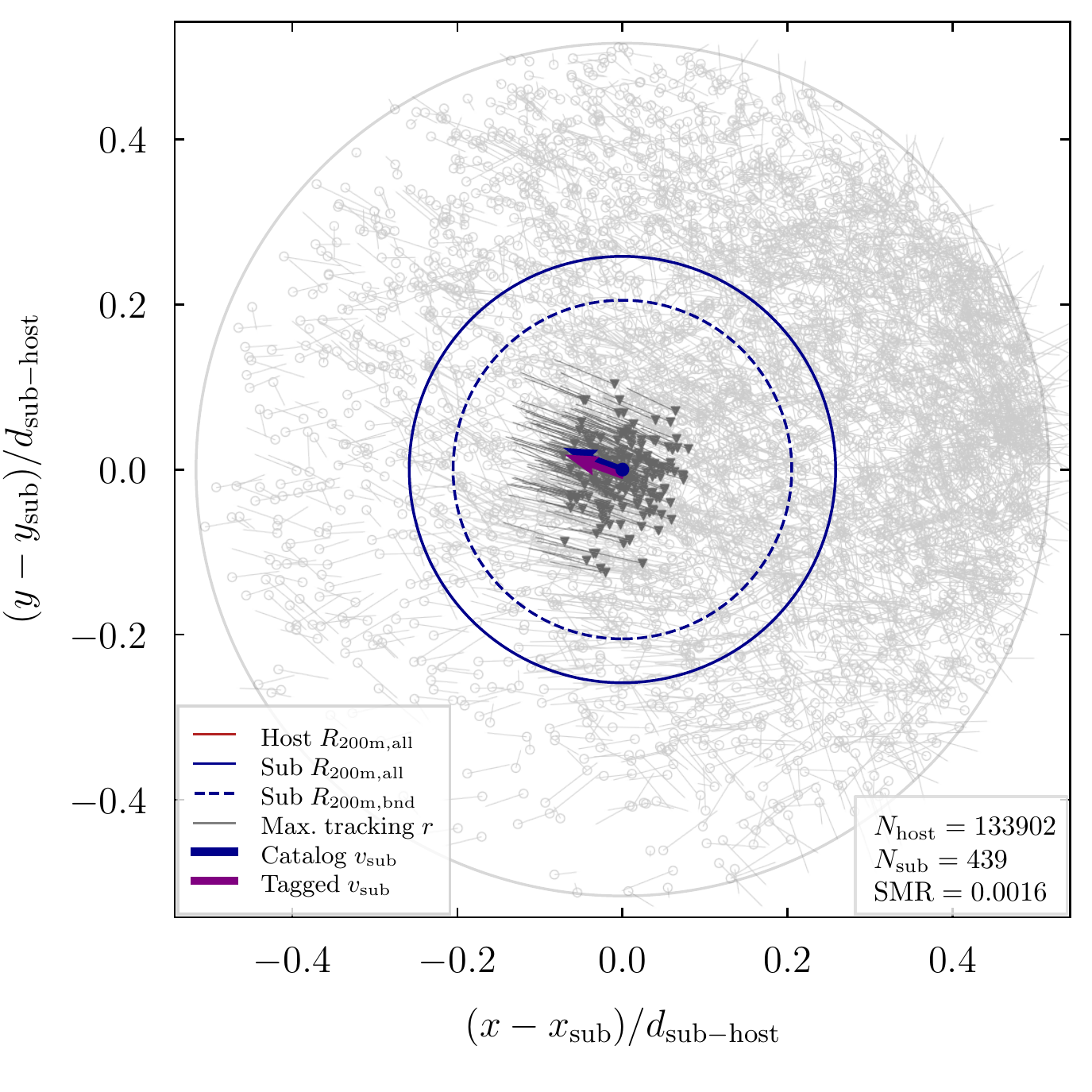}
\includegraphics[trim =  0mm 5mm 2mm 0mm, clip, scale=\panelsize]{\figdir/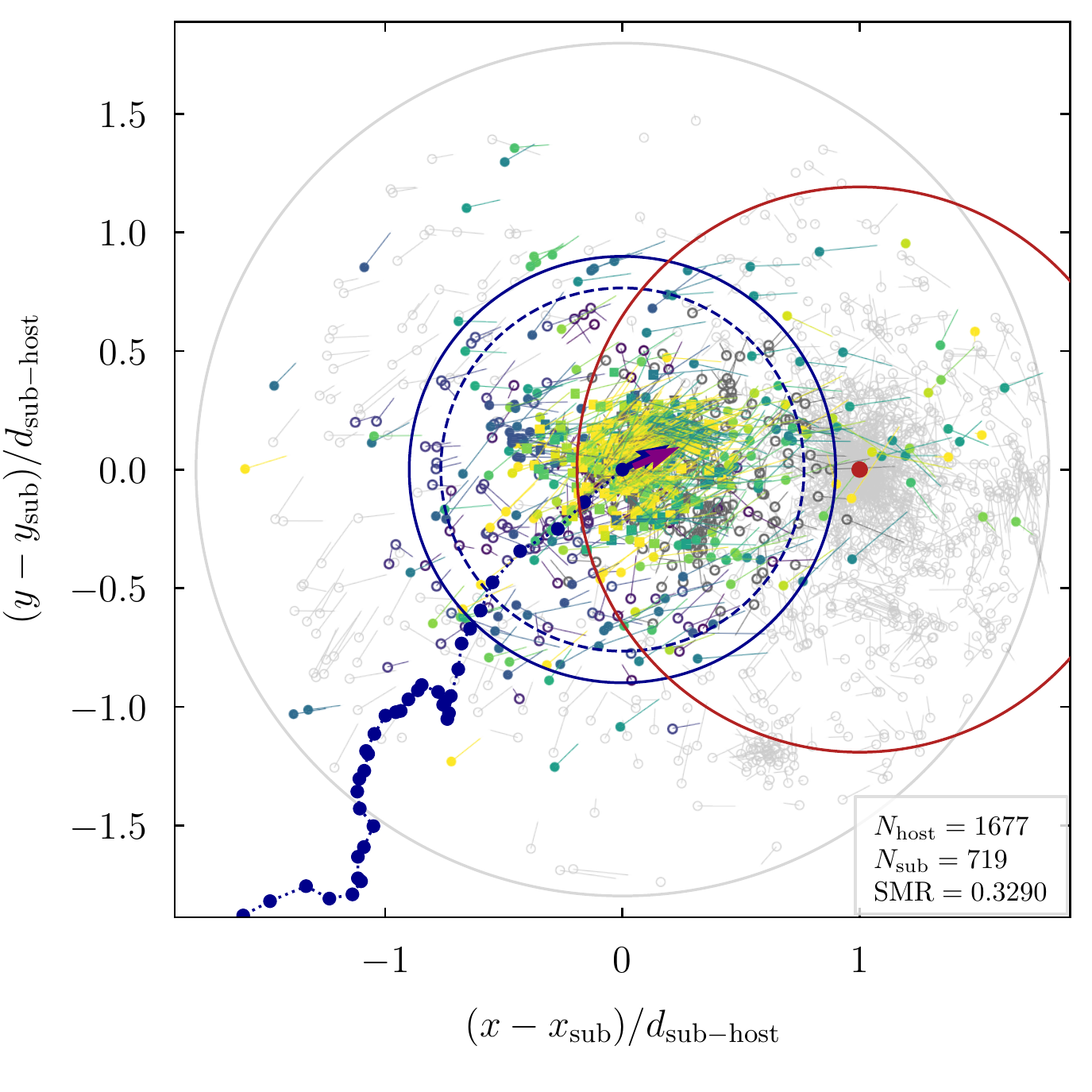}
\includegraphics[trim =  21mm 5mm 2mm 0mm, clip, scale=\panelsize]{\figdir/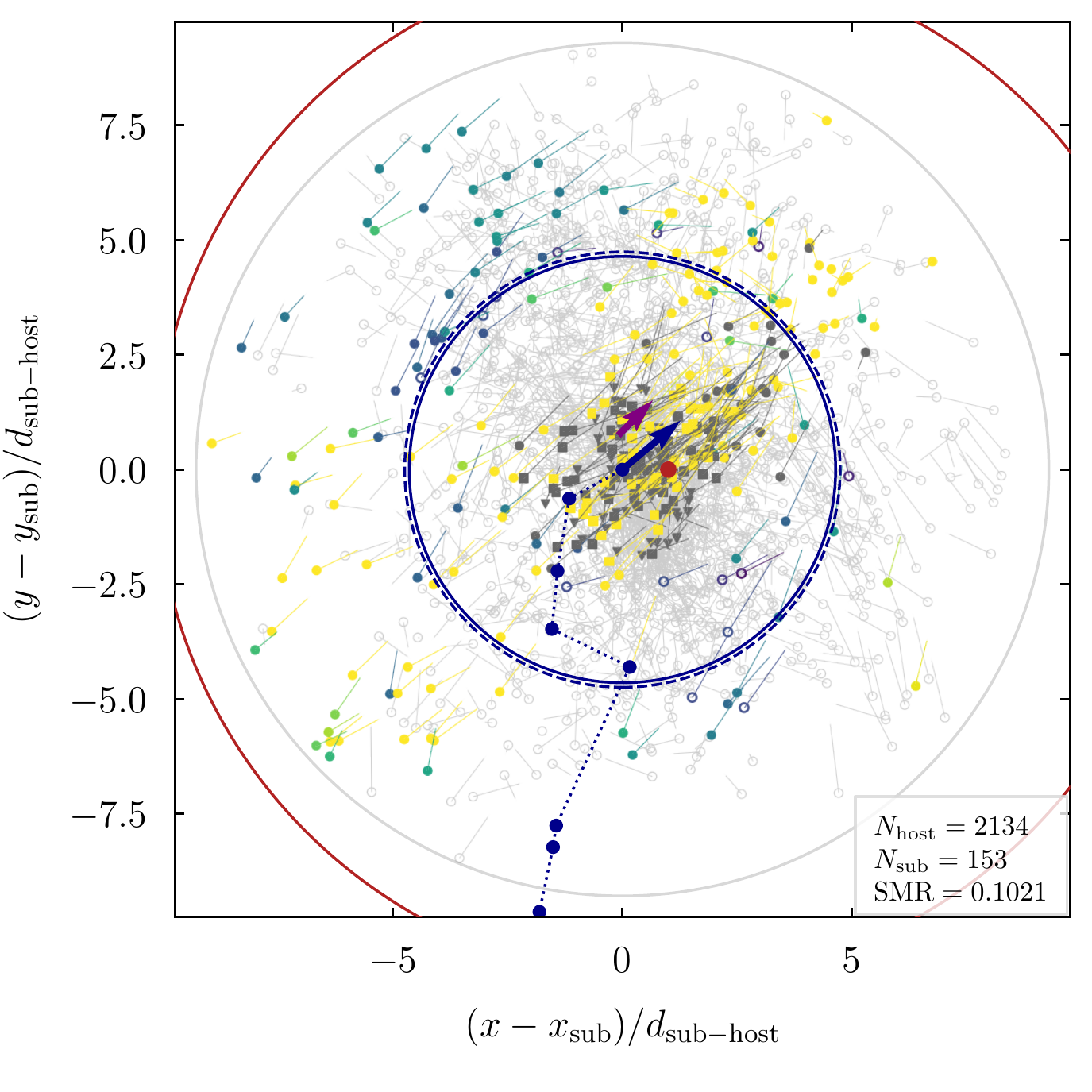}
\caption{Visual illustration of the algorithm to tag particles as subhalo members. Each panel shows a different subhalo in our test simulation at the snapshot after infall, selected to illustrate a few typical scenarios. Each point represents a dark matter particle. The plots are rotated into a frame where the subhalo and host centres are at $(0,0,0)$ and $(1,0,0)$, respectively, and where the velocity of the subhalo lies in the $x$-$y$ plane. The red and blue circles show the host and subhalo radii at the time of infall. The dotted lines show the subhalo's previous trajectory in the host frame. The gray circle at $2 \times R_{\rm 200m,sub}$ is the radius within which we consider particles for tagging; other particles are not shown. The lines next to particles correspond to their velocity in the host frame, scaled such that the host's $V_{\rm 200m}$ corresponds to $1/20$ of the plot width. The dark blue and purple arrows show the overall velocity of the subhalo according to the halo finder and the mean of the tagged particles. Filled symbols denote particles tagged as belonging to the subhalo, empty symbols denote non-tagged particles. Most particles are tagged because they have a record of previously having entered the subhalo; those are shown as coloured symbols, where colour indicates the relative distance from the host where particles first fell into the subhalo (yellow indicating far away and thus members, purple indicating nearby and thus not members). Squares, triangles, and circles indicate whether particles are also gravitationally bound (see legend). {\it Left top:} A subhalo near the host centre has already undergone significant tidal disruption, highlighting the need to tag gravitationally unbound particles outside the subhalo radius. {\it Right top:} This subhalo has no records of prior accretion because it formed just outside the host. Boundness is the only criterion, and the results agree well with \rockstar. {\it Left bottom:} A major merger with a non-trivial distribution of tagged and untagged particles at all radii. {\it Right bottom:} A poorly resolved merger where the subhalo spatially overlaps with the host centre one snapshot after infall. Subhalo membership is determined by a mixture of boundness and particles' histories.}
\label{fig:subtagging}
\end{figure*}

The most important, and arguably arbitrary, step in our algorithm is to decide which particles belong to a subhalo. We determine subhalo membership at infall, i.e., when the subhalo first crosses its $R_{\rm 200m,host}$. At that point, we consider all particles that lie within $2\ R_{\rm 200m,sub}$. Whether or not those particles are deemed to belong to the subhalo is decided by a combination of three user-defined criteria: the time since a particle fell into the subhalo, the distance from the host centre at which this infall happened, and whether the particle is gravitationally bound to the subhalo. We have extensively experimented with the free parameters and with the relative importance of the three criteria. The infall time and infall distance criteria have fairly similar effects. The default criterion used in this work is to require a particle to have entered the subhalo at least $2 R_{\rm 200m,host}$ from the host centre and to ignore the time since infall.

To identify particles that do physically become part of the subhalo after it crosses $2 R_{\rm 200m,host}$, we use a third criterion, namely whether a particle is strongly bound to the subhalo. This final step presents us with significant algorithmic uncertainty \citep[e.g.,][]{behroozi_13_unbound}. We consider particle $i$ to be bound if its gravitational potential exceeds its kinetic energy,
\begin{equation}
\label{eq:bound}
-\Phi_{\rm i} = G \sum_{j \neq i}^{N_{\rm ptl}} \frac{1}{r_{\rm ij} + \epsilon} \geq f_{\rm bnd} \frac{v_{\rm i}^2}{2}  \,,
\end{equation}
where $N_{\rm ptl}$ is the number of particles to consider, $f_{\rm bnd}$ is a threshold for the binding-to-kinetic energy ratio, and $\epsilon$ is the force softening length of the simulation. We solve for the potential using an adapted tree potential algorithm from \rockstar, which constitutes a modified \citet{barnes_86} tree with binary rather than octagonal splitting \citep{behroozi_13_rockstar}. If a tree node is so close that the distribution of the particles within it would cause a significant deviation from its combined contribution, the potential for that node is computed by direct summation. With the error tolerance adopted here and a maximum of $10$ particles per node, the potential of a random distribution of \num{10000} particles is computed with a maximum error of $0.5\%$ and a typical error of $0.2\%$. We have also compared the tree potential with direct summation for some randomly selected subhaloes and found similar accuracy.

The most important free parameter in \eqmn{eq:bound} is which particles we include in the calculation. This freedom means that the question of whether a particle is gravitationally bound is fundamentally ill-posed because it sensitively depends on an initial selection of particles (for example, the FOF group in the case of \rockstar or \subfind). If this selection is generous, e.g., all particles within a few times $\rtom$, virtually all particles will be bound; if only central particles are included, virtually all particles will be unbound. Furthermore, one can do a single pass of \eqmn{eq:bound} or iterate to weed out particles that become unbound as the particle set is diminished. Given that the purpose is to check for particles that were recently added but that are strongly bound, we set $f_{\rm bnd} = 1$ but consider only particles within $0.5\ \rtom$ of the subhalo centre (using the last all-particle $\rtom$ measured before infall). These parameters result in only a fraction of the particles within $\rtom$ being considered bound in most cases. 

Fig.~\ref{fig:subtagging} shows examples of our algorithm to tag subhalo particles. The distributions of tagged and untagged particles are strongly overlapping in space in almost all examples. The all-particle and bound-only $\rtom$ (solid and dashed blue circles) can differ somewhat for haloes that are about to enter a larger halo. We use the all-particle radius for all calculations described in this section because it is more conservative (includes more particles) and because it does not rely on halo finder calculations, but the impact on how many particles are assigned to subhaloes is small.

\bsp
\label{lastpage}
\end{document}